\documentclass[10pt,twocolumn,letterpaper]{article}

\usepackage{zhanggroup}
\usepackage{graphicx}
\usepackage{amsmath}
\usepackage{amssymb}
\usepackage{booktabs}
\usepackage{tikz}
\usepackage{amsfonts}
\usepackage{xspace} 
\usepackage{mathrsfs}
\usepackage{amsmath}
\usepackage{amsthm}
\usepackage{subcaption}
\usepackage{booktabs}
\usepackage{multirow}
\usepackage{graphicx}
\usepackage{url}
\usepackage{caption,subcaption}
\usepackage{bbm}
\usepackage[linesnumbered,ruled]{algorithm2e}
\usepackage{adjustbox}
\usepackage{booktabs}
\usepackage{makecell}
\usepackage{tcolorbox}
\usepackage{multirow}
\usepackage{hyperref}
\usepackage[absolute]{textpos}
\hypersetup{
  colorlinks,
  linkcolor={blue!70!green},
  citecolor={green!70!blue},
  urlcolor={orange!70!red}
}

\newcommand{\mypara}[1]{\noindent{\bf {#1}.} \xspace}

\newcommand{\ContSteal}{{Cont-Steal}\xspace}

\begin{document}

\begin{textblock}{13}(1.5,1)
\centering
To Appear in the IEEE/CVF Conference on Computer Vision and Pattern Recognition, June 2023.
\end{textblock}

\title{\Large \bf Can't Steal? Cont-Steal! Contrastive Stealing Attacks \\ Against Image Encoders}

\date{}

\author{
Zeyang Sha\textsuperscript{1}\ \ \
Xinlei He\textsuperscript{1}\ \ \
Ning Yu\textsuperscript{2}\ \ \
Michael Backes\textsuperscript{1}\ \ \
Yang Zhang\textsuperscript{1}
\\
\\
\textsuperscript{1}\textit{CISPA Helmholtz Center for Information Security}\ \ \ 
\textsuperscript{2}\textit{Salesforce Research}
}

\maketitle

\begin{abstract}

Self-supervised representation learning techniques have been developing rapidly to make full use of unlabeled images.
They encode images into rich features that are oblivious to downstream tasks.
Behind their revolutionary representation power, the requirements for dedicated model designs and a massive amount of computation resources expose image encoders to the risks of potential model stealing attacks - a cheap way to mimic the well-trained encoder performance while circumventing the demanding requirements.
Yet conventional attacks only target supervised classifiers given their predicted labels and/or posteriors, which leaves the vulnerability of unsupervised encoders unexplored.

In this paper, we first instantiate the conventional stealing attacks against encoders and demonstrate their severer vulnerability compared with downstream classifiers.
To better leverage the rich representation of encoders, we further propose \ContSteal, a contrastive-learning-based attack, and validate its improved stealing effectiveness in various experiment settings.
As a takeaway, we appeal to our community's attention to the intellectual property protection of representation learning techniques, especially to the defenses against encoder stealing attacks like ours.
\footnote{See our code in \url{https://github.com/zeyangsha/Cont-Steal}.}

\end{abstract}

\section{Introduction}
\label{sec:introduction}

Recent years have witnessed the great success of applying deep learning (DL) to computer vision tasks.
Different from supervised DL models, self-supervised learning which transforms unlabeled data samples into rich representations, has gained more and more popularity.

\begin{figure}[!t]
\includegraphics[width=0.95\linewidth]{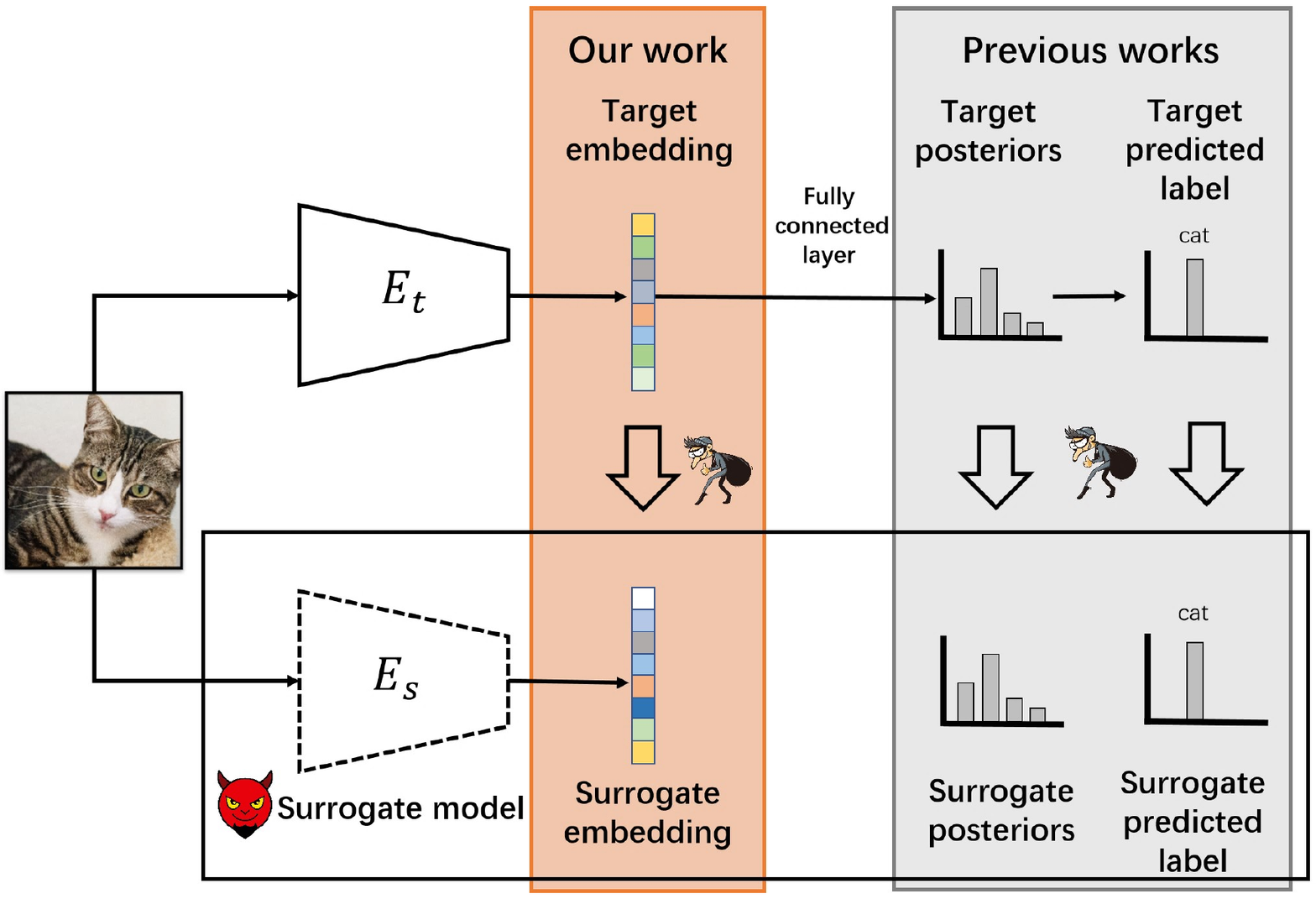}
\caption{Model stealing attacks against classifiers (previous) v.s. model stealing attacks against encoders (ours). Previous works aim to steal a whole classifier using the predicted label or posteriors of a target model. In our work, we aim to steal the target encoder using its embeddings. The target encoder ($E_t$) is pre-trained and fixed, as shown in the solid frame. The surrogate encoder ($E_s$) is trainable by the adversary, as shown in the dashed frame.}
\label{fig:overview}
\end{figure}

Behind its powerful representation, it is non-trivial to obtain a state-of-the-art image encoder.
For instance, SimCLR~\cite{CKNH20} uses 128 TPU v3 cores to pre-train a ResNet-50 encoder with a batch size of 4096.
Therefore, many big companies provide cloud-based self-supervised learning encoder services for users.
For instance, Cohere,\footnote{\url{https://cohere.ai/}} OpenAI,\footnote{\url{https://beta.openai.com/docs/api-reference/embeddings/}} 
and  Clarifai\footnote{\url{https://www.clarifai.com/models/general-image-embedding/}} provide the embedding API of images and texts for commercial usage.
There are many works \cite{LJQG21,CHZ22,JLG22} exploring security issues of encoder-based API.
Therefore, it is a very important and urgent problem.

These kinds of service leave the possibility of \textit{model stealing attacks}~\cite{TZJRR16,WG18,CCGJY18,OSF19,JCBKP20,KTPPI20,WYPY20,SHHZ22}.
In these attacks, the adversary aims to steal the parameters or functionalities of target models with only query access to them.
A successful model stealing attack does not only threaten the intellectual property of the target model, but also serves as a stepping stone for further attacks such as adversarial examples~\cite{BCMNSLGR13,GSS15,CW17,PMJFCS16,TKPGBM17,WWTDLZ19}, backdoor attacks~\cite{SBZ20,CSBMSWZ21,JLG22,SHBHZ22}, and membership inference attacks~\cite{SSSS17,NSH18,NSH19,SZHBFB19,SDSOJ19,SM21,HZ21,HWWBSZ21,HYYBGC21,LJQG21}.
So far, model stealing attacks concentrate on the supervised classifiers, i.e., the model responses are prediction posteriors or labels for a specific downstream task.
The vulnerability of unsupervised image encoders is unfortunately unexplored.

\mypara{Our Work} 
To fill this gap, we pioneer the systematic investigation of model stealing attacks against image encoders.
In this work, the adversary's goal is to steal the functionalities of the target model.
See \autoref{fig:overview} for an overview and a comparison with previous works.
More specifically, we focus on encoders trained by \textit{contrastive learning}, which is one of the most cutting-edge unsupervised representation learning strategies that unleash the information of unlabeled data.

We first instantiate the conventional stealing attacks against encoders and expose their vulnerability.
Given an input image, the target encoder outputs its representation (referred to as embedding).
Similar to model stealing attacks against classifiers, we consider the embedding as the ``ground truth'' label to guide the training procedure of a surrogate encoder on the adversary side.
To measure the effectiveness of stealing attacks, we train an extra linear layer for the target and surrogate encoders towards the same downstream classification task.
Preferably, the surrogate model should achieve both high classification accuracy and high agreement with the target predictions.

We evaluate our attacks on five datasets against four contrastive learning encoders.
Our results demonstrate that the conventional attacks are more effective against encoders than against downstream classifiers.
For instance, when we steal the downstream classifier pre-trained by SimCLR on CIFAR10 (with posteriors as its responses) using STL10 as the surrogate dataset, the adversary can only achieve an accuracy of 0.359.
The accuracy, however, increases to 0.500 instead when we steal its encoder (with the embedding as its response).

Despite its encouraging performance, conventional attacks are not the most suitable ones against encoders.
This is because they treat each image-embedding pair individually without interacting across pairs.
Different embeddings are beneficial to each other as they can serve as anchors to better locate the position of the other embeddings in their space.
Contrastive learning~\cite{OLV18,WXYL18,HFWXG20,CKNH20,GSATRBDPGAPKMV20,CH21,KTWSTIMLK20} is a straightforward idea to achieve this goal.
It is formulated to enforce the embeddings of different augmentations of the same images closer and those of different images further.

In a similar spirit, we propose \ContSteal, a contrastive-learning-based model stealing attack against the encoder.
The goal of \ContSteal is to enforce the surrogate embedding of an image close to its target embedding (defined as a positive pair) and also push away embeddings of different images irrespective of being generated by the target or the surrogate encoders (defined as negative pairs).

The comprehensive evaluation shows that \ContSteal outperforms the conventional model stealing attacks to a large extent.
For instance, when CIFAR10 is the target dataset, \ContSteal achieves an accuracy of 0.714 on the SimCLR encoder pretrained on CIFAR10 with the surrogate dataset and downstream dataset being STL10, while the conventional attack only achieves 0.457 accuracy.
Also, \ContSteal is more query-efficient and dataset-independent (see \autoref{fig:heatmap} for more details).
This is because \ContSteal leverages higher-order information across samples to mimic the functionality of the target encoder.
To mitigate the attacks, we evaluate different defense mechanisms including noise, top-$k$, rounding, and watermark.
Our evaluations show that in most cases, these mechanisms cannot effectively defend against \ContSteal.
Among them, top-$k$ can reduce the attack performance to the largest extent.
However, it also strongly limits the target model's utility.

As a takeaway, our attack further exposes the severe vulnerability of pre-trained encoders.
We appeal to our community's attention to the intellectual property protection of representation learning techniques, especially to the defenses against encoder stealing attacks like ours.

\section{Threat Model}
\label{section:threat_model}

In this work, for the encoder pre-trained with images, we consider image classification as the downstream task.
We refer to the encoder as the target encoder.
Then we treat both the encoder and the linear layer trained for the downstream task together as the target model.
We first introduce the adversary's goal and then characterize different background knowledge that the adversary might have.

\mypara{Adversary's Goal}
Following previous work~\cite{JCBKP20,KTPPI20,SHHZ22}, we taxonomize the adversary's goal into two dimensions, i.e., theft and utility.
The theft adversary aims to build a surrogate encoder that has similar performance on the downstream tasks as the target encoder.
Different from the thief adversary, the goal of the utility adversary is to construct a surrogate encoder that behaves normally on different downstream tasks.
In this case, the surrogate encoder not only faithfully ``copies'' the behaviors of the target encoder, but also serves as a stepping stone to conduct other attacks.

\mypara{Adversary's Background Knowledge}
We categorize the adversary's background knowledge into two dimensions, i.e., the knowledge of the target encoder and the distribution of the surrogate dataset.

Regarding knowledge of the target encoder, we assume that the adversary only has black-box access to it, which means that they can only query the target encoder with an input image and obtain the corresponding output, i.e., the embedding of the input image.

Regarding the surrogate dataset that is used to train the surrogate encoder, we consider two cases.
First, we assume the adversary has the same training dataset as the target encoder.
However, such an assumption may be hard to achieve as such datasets are usually private and protected by the model owner.
In a more extreme case, we assume that the adversary has totally no information about the target encoder's training dataset, which means that they can only use a different distribution dataset to conduct the model stealing attacks.
We later show that the adversary can still launch effective model stealing attacks against the target encoder given a surrogate dataset that is distributed differently compared to the target dataset.

For the model architecture that is used to train the surrogate encoder, we consider two cases.
First, we assume the adversary is aware of the target encoder's architecture and can train the same architecture surrogate encoder.
Then we relax our assumption that the adversary uses different architectures to train the surrogate encoder.
Our evaluation shows that the choice of architecture does not have much impact on the attack performance (see \autoref{table:noise}), which makes the attack more realistic.

Note that we also compare our attacks against the encoders to the traditional model stealing attacks that focus on the whole classifier (which has an encoder and a linear layer).
If the attack targets a whole classifier, we assume the adversary may obtain the posteriors or the predicted label for an input image.

\section{Model Stealing Attacks}
\label{section:contrastive_stealing}

In this section, we first describe the conventional attacks against the encoders.
Then, we propose a novel contrastive stealing framework, \ContSteal, to steal the encoders more effectively.

\subsection{Conventional Attacks Against Encoders}
\label{subsection:normal_stealing}

The adversary takes two steps to conduct the model stealing attacks against the target encoder and one step for further evaluation.

\mypara{Obtain the Surrogate Dataset}
To conduct model stealing attacks, the adversary first needs to obtain a surrogate dataset.
Based on the knowledge of the target classifier's training dataset (target dataset), we consider two cases.
If the adversary has full knowledge of the target dataset, they can directly leverage the target dataset itself as the surrogate dataset.
Or the adversary has no knowledge of the target dataset, which means that they can only construct the surrogate dataset, which is distributed differently from the target dataset.

\mypara{Train the Surrogate Encoder}
Slightly different from the classifier, the response of the encoder is an embedding, which is a feature vector.
In this case, the adversary can still leverage a similar loss function to optimize the surrogate encoder, which can be defined as follows:

\begin{equation}
\begin{aligned}
L_{MS} & = \sum_{k=1}^N l(E_T(x_k), E_S(x_k))
\end{aligned}
\end{equation}
where $E_T(\cdot)$/$E_S(\cdot)$ is the target/surrogate encoder, $N$ is the total number of samples on the surrogate dataset, and $l(\cdot)$ is the MSE loss.

\mypara{Apply the Surrogate Encoder to Downstream Tasks}
To evaluate the effectiveness of model stealing attacks against the encoder, the adversary can leverage the same downstream task to both the target and surrogate encoders.
Concretely, the adversary trains an extra linear layer for the target and surrogate encoders, respectively.
Note that we refer to the target/surrogate encoder and the extra linear layer as the target/surrogate classifiers.
Then, the adversary quantifies the attack effectiveness by measuring the performance of the target/surrogate classifier on the downstream tasks.

\subsection{\ContSteal Attacks Against Encoders}
\label{subsection:contrastive_stealing}

To better leverage the rich information from the embeddings, we propose \ContSteal, a contrastive learning-based model stealing attacks against encoders, which leverages contrastive learning to enhance the stealing performance.
Concretely, \ContSteal aims to enforce the surrogate embedding of an image to get close to its target embedding (defined as a positive pair), and also push away embeddings of different images regardless of being generated by the target or the surrogate encoders (defined as negative pairs).
There are three steps for the adversary to conduct contrastive stealing attacks against encoders and one step for further evaluation.

\mypara{Obtain the Surrogate Dataset}
The adversary follows the same strategy as \autoref{subsection:normal_stealing} to obtain the surrogate dataset.

\mypara{Data Augmentation}
Our proposed \ContSteal leverages data augmentation to transform an input image into its two augmented views.
In this paper, we leverage RandAugment~\cite{CZSL20} as the augmentation method, which is made up of a group of advanced augmentation operations.
Concretely, we set $n=2$ and $m=14$ following Cubuk et al.~\cite{CZSL20} where $n$ denotes the number of transformations to a given sample and $m$ represents the magnitude of global distortion.

\mypara{Train the Surrogate Encoder}
Instead of querying the encoders with the original images, the adversary queries the encoders with the augmented views of them.
Concretely, for an input image $x_i$, we generate two augmented views of it, i.e., $\widetilde{x}_{i,s}$ and $\widetilde{x}_{i,t}$, where $\widetilde{x}_{i,s}$/$\widetilde{x}_{i,t}$ is used to query the surrogate/target encoder.
We consider $(\widetilde{x}_{i,s},\widetilde{x}_{j,t})$ as a positive pair if $i=j$, and otherwise a negative pair.

Given a mini-batch of $N$ samples, we generate $N$ augmented views as the input of the target encoder and another $N$ augmented views as the input of the surrogate encoders.
Concretely, the loss of \ContSteal can be formulated as follows:

\begin{small}
\begin{align}
&D^{+}_{encoder}(i) = exp(sim(E_S(\widetilde{x}_{i,s}),E_T(\widetilde{x}_{i,t}))/ \tau)), \\
&D^{-}_{encoder}(i) = \sum_{k=1}^{N} (exp(sim(E_S(\widetilde{x}_{i,s}),E_T(\widetilde{x}_{k,t}))/ \tau)), \\
&D^{-}_{self}(i) = \sum_{k=1}^{N} \mathbbm{1}_{[k \neq i]} (exp(sim(E_S(\widetilde{x}_{i,s}),E_S(\widetilde{x}_{k,s
}))/ \tau)), \\
&l(i) = -log\frac{ D^{+}_{encoder}(i)}{D^{-}_{encoder}(i) +D^{-}_{self}(i)},\\
&L_{\ContSteal} = \frac{\sum_{k=1}^Nl(k)}{N},
\end{align}
\end{small}

where $E_S(\cdot)$ and $E_T(\cdot)$ denotes the surrogate and target encoder, $sim(u,v) = u^Tv/||u||||v||$ represents the cosine similarity between $u$ and $v$, and $\tau$ is parameter to control the temperature.

\begin{figure}[!t]
\centering
\includegraphics[width=0.95\linewidth]{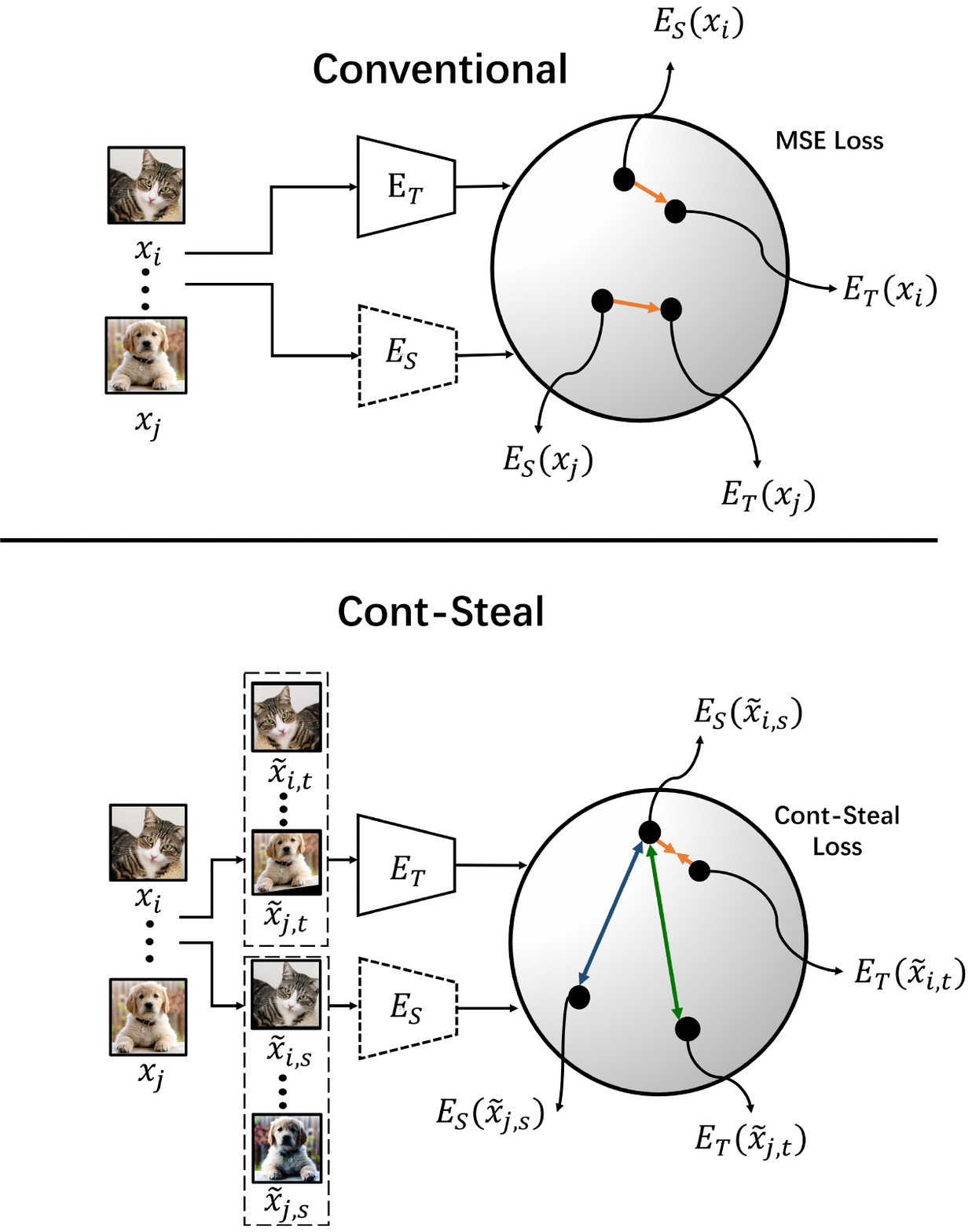}
\caption{Conventional attack (top) vs. \ContSteal (bottom) against encoders.
Conventional attack applies MSE loss to approximate target embeddings for each sample individually. \ContSteal (bottom) introduces data augmentation and interacts across multiple samples: associating target/surrogate embeddings of the same images closer and repulsing those of different images farther away. The target encoder ($E_t$) is pre-trained and fixed, as shown in the solid frame. The surrogate encoder ($E_s$) is trainable by the adversary, as shown in the dashed 
frame.}
\label{fig:contrastive_loss}
\end{figure}

As illustrated in \autoref{fig:contrastive_loss}, the conventional attack treats each embedding individually without interacting across pairs.
However, different embeddings are beneficial to each other as they can serve as anchors to better locate the position of the other embeddings in their space.
\ContSteal maximizes the similarity of embeddings generated from the target and surrogate encoders for a positive pair $(\widetilde{x}_{i,s},\widetilde{x}_{i,t})$ (orange arrows in \autoref{fig:contrastive_loss}).
For the embedding generated from the target and surrogate encoders for any pair $(\widetilde{x}_{i,s},\widetilde{x}_{j,t})$, contrastive stealing aims to make them more distant (green arrows in \autoref{fig:contrastive_loss}).
Besides, as pointed out by Chen et al.~\cite{CKNH20}, contrastive learning benefits larger negative samples.
To achieve this goal, we also consider the embeddings generated from the surrogate encoder for augmented views of different images, i.e., $(\widetilde{x}_{i,s},\widetilde{x}_{j,s})$, as negative pairs minimize their similarity (blue arrows in \autoref{fig:contrastive_loss}).
We later show that such design can enhance the performance of contrastive stealing (see \autoref{table:half-contrastive}).

\mypara{Apply the Surrogate Encoder to Downstream Tasks}
We follow \autoref{subsection:normal_stealing} to evaluate the effectiveness of model stealing on downstream tasks.

\section{Experiments}
\label{section:evaluation}

In this section, we first describe the experimental setup in \autoref{subsection:exp_setup}.
Then we show the performance of the target encoders on the downstream tasks.
Next, we summarize the performance of conventional attacks against classifiers and encoders in \autoref{subsection:encoder}.
Lastly, we evaluate the performance of \ContSteal and conduct ablation studies to demonstrate its effectiveness under different settings in \autoref{subsection:evaluation_contsteal}.

\subsection{Experimental Setup}
\label{subsection:exp_setup}

Our encoders are pre-trained on CIFAR10 ~\cite{CIFAR}, and ImageNet ~\cite{DDSLLF09}.
We use four different kinds of contrastive methods: SimCLR ~\cite{CKNH20}, MoCo ~\cite{HFWXG20}, BYOL ~\cite{GSATRBDPGAPKMV20} and SimSiam ~\cite{CH21} to train a ResNet18 ~\cite{HZRS16} as our target encoders.
Our implementation is based on a PyTorch framework of contrastive learning.\footnote{\url{https://github.com/vturrisi/solo-learn/}}
Then, these well pre-trained encoders will be applied to train downstream classifiers on CIFAR10 ~\cite{CIFAR}, STL10 ~\cite{CNL11}, Fashion-MNIST ~\cite{XRV17}, and SVHN ~\cite{NWCBWN11}.
In the experiments in the model stealing section, we use CIFAR10, STL10, Fashion-MNIST, and SVHN to conduct the attack.

Agreement and accuracy are used as metrics to evaluate the model stealing attack’s performance.
The agreement will evaluate the similarity of surrogate encoders and target encoders in downstream tasks.
The accuracy will evaluate the utility of surrogate models on downstream tasks.
For each metric, a larger value is more desirable.
During the stealing process, we set the batch size as 128 and the learning rate as 0.001.
We show more results on the impact of hyperparameters in Supplementary Material in \autoref{appendix:abaltion}.

\begin{figure}[!t]
\centering
\begin{subfigure}{0.4\columnwidth}
\includegraphics[width=\columnwidth]{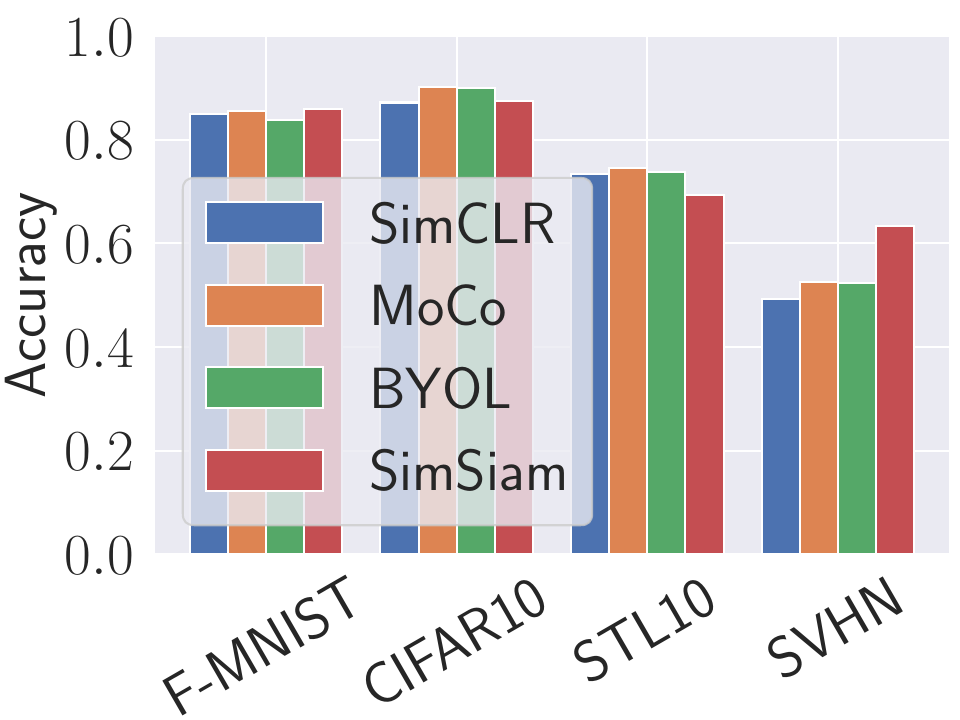}
\caption{CIFAR10}
\label{fig:target_cifar10}
\end{subfigure}
\begin{subfigure}{0.4\columnwidth}
\includegraphics[width=\columnwidth]{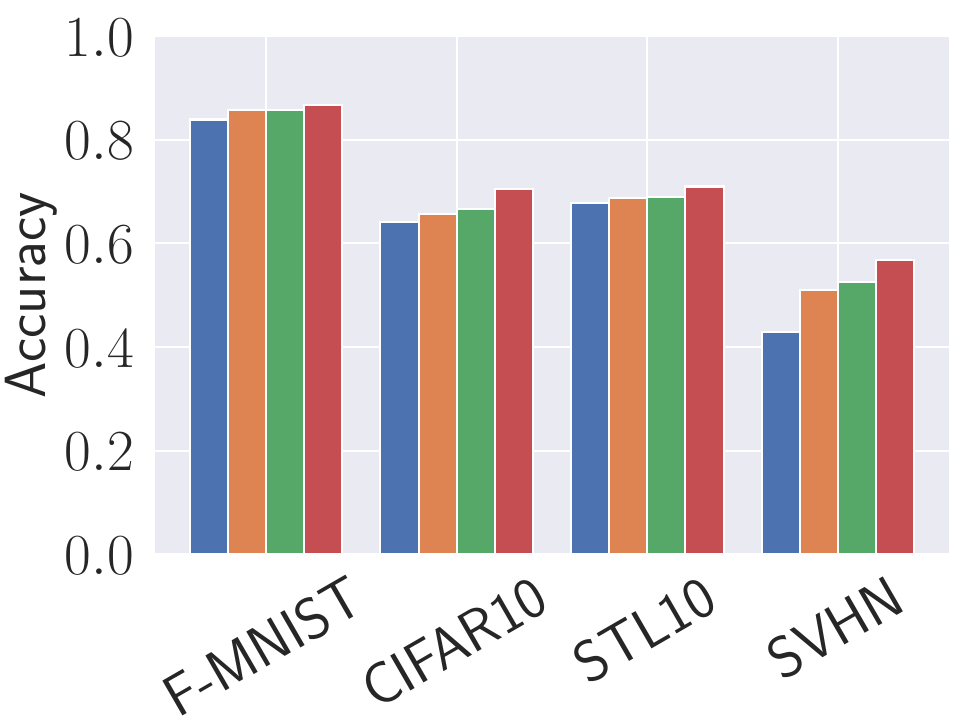}
\caption{ImageNet100}
\label{fig:target_imagenet}
\end{subfigure}
\caption{The performance of target classifiers composed by target encoder and an extra linear layer. 
The encoders are pre-trained on CIFAR10 (a) and ImageNet100 (b).
The x-axis represents different downstream datasets for the target encoder and classifier. 
The y-axis represents the target model's accuracy on downstream tasks.}
\label{fig:target_performance}
\end{figure}

\subsection{Performance of Conventional Attacks}
\label{subsection:encoder}
\begin{figure*}[!t]
\centering
\begin{subfigure}{0.44\columnwidth}
\includegraphics[width=\columnwidth]{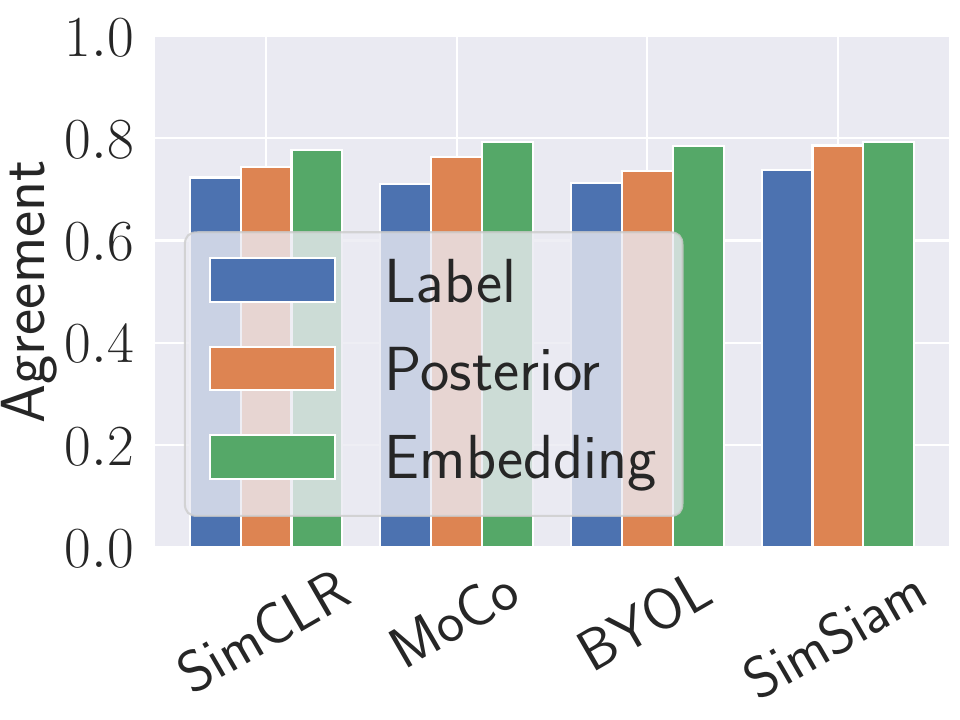}
\caption{CIFAR10}
\label{fig:agreement_normal_target_cifar10_cifar10_cifar10}
\end{subfigure}
\begin{subfigure}{0.44\columnwidth}
\includegraphics[width=\columnwidth]{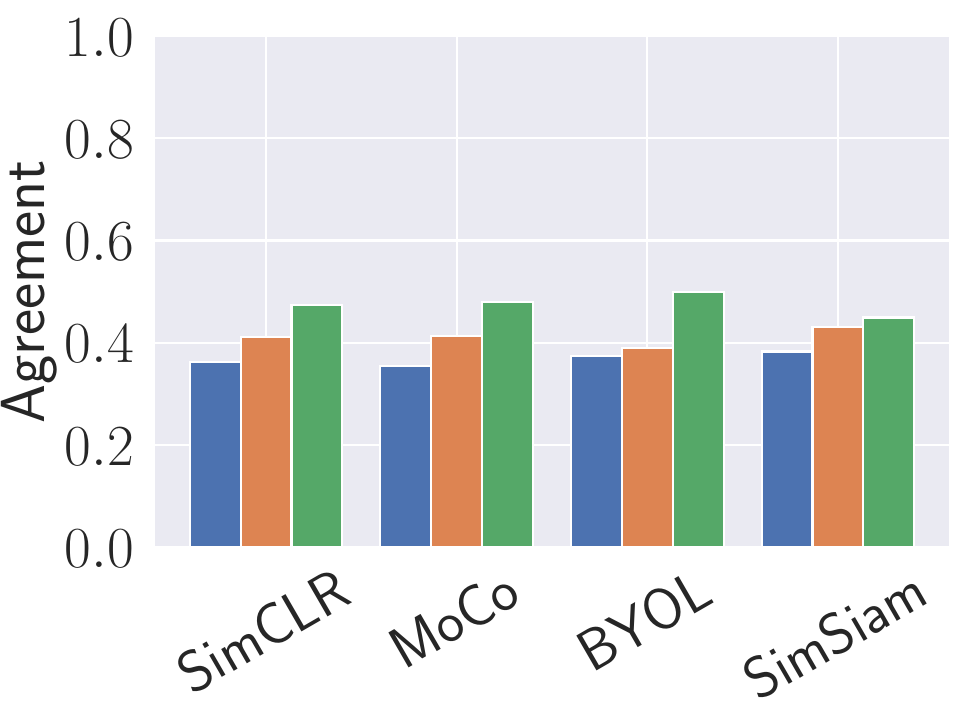}
\caption{STL10}
\label{fig:agreement_normal_target_cifar10_cifar10_stl10}
\end{subfigure}
\begin{subfigure}{0.44\columnwidth}
\includegraphics[width=\columnwidth]{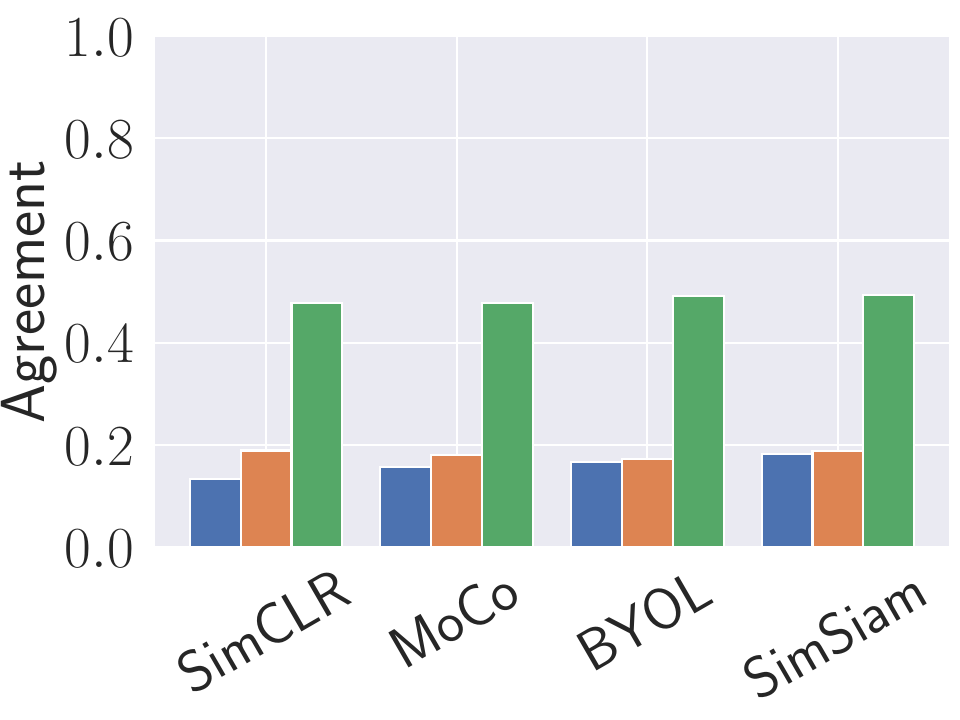}
\caption{F-MNIST}
\label{fig:agreement_normal_target_cifar10_cifar10_mnist}
\end{subfigure}
\begin{subfigure}{0.44\columnwidth}
\includegraphics[width=\columnwidth]{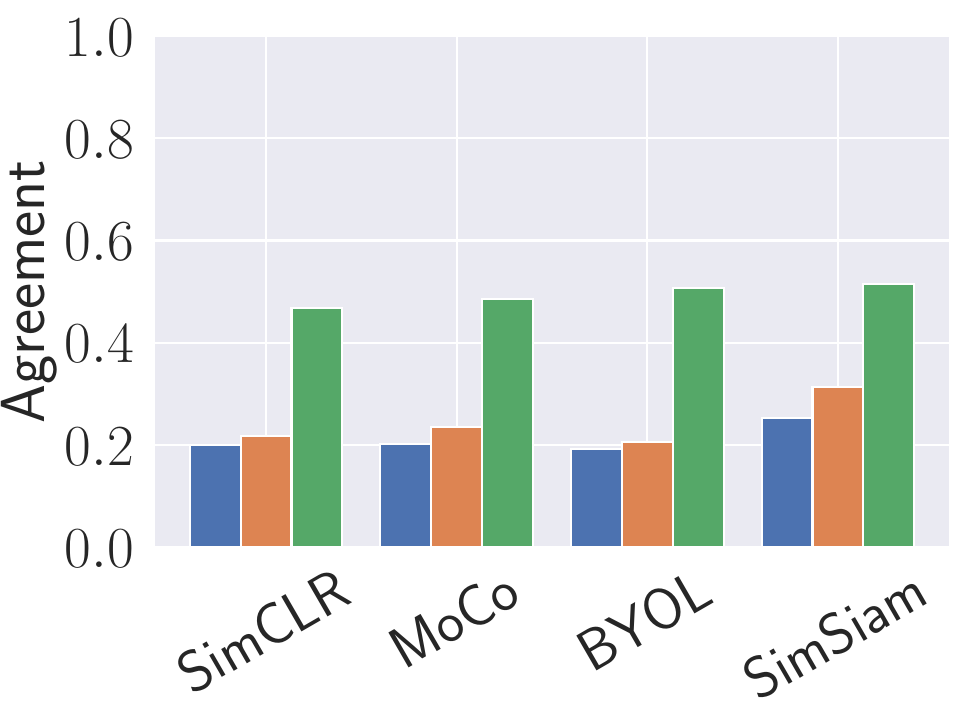}
\caption{SVHN}
\label{fig:agreement_normal_target_cifar10_cifar10_svhn}
\end{subfigure}
\begin{subfigure}{0.44\columnwidth}
\includegraphics[width=\columnwidth]{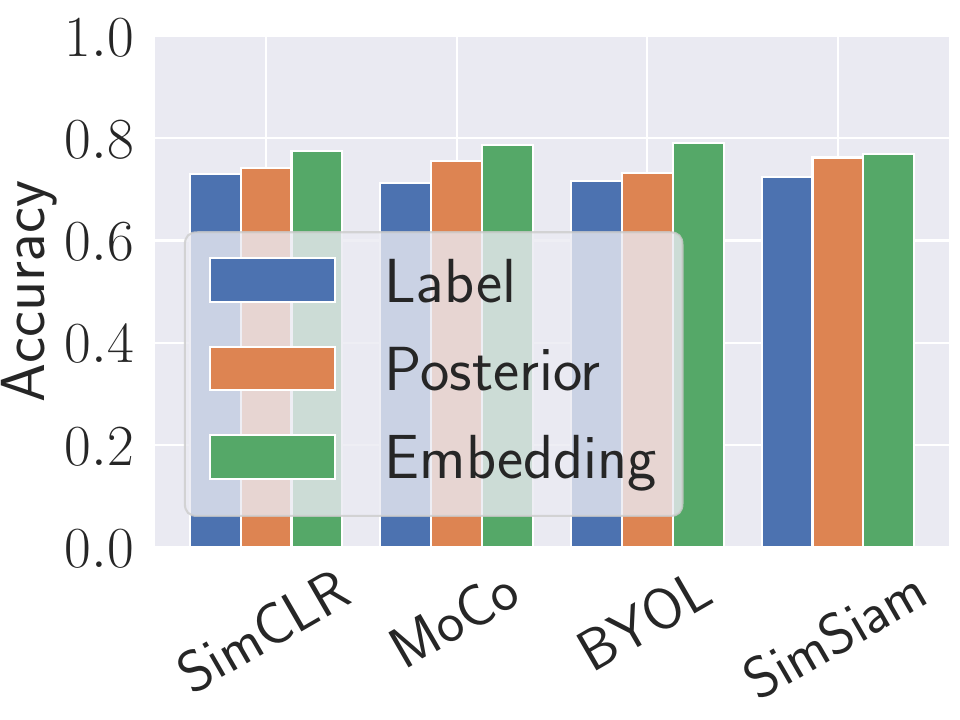}
\caption{CIFAR10}
\label{fig:accuracy_normal_target_cifar10_cifar10_cifar10}
\end{subfigure}
\begin{subfigure}{0.44\columnwidth}
\includegraphics[width=\columnwidth]{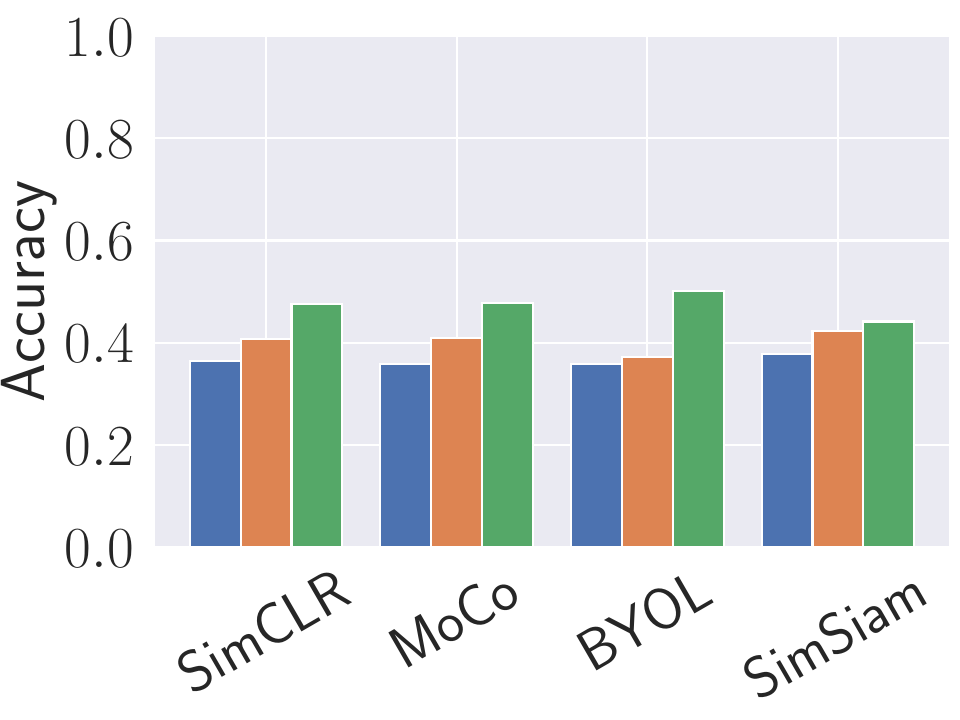}
\caption{STL10}
\label{fig:accuracy_normal_target_cifar10_cifar10_stl10}
\end{subfigure}
\begin{subfigure}{0.44\columnwidth}
\includegraphics[width=\columnwidth]{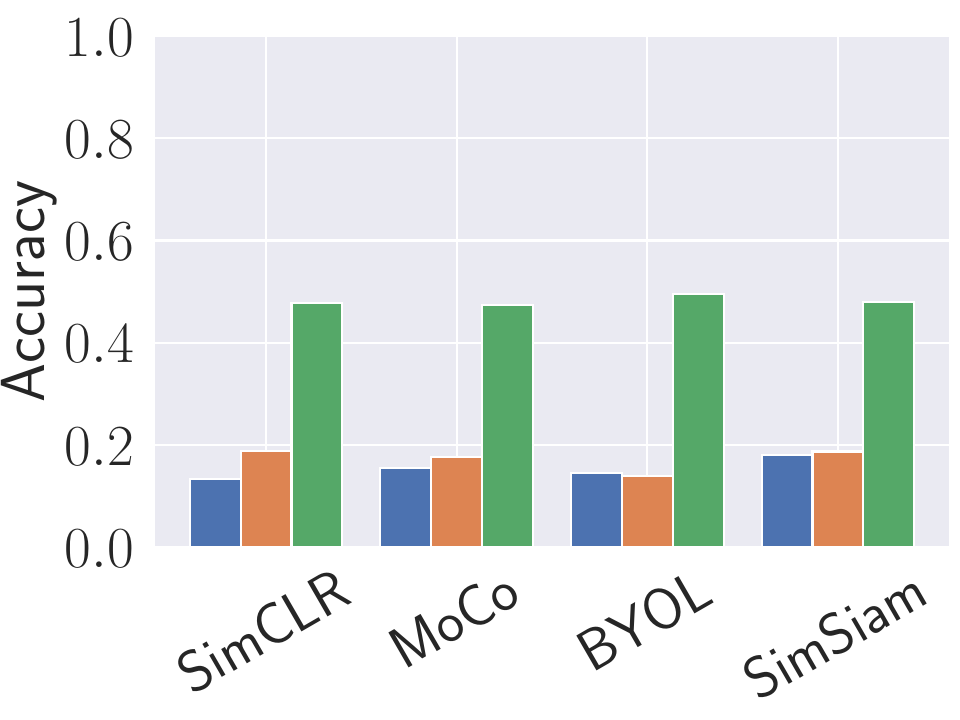}
\caption{F-MNIST}
\label{fig:accuracy_normal_target_cifar10_cifar10_mnist}
\end{subfigure}
\begin{subfigure}{0.44\columnwidth}
\includegraphics[width=\columnwidth]{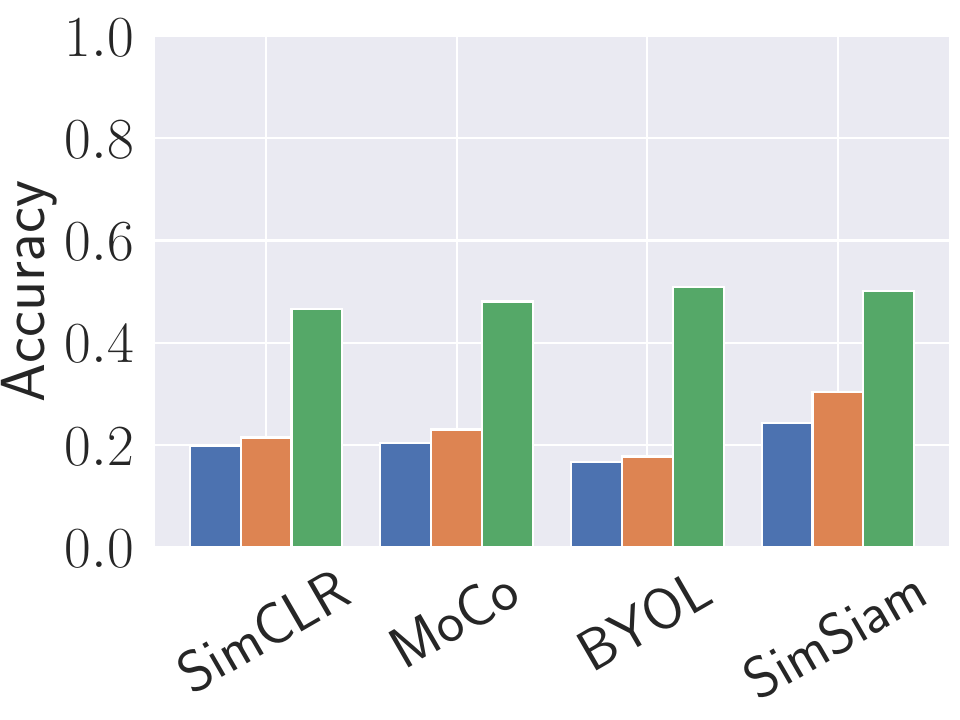}
\caption{SVHN}
\label{fig:accuracy_normal_target_cifar10_cifar10_svhn}
\end{subfigure}
\caption{The performance of model stealing attack against target encoders and downstream classifiers both trained on CIFAR10.
Target models can output predicted labels, posteriors, or embeddings.
The adversary uses CIFAR10, STL10, Fashion-MNIST (F-MNIST), and SVHN to conduct model stealing attacks.
The x-axis represents different kinds of target models.
The first line's y-axis represents the agreement of the model stealing attack. 
The second line's y-axis represents the accuracy of the model stealing attack.}
\label{fig:encoder_cifar10}
\end{figure*}

We first show the target encoder's performance in various downstream tasks.
The results are summarized in \autoref{fig:target_performance}.
We conduct our experiments to explore whether the encoders are more vulnerable to model stealing attacks.
We show our results of target encoders and downstream classifiers both trained on CIFAR10 in \autoref{fig:encoder_cifar10}.
In all cases, the adversary can get better attack performance by stealing encoders rather than classifiers.
This gap becomes especially apparent when the adversary has absolutely no knowledge of the train data.
This is because the rich information in embeddings can better facilitate the learning process of surrogate encoders.
For instance, when the surrogate dataset is CIFAR10 (the same as the target downstream dataset), stealing SimCLR's embeddings can achieve $0.785$ agreement, while stealing predicted labels can achieve $0.712$ agreement.
However, when the surrogate dataset is totally different from the downstream target dataset, e.g., SVHN, stealing embeddings from SimCLR can still achieve $0.507$ agreement while the agreement of stealing predicted labels drops to $0.192$.
We show more results in Supplementary Material \autoref{subsection:appendix_conventional} due to the page limitation.

\begin{figure}[!t]
\begin{subfigure}{\columnwidth}
\centering
\includegraphics[width=0.85\columnwidth]{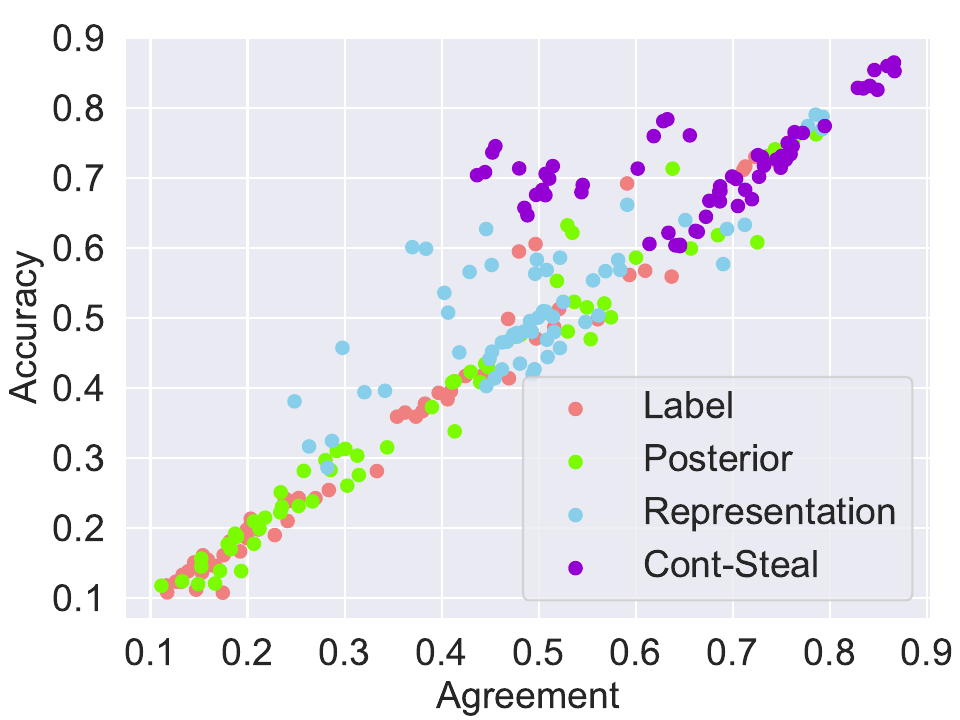}
\end{subfigure}
\caption{The relationship between accuracy and agreement.
The x-axis is the agreement number, and the y-axis is the accuracy number.}
\label{fig:acc_agg_gap}
\end{figure}

We also find that all model stealing attacks' accuracy and agreement are highly correlated.
As shown in \autoref{fig:acc_agg_gap}, the agreement is highly correlated with the accuracy.
This indicates that besides accuracy, the agreement can also be used as a metric to evaluate the performance of model stealing attacks.
We show the result on \autoref{fig:acc_agg_gap}.
It can be obviously seen that agreement is highly related to accuracy.
We use the linear regression method to describe the relationship between agreement and accuracy and find that the relation function is $y=0.940x$.

\begin{figure}[!t]
\centering
\begin{subfigure}{0.3\columnwidth}
\includegraphics[width=\columnwidth]{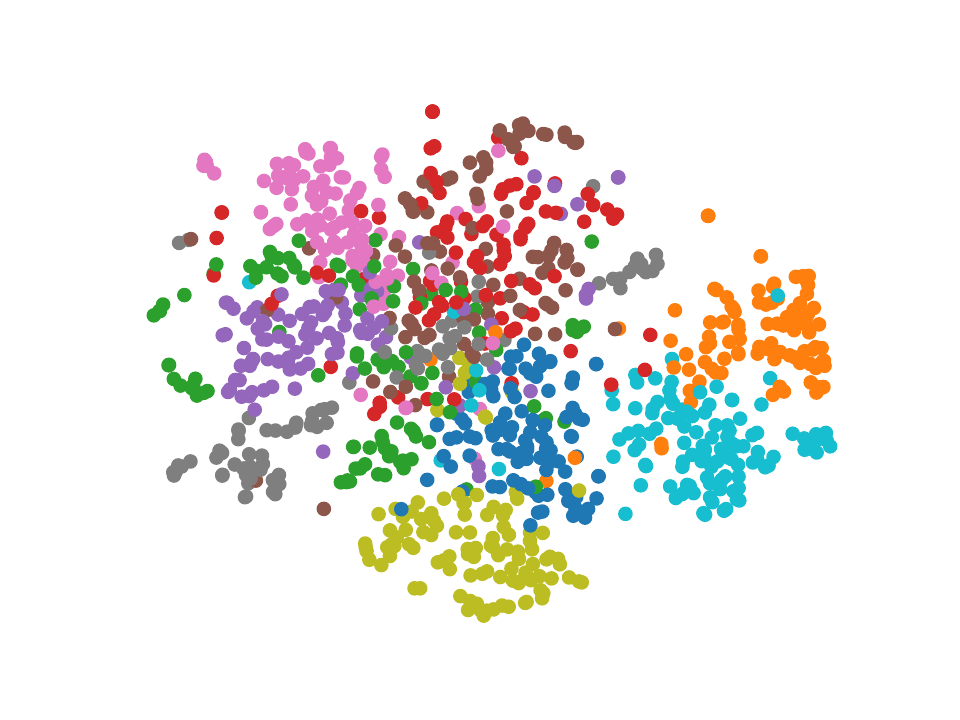}
\caption{Target encoder}
\label{fig:embedding2_target}
\end{subfigure}
\begin{subfigure}{0.3\columnwidth}
\includegraphics[width=\columnwidth]{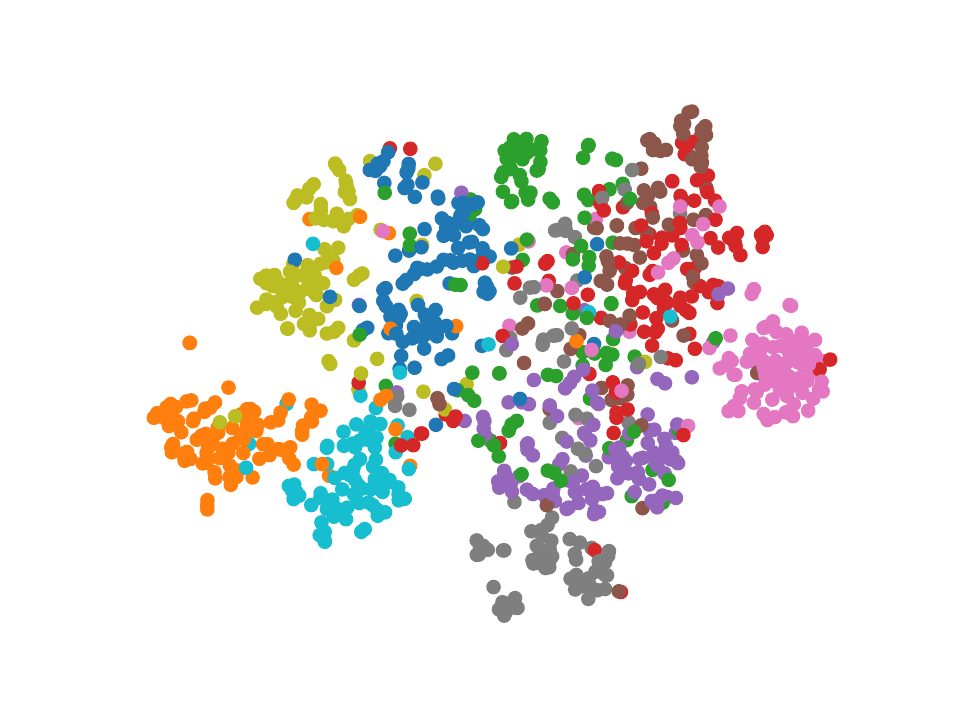}
\caption{\ContSteal}
\label{fig:embedding2_contrastive}
\end{subfigure}
\begin{subfigure}{0.3\columnwidth}
\includegraphics[width=\columnwidth]{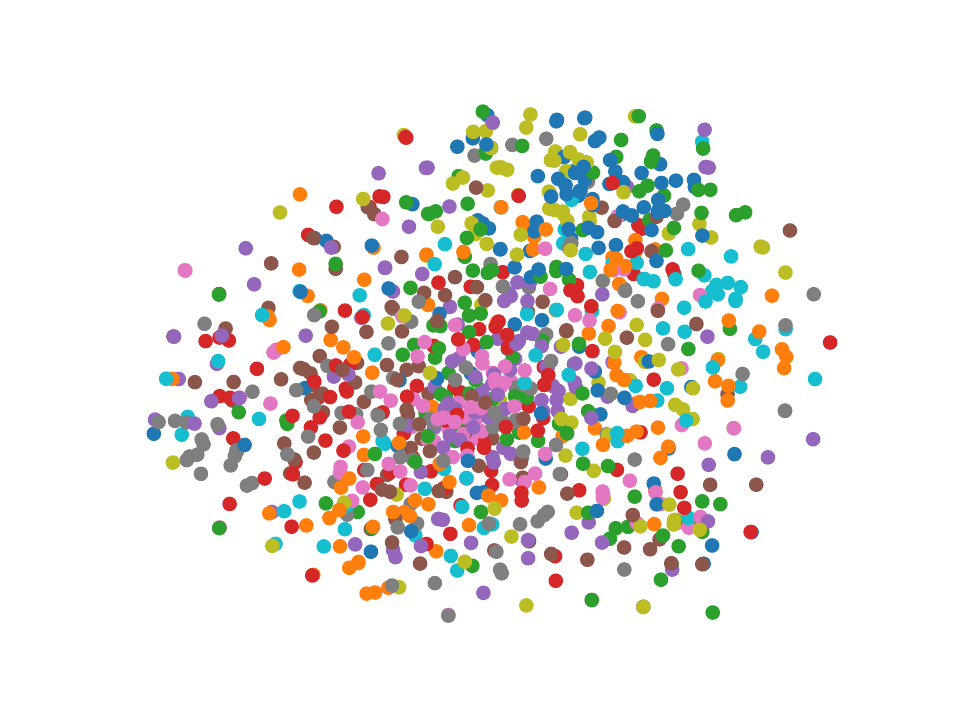}
\caption{Conventional }
\label{fig:embedding2_normal}
\end{subfigure}
\caption{The t-SNE projection of $1,000$ randomly selected samples' embeddings from target encoder, surrogate encoder under \ContSteal, and surrogate encoder under the conventional attack, respectively. 
Note that the target encoder is pre-trained by SimCLR on CIFAR10.
}
\label{fig:embedding}
\end{figure}

\begin{figure*}[!t]
\centering
\begin{subfigure}{0.44\columnwidth}
\includegraphics[width=\columnwidth]{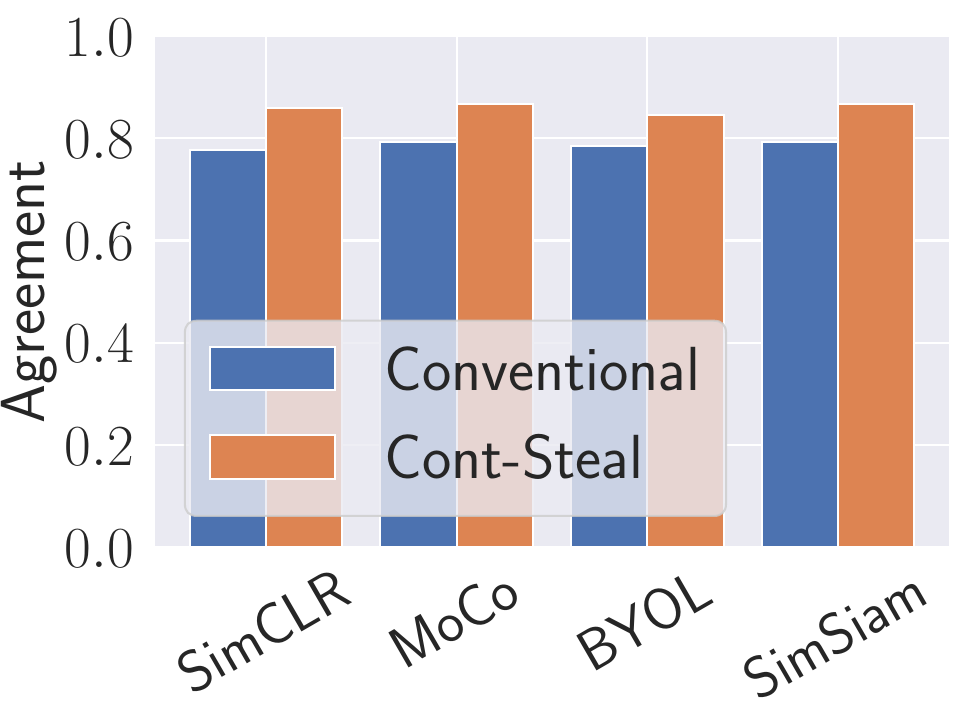}
\caption{CIFAR10}
\label{fig:contrastive_agreement_cifar10_cifar10}
\end{subfigure}
\begin{subfigure}{0.44\columnwidth}
\includegraphics[width=\columnwidth]{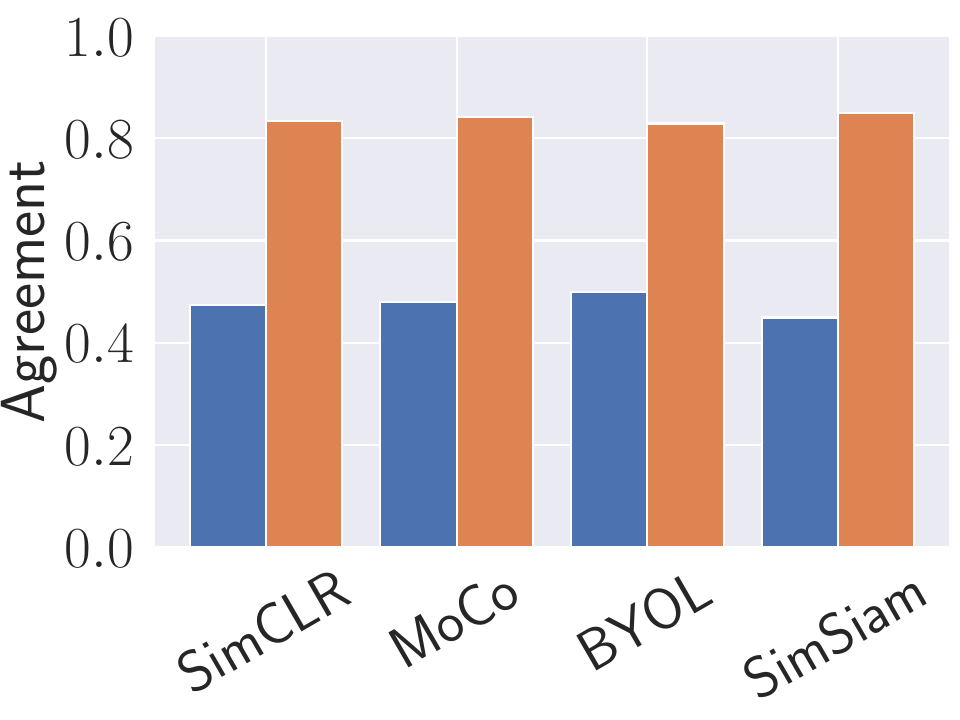}
\caption{STL10}
\label{fig:contrastive_agreement_cifar10_wild}
\end{subfigure}
\begin{subfigure}{0.44\columnwidth}
\includegraphics[width=\columnwidth]{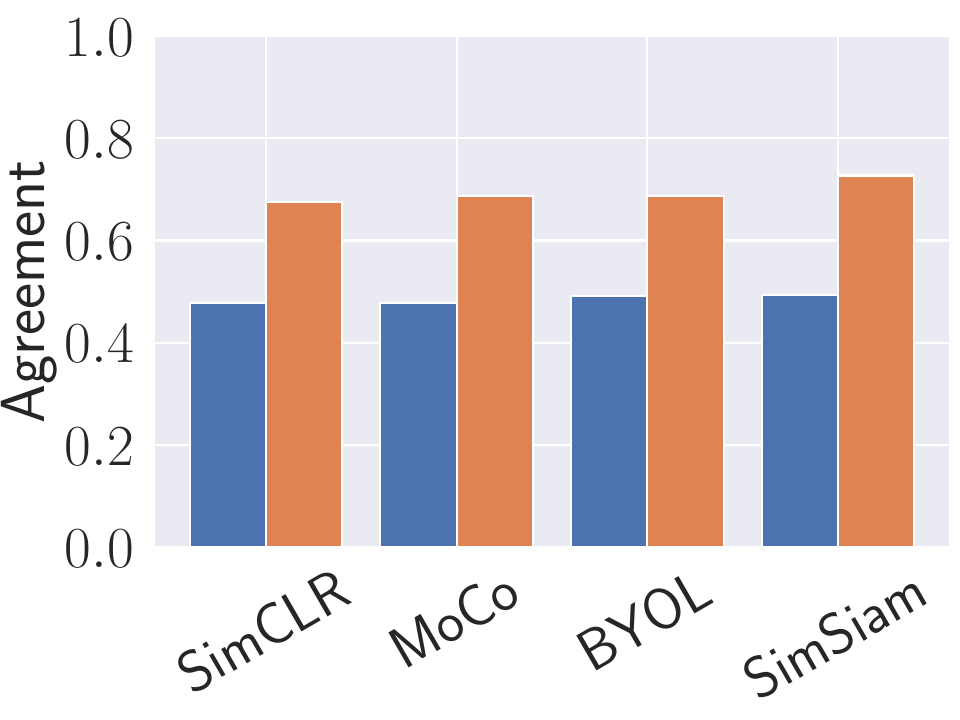}
\caption{F-MNIST}
\label{fig:contrastive_agreement_cifar10_cifar100}
\end{subfigure}
\begin{subfigure}{0.44\columnwidth}
\includegraphics[width=\columnwidth]{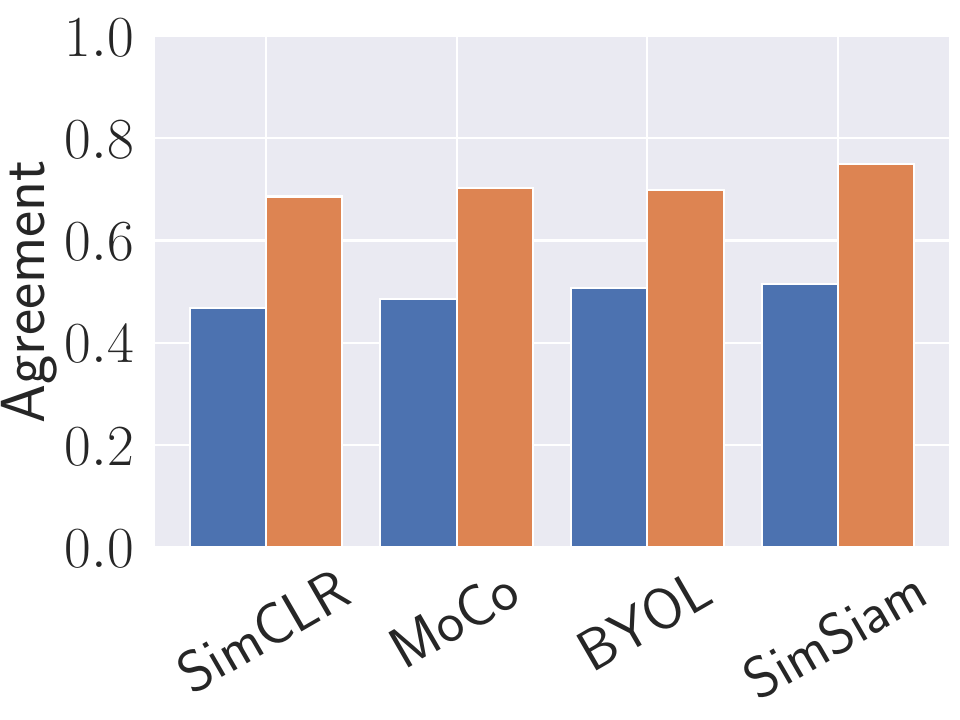}
\caption{SVHN}
\label{fig:contrastive_agreement_cifar10_stl10}
\end{subfigure}
\begin{subfigure}{0.44\columnwidth}
\includegraphics[width=\columnwidth]{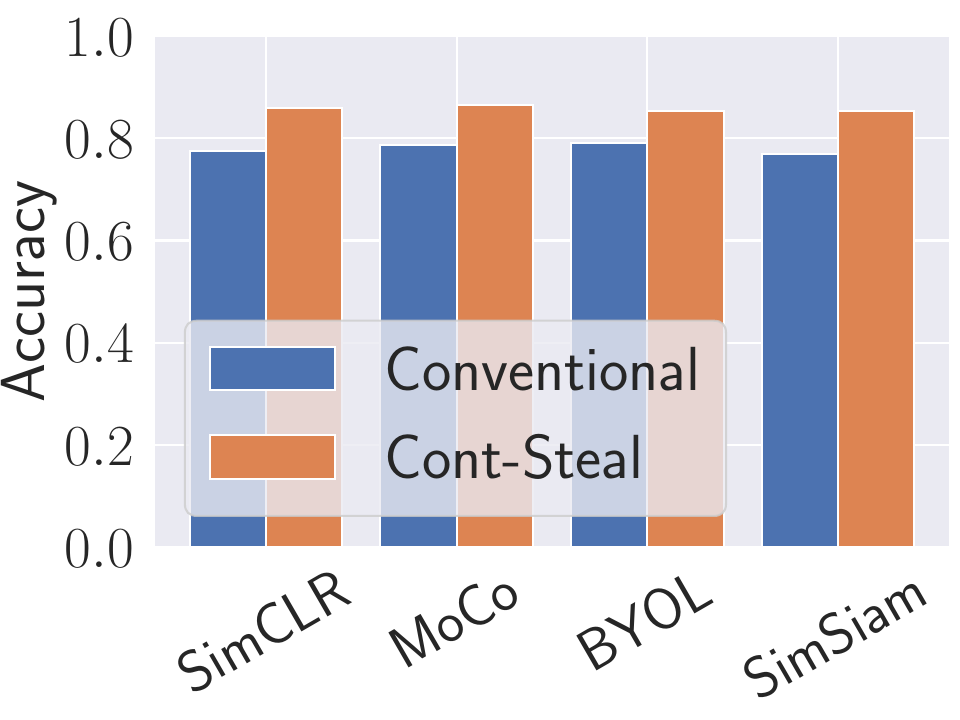}
\caption{CIFAR10}
\label{fig:contrastive_accuracy_cifar10_cifar10}
\end{subfigure}
\begin{subfigure}{0.44\columnwidth}
\includegraphics[width=\columnwidth]{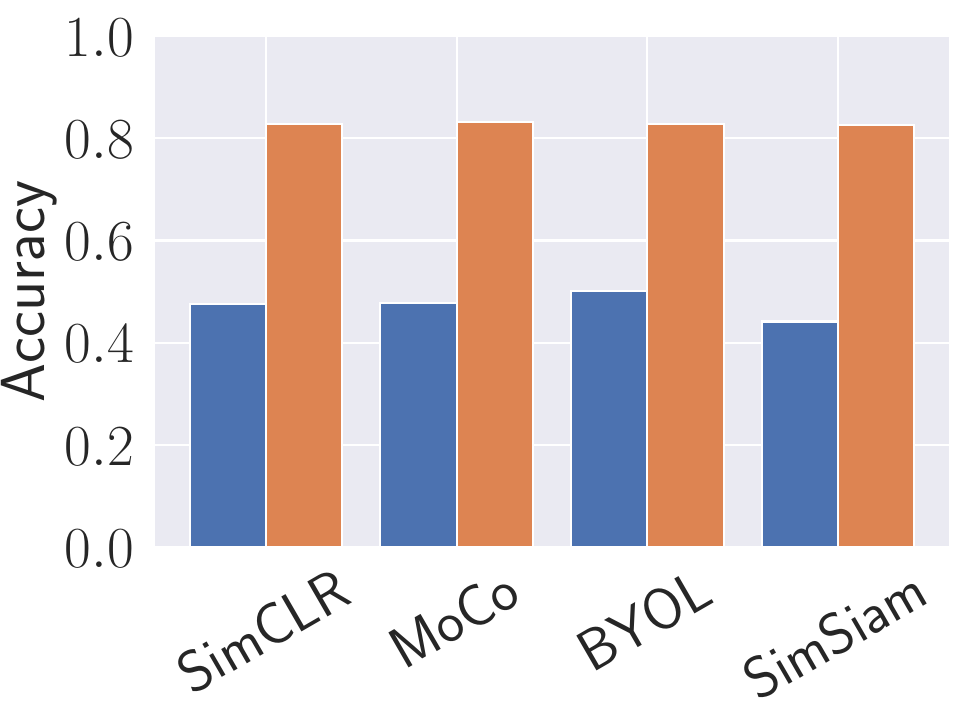}
\caption{STL10}
\label{fig:contrastive_accuracy_cifar10_wild}
\end{subfigure}
\begin{subfigure}{0.44\columnwidth}
\includegraphics[width=\columnwidth]{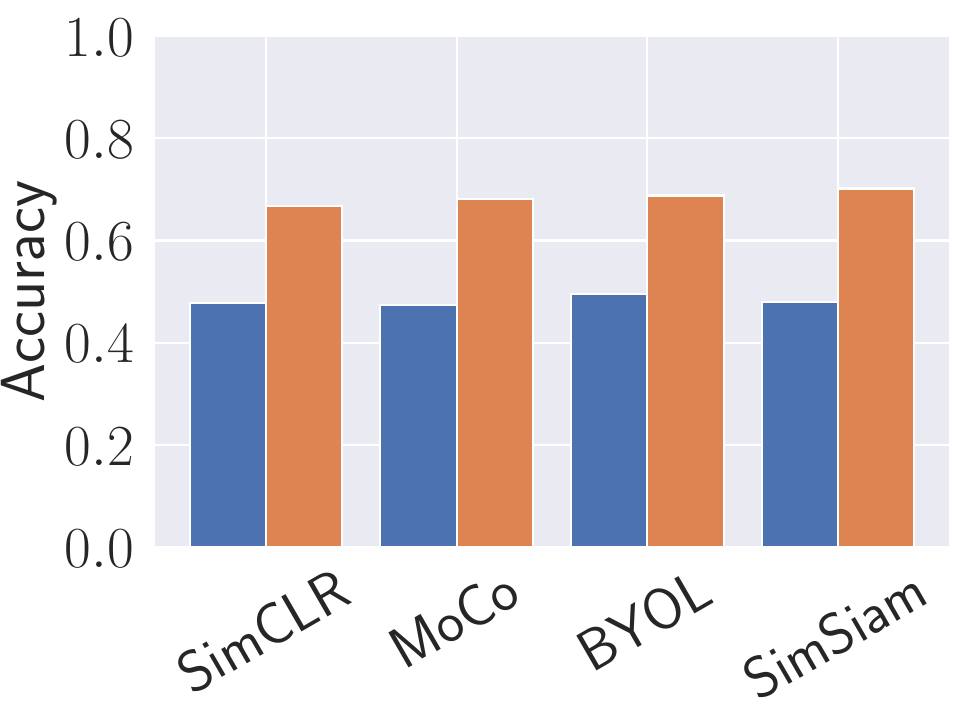}
\caption{F-MNIST}
\label{fig:contrastive_accuracy_cifar10_cifar100}
\end{subfigure}
\begin{subfigure}{0.44\columnwidth}
\includegraphics[width=\columnwidth]{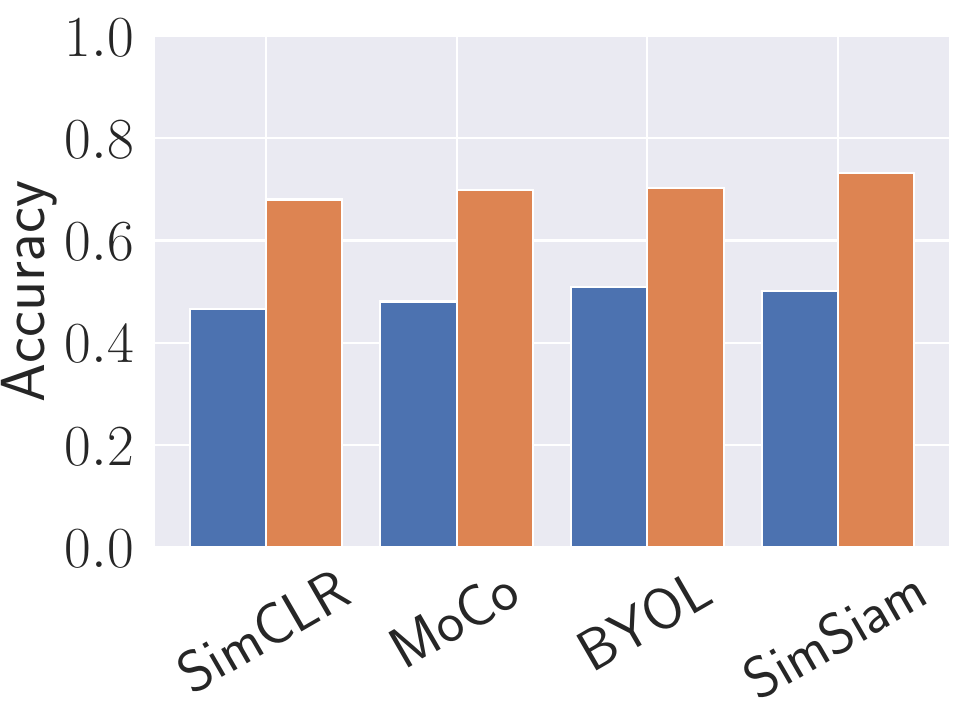}
\caption{SVHN}
\label{fig:contrastive_accuracy_cifar10_stl10}
\end{subfigure}
\caption{The performance of \ContSteal and conventional attack against target encoders trained on CIFAR10.
The adversary uses CIFAR10, STL10, F-MNIST, and SVHN to conduct model stealing attacks.
The adversary uses CIFAR10 as the downstream task to evaluate the attack performance.
The x-axis represents different kinds of the target model.
The first line's y-axis represents the agreement of the model stealing attack. 
The second line's y-axis represents the accuracy of the model stealing attack.}
\label{fig:contrastive_cifar10}
\end{figure*}

\subsection{Performance of \ContSteal}
\label{subsection:evaluation_contsteal}

As shown in \autoref{subsection:encoder}, encoders are more vulnerable to model stealing attacks since the embedding usually contains richer information compared to the predicted label or posteriors.
We then show that our proposed \ContSteal can achieve better attack performance by making deeper use of embeddings' information.

\autoref{fig:contrastive_cifar10} shows the attack performance when the target pre-training dataset is CIFAR10.
Note that we also show the attack performance on other settings in the appendix.
We discover that compared to conventional attacks against encoders, \ContSteal can consistently achieve better performance.
For instance, as shown in \autoref{fig:contrastive_agreement_cifar10_stl10}, when the target encoder is MoCo trained on CIFAR10, if the adversary uses STL10 to conduct model stealing attacks against encoders, the surrogate encoder can achieve $0.841$ agreement in CIFAR10 downstream tasks with the Cont-Steal but only $0.479$ with conventional attacks.
Another finding is that compared to the same distribution surrogate dataset, our \ContSteal can better enhance the performance when the surrogate dataset comes from a different distribution from the pre-trained dataset.
For instance, when the target encoder is SimCLR trained on CIFAR10, \ContSteal outperforms conventional attack by $0.055$ agreement when the surrogate dataset is also CIFAR10, while the improvement increases to $0.207$ and $0.214$ when the surrogate dataset is STL10.
We show more comparing results in Supplementary Material \autoref{subsection:appendix_contsteal}.
Note that our \ContSteal also has great performance on other recent state-of-the-art visual models (ViT~\cite{DBKWZUDMHGUH21}, MAE~\cite{HCXLDG21}, and CLIP~\cite{RKHRGASAMCKS21}), as we show in \autoref{appendix:other-model-performance}, and can have better performance than other recent similar attacks~\cite{LJLG22,DDKGP22} shown in \autoref{subsection:compare}.

To better understand why \ContSteal can always achieve better performance, we extract samples' embeddings generated by different encoders, i.e., the target encoder, surrogate encoder trained with the conventional attack, and surrogate encoder trained with the \ContSteal, and project them into a 2-dimensional space using t-SNE.
From the results summarized in \autoref{fig:embedding}, we find that \ContSteal can effectively mimic the pattern of the embeddings as the target encoder.
However, the conventional attack fails to capture such patterns for a number of input samples, e.g., the outer circle in \autoref{fig:embedding_normal}.
This further demonstrates that \ContSteal benefits from jointly considering different embeddings as they can serve as anchors to better locate the position of the other embeddings in their space.
We also show some ablation study results on Supplementary Material \autoref{appendix:abaltion} to show that with the less surrogate dataset, less training epoch, and different model architecture, \ContSteal can still achieve much better results than conventional steal.
Also, we show further attacks based on the stole models on Supplementary Material \autoref{appendix:further} to show that \ContSteal can be used as a springboard for other attacks.

\begin{table}[!t]
\centering
\caption{The monetary and (training) time costs for normal training and \ContSteal attack.
\ContSteal's monetary cost contains two parts: query cost and training cost.
Note that we ignore the query time cost of \ContSteal as it normally has a smaller value than the training time cost.
}
\label{table:cost}
\scalebox{0.7}
{
\begin{tabular}{l c  c  c  c}
\toprule
~ & \multicolumn{2}{c}{\textbf{Monetary Cost}} & \multicolumn{2}{c}{\textbf{Time Cost}} \\
\midrule
\textbf{Model} & \textbf{Normal (\$)} & \textbf{\ContSteal (\$)} & \textbf{Normal (h)} & \textbf{\ContSteal (h)}  \\
\midrule
SimCLR & 58.68 & 11.83 (1.83 + 10) & 20.01 &  0.62\\
\midrule 
MoCo & 54.83 & 12.13 (2.13 + 10) & 18.69 & 0.73 \\
\midrule
BYOL & 61.46 & 12.08 (2.08 + 10) & 20.96 & 0.71 \\
\midrule
SimSiam & 57.14 & 12.00 (2.00 + 10) & 19.46 & 0.68 \\
\bottomrule
\end{tabular}
}
\end{table}

\subsection{Cost Analysis}

As we mentioned before, pre-train a state-of-the-art encoder is time-consuming and resource-demanding.
We wonder if the model stealing attacks can steal the functionality of the encoder with much less cost.
To this end, we evaluate the time and monetary cost of training an encoder from scratch or stealing a pre-trained encoder via \ContSteal.
The monetary cost of model stealing includes querying the target model and training the surrogate model.
We refer to the query price as $\$1$ for 1,000 queries based on AWS.\footnote{\url{https://aws.amazon.com/rekognition/pricing/}}
Our experiment is conducted on 1 NVIDIA A100 whose price is $\$2.934$ per hour based on google cloud.\footnote{\url{https://cloud.google.com/compute/gpus-pricing/}}

The monetary and time cost is shown in \autoref{table:cost}.
We observe that \ContSteal can obtain a surrogate encoder with much less money and time cost than training the encoder from scratch.
For instance, a ResNet18 trained by SimCLR on CIFAR10 takes 20.01 hours and 58.68\$ on 1 NVIDIA A100 GPU, while \ContSteal only takes 0.62 hours and 11.83\$ to steal an encoder that performs similarly on downstream tasks.
The results demonstrate that \ContSteal is able to construct surrogate encoders that perform similarly to the target encoders but with much less time and monetary cost.

\subsection{Defenses}
\label{sec:defense}

In this section, we will consider different defenses against model stealing attacks on encoders to evaluate the robustness of our proposed attack.
We divided all defenses into three categories: perturbation-based defense~\cite{OSF19} and watermark-based defense~\cite{ABCPK18}.

\mypara{Perturbation-based Defense}
In this defense setting, the defender aims to perturb the output of the target model to limit the information the adversary can obtain.
The common practice of this kind of defense includes adding noise~\cite{OSF19}, top-$k$~\cite{OSF19}, and feature rounding~\cite{TZJRR16}.

\begin{figure}[!t]
\centering
\begin{subfigure}{0.32\columnwidth}
\includegraphics[width=\columnwidth]{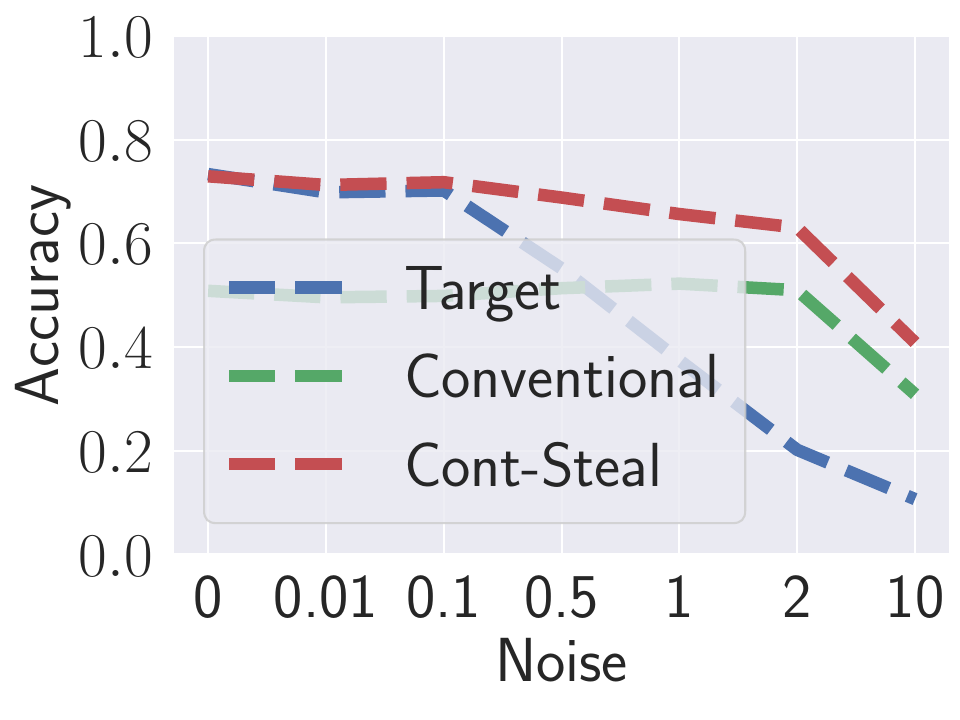}
\caption{Adding noise}
\label{fig:embedding_target}
\end{subfigure}
\begin{subfigure}{0.32\columnwidth}
\includegraphics[width=\columnwidth]{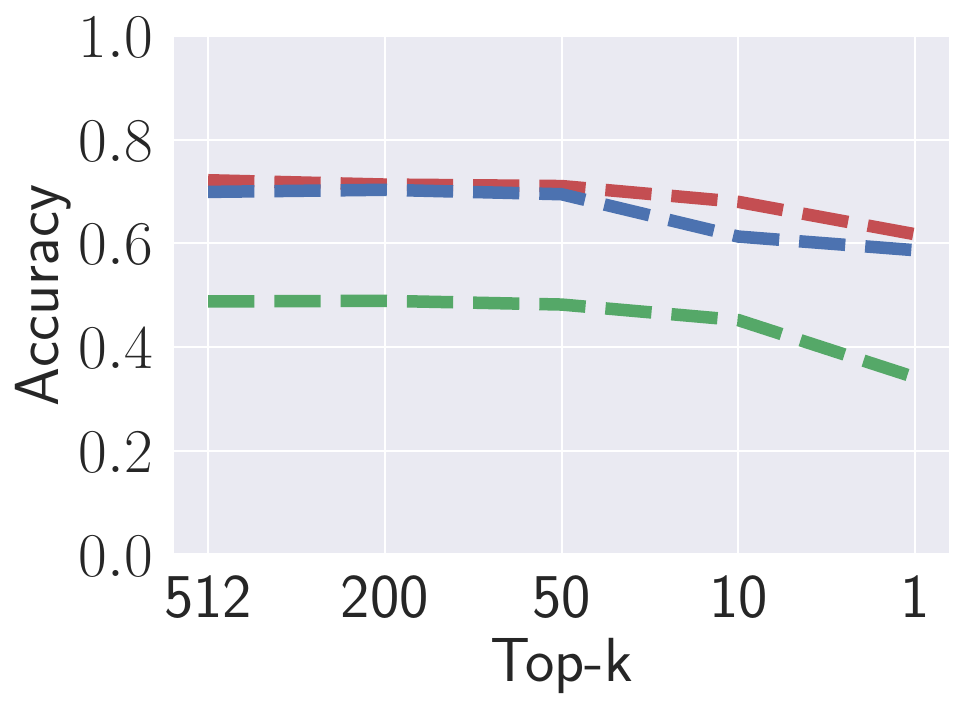}
\caption{Top-$k$}
\label{fig:embedding_contrastive}
\end{subfigure}
\begin{subfigure}{0.32\columnwidth}
\includegraphics[width=\columnwidth]{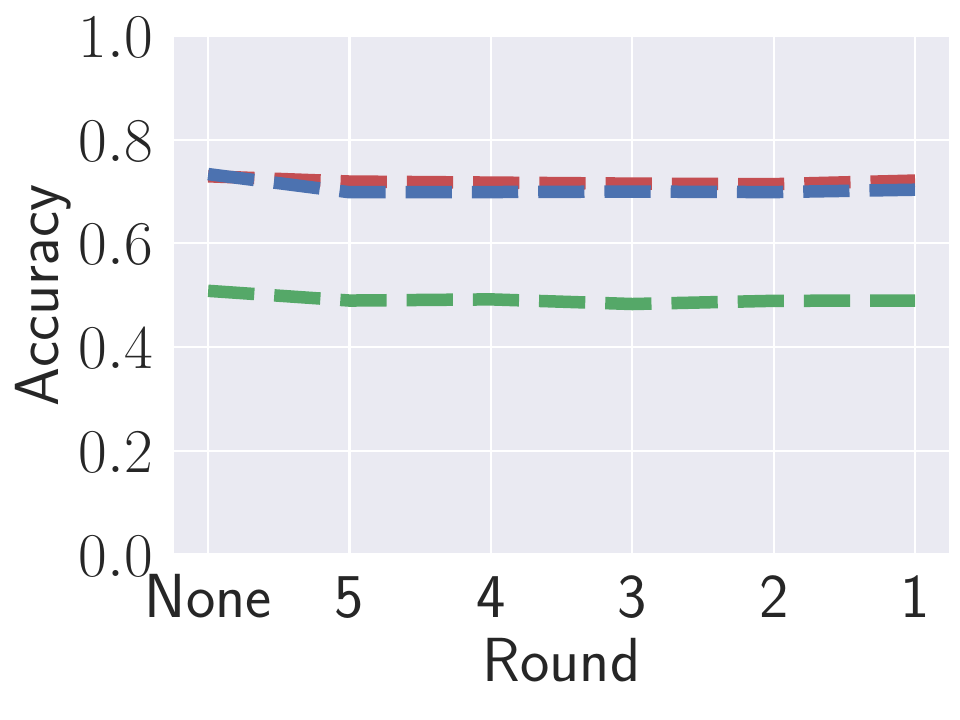}
\caption{Rounding}
\label{fig:embedding_normal}
\end{subfigure}
\caption{
The performance of different defense methods.
Target encoders are trained on CIFAR10.
The downstream dataset and surrogate dataset are both STL10.
The x-axis represents different defense levels.
The y-axis represents the model's accuracy.
}
\label{fig:pertubation}
\end{figure}

Adding noise means that the defender will introduce noise value to the original output of the model.
In our case, we consider adding Gaussian noise to the embeddings generated by the target encoder.
We set the mean value to 0, and different noise levels represent different standard deviations of the Gaussian distribution.
For Top-$k$, the defender will only output the first $k$ largest number of each embedding (and set the rest as 0).
In this way, the high-dimensional information of the image contained in embeddings can be appropriately reduced.
Regarding feature rounding, the defender will truncate the values in the embedding to a specific digit.
As a case study, we consider a ResNet18 encoder pre-trained on CIFAR10 with SimCLR and take STL10 to train its downstream classifier.
The experimental results are summarized in \autoref{fig:pertubation}.
We can observe that while adding noise and top-$k$ can reduce the model stealing attacks' performance, it may also degrade the target model performance to a large extent.
For instance, when the noise increases from 0 to 10, the attack performance of \ContSteal decreases from 0.729 to 0.410, while the target encoder's performance drops from 0.734 to 0.098.
On the other hand, rounding only has a limited effect on both target model performance and attack performance.
This indicates that perturbation-based defense cannot defend against the encoder's model stealing attack effectively since they cannot reach a good trade-off between attack performance and model utility.

\begin{table}[!htbp]
\centering
\caption{Watermark defense.
Pretrain dataset and surrogate dataset are both CIFAR10.
Watermark leverages a watermark rate (wr) to verify the ownership of target models. A higher wr denotes a better verification performance.}
\label{table:watermark}
\scalebox{0.7}
{
\begin{tabular}{lccc}
\toprule
\textbf{Dataset} & \textbf{Target model (acc/wr)} & \textbf{\ContSteal (acc/wr)} & \textbf{Baseline (acc/wr)} \\
\midrule
CIFAR10 & 0.864 / 0.998 & 0.769 / 0.130 & 0.871 / 0.095\\
\midrule
STL10  & 0.721 / 0.999  & 0.702 / 0.034 & 0.733 / 0.111\\
\midrule
SVHN  & 0.501 / 0.999 & 0.535 / 0.303 & 0.492 / 0.103\\
\midrule
F-MNIST  & 0.857 / 0.999 & 0.813 / 0.061 & 0.850 / 0.099\\
\bottomrule
\end{tabular}
}
\end{table}

\mypara{Watermark-based Defense}
Watermark-based defense is also one of the most popular defense methods against model stealing attacks~\cite{ABCPK18}.
Watermark provides copyright protection by adding some specific identification to the target model.
If the surrogate model is stolen from the watermarked target model, then ideally, it will contain the same watermark as well.
Adi et al.\cite{ABCPK18} show that backdoor technology can be used as the watermark to protect the model.
In that sense, BadEncoder~\cite{JLG22}, a backdoor mechanism against the encoder, can be leveraged as a watermarking technology for our target encoder as well.
The defenders first will train the watermarked (backdoored) encoder, where images with a certain trigger will cause misclassification.
Then, if they find the surrogate model can also misclassify images with the same trigger, the defenders can claim ownership of the surrogate model.

In our experiments, we leverage BadEncoder to watermark the encoder pre-trained on CIFAR10 by SimCLR, and leverage different downstream datasets to perform different tasks.
We assume a strong adversary that has the same downstream dataset as the surrogate dataset.
Also, we consider the baseline cases where the trigger samples are fed into the clean model to calculate the watermark rate (wr).
As shown in \autoref{table:watermark}), the watermark cannot be preserved as the surrogate models constructed by \ContSteal have similar wr as the baseline model.
For instance, when the downstream dataset is CIFAR10, \ContSteal builds a surrogate model with 0.769 accuracy while only 0.130 wr, which is close to the baseline model.
This indicates that \ContSteal can bypass the watermarking technique as it can reach similar utility while reducing the wr to a large extent.
Note that there is also another work to protect contrastive learning models from model stealing attacks using dataset inference~\cite{DDKDGCBP22}.
We show in Supplementary Material in \autoref{subsection:more-defense} that this kind of defense can be easily bypassed by \ContSteal.

\section{Related Work}
\label{sec:related_work}

\mypara{Contrastive Learning}
Contrastive learning is one of the most popular methods to train encoders.
Current works~\cite{OLV18,WXYL18,HFWXG20,CKNH20,GSATRBDPGAPKMV20,CH21} propose different advanced contrastive learning algorithms.
SimCLR, MoCo, BYOL, SimSiam are currently the mainstream frameworks of contrastive learning.
Thus, we concentrate on them in this paper.
There are many works on evaluating the security and privacy risks of contrastive learning.
Previous works~\cite{HZ21,JLG22,LJQG21} propose membership inference attacks, attribution inference attacks, and backdoor attacks on contrastive learning.
All proposed attacks show that contrastive-based models are vulnerable to popular attacks.
Therefore, the security issues of self-supervised learning deserve more attention.

\mypara{Model Stealing Attack}
In model stealing, the adversary's goal is to steal part of the target model.
Tram{\`e}r et al.~\cite{TZJRR16} proposed the first model stealing attack against black-box machine learning API to steal its parameters.
Wang et al.~\cite{WG18} proposed the first hyperparameter stealing attacks against ML models.
Oh et al.~\cite{OASF18} also tried to steal machine learning model's architectures and hyperparameters.
Orekondy et al.~\cite{OSF19} proposed knockoff nets, which aim at stealing the functionality of black-box models.
Krishna et al.~\cite{KTPPI20} formulated a model stealing attack against BERT-based API.
Besides, Wu et al.~\cite{WYPY20} and Shen et al.~\cite{SHHZ22} perform model stealing attacks against Graph Neural Networks.
These works often have relatively strong assumptions, such as the model family is known and the victim's data is partly available while we conduct model stealing attacks against encoders and relax the above assumption.

\section{Conclusion}
\label{sec:conclusion}

In this paper, we conduct the first model stealing risk assessment towards image encoders.
Our evaluation shows that the encoder is more vulnerable to model stealing attacks compared to the classifier.
This is because the embedding provided by the encoder contains richer information than the posteriors or predicted labels from whole classifiers.

To better unleash the power from the embeddings, we propose \ContSteal, a contrastive learning-based model stealing method against encoders.
Concretely, \ContSteal introduces different types of negative pairs as ``anchors'' to better navigate the surrogate encoder and learn the functionality of the target encoder.
Extensive evaluations show that \ContSteal consistently performs better than conventional attacks against encoders.
And such an advantage is further amplified when the adversary has no information on the target dataset, a limited amount of data, and restricted query budgets.
Our work points out that the threat of model stealing attacks against encoders is largely underestimated, which prompts the need for more effective intellectual property protection of representation learning techniques.

\mypara{Acknowledgments}
We thank all anonymous reviewers for their constructive comments.
This work is partially funded by the Helmholtz Association within the project ``Trustworthy Federated Data Analytics'' (TFDA) (funding number ZT-I-OO1 4) and by the European Health and Digital Executive Agency (HADEA) within the project ``Understanding the individual host response against Hepatitis D Virus to develop a personalized approach for the management of hepatitis D'' (D-Solve) (grant agreement number 101057917).

\bibliographystyle{plain}
\bibliography{normal_generated_py3}

\newpage
\clearpage
\appendix

\section{Supplementary Material}

\subsection{Training Algorithm of \ContSteal}

\begin{algorithm}
\SetKwData{Left}{left}\SetKwData{This}{this}\SetKwData{Up}{up}
\SetKwFunction{Union}{Union}\SetKwFunction{FindCompress}{FindCompress}
\SetKwInOut{Input}{input}\SetKwInOut{Output}{output}
\Input{Surrogate training dataset $D_{surrogate}^{train}$, target encoder $E_{t}$, surrogate encoder $E_{s}$}
\BlankLine
\emph{Initialize $E_{s}$'s parameters}\;
\For{$each$ $epoch$}{
\For{$each$ $batch$}{
Sample a batch with N training data samples {$x_1,x_2,\cdots,x_N$} from $D_{surrogate}^{train}$

Generate augmented data samples: {$(\widetilde{x}_{1,t},\widetilde{x}_{1,s}), (\widetilde{x}_{2,t},\widetilde{x}_{2,s}),\cdots,(\widetilde{x}_{N,t},\widetilde{x}_{N,s})$}, where $\widetilde{x}_{k,t}$ and $\widetilde{x}_{k,s}$ are the two augmented views of $x_k$

Feed $\widetilde{x}_{k,t}$ to $E_{t}$ and $\widetilde{x}_k$ to $E_{s}$ to calculate the contrastive steal loss: $ L_{\ContSteal} = \frac{\sum_{k=1}^Nl(k)}{N} $

Optimize $E_{s}$'s parameters with the contrastive steal loss $ L_{\ContSteal}$
}
}
\Return{Surrogate encoder $E_{s}$}
\caption{The training process of \ContSteal.}
\label{algo:contrastive_steal}
\end{algorithm}

\autoref{algo:contrastive_steal} presents the training process of contrastive stealing.
In each batch, given $N$ training samples, we first generate $2N$ augmented views and feed the target encoder and surrogate encoder with different views generated by the same samples.
Then, we optimize the surrogate encoder by minimizing $L_{\ContSteal}$.

\subsection{Ablation Studies on Adversary Training Process}
\label{appendix:abaltion}

\begin{figure*}[t]
\centering
\begin{subfigure}{0.98\columnwidth}
\includegraphics[width=\columnwidth]{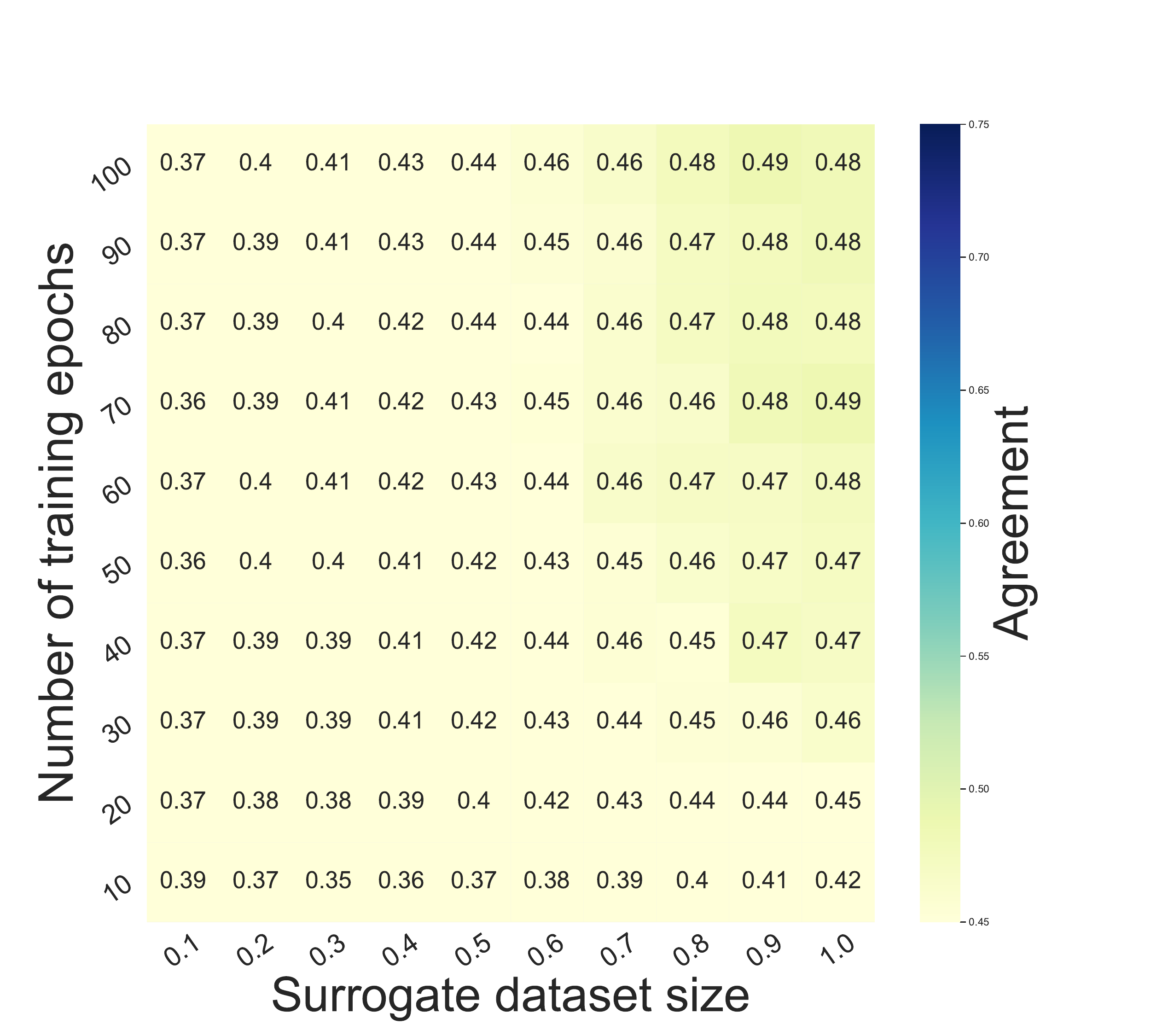}
\caption{Conventional}
\label{fig:heatmap_conventional}
\end{subfigure}
\begin{subfigure}{0.98\columnwidth}
\includegraphics[width=\columnwidth]{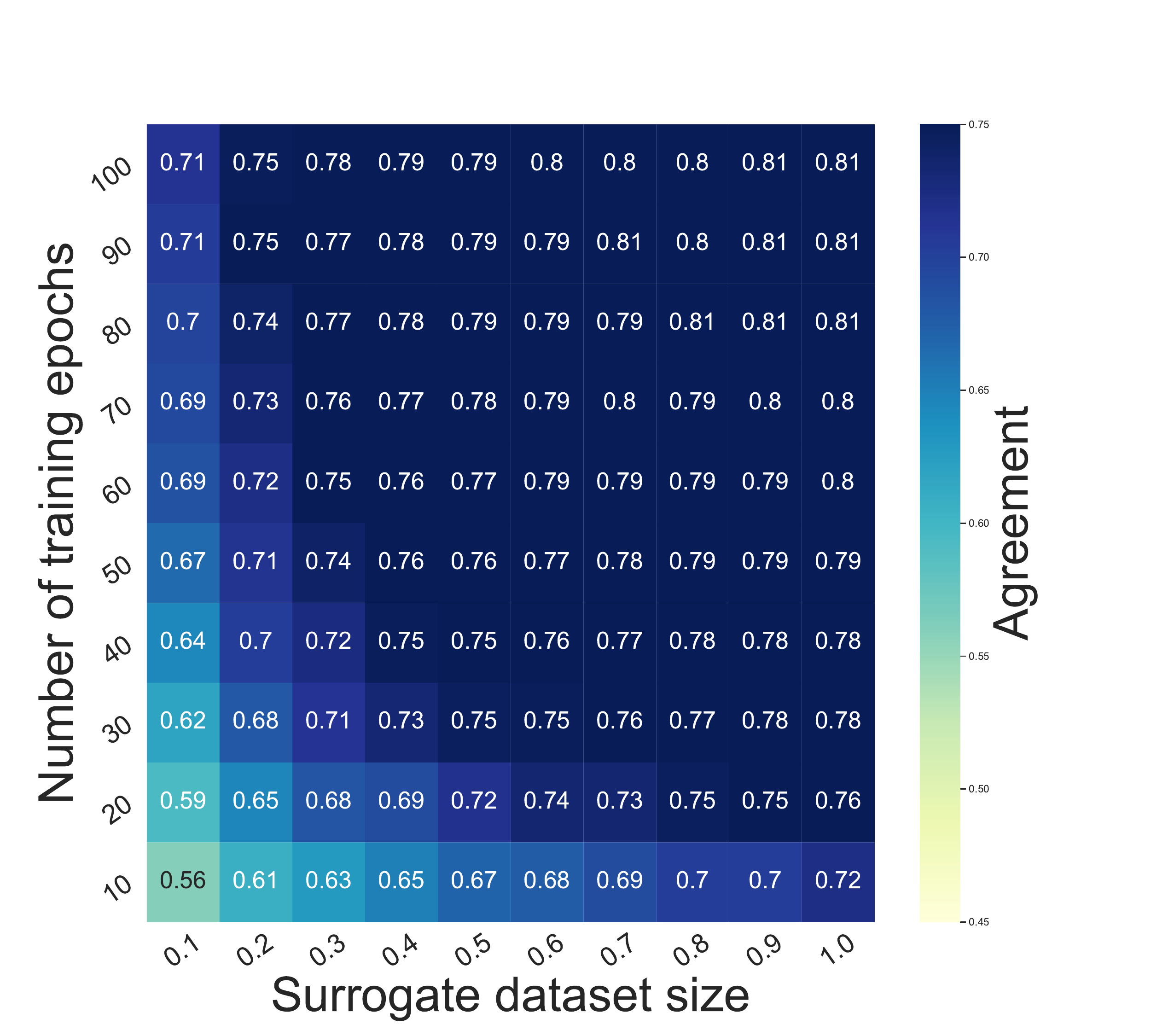}
\caption{\ContSteal}
\label{fig:heatmap_contrastive}
\end{subfigure}
\caption{Heatmap of the agreement scores of model stealing attacks.
The target model's encoder and downstream classifier are both ResNet18 trained by SimCLR on CIFAR10.
The surrogate dataset is STL10. 
Surrogate dataset's size refers to the proportion of surrogate data we used for the whole surrogate dataset.
We show the performance of 100 combinations of different training epochs and the surrogate gate dataset's size.
}
\label{fig:heatmap}
\end{figure*}

\begin{table}[!t]
\centering
\caption{\ContSteal attack performance of different surrogate architectures.
Target encoders (ResNet18) and downstream classifiers are trained on CIFAR10.
The surrogate dataset is also CIFAR10.
}
\label{table:noise}
\scalebox{0.7}
{
\begin{tabular}{l l c c}
\toprule
\textbf{Framework} & \textbf{Architectures} & \textbf{Agreement} & \textbf{Accuracy} \\
\midrule
\multirow{3}*{SimCLR} & ResNet18  & 0.835 & 0.839  \\
~ &  ResNet34  & 0.837 & 0.842\\
~ &  ResNet50  & 0.844 & 0.840 \\
~ &  DenseNet161 & 0.831 & 0.828\\
~ &  MobileNetV2 & 0.815 & 0.811\\
\midrule
\multirow{3}*{MoCo} & ResNet18  & 0.857 & 0.849  \\
~ &  ResNet34  & 0.858 & 0.849\\
~ &  ResNet50  & 0.867 & 0.856 \\
~ &  DenseNet161 & 0.813 & 0.811\\
~ &  MobileNetV2 & 0.796 & 0.801\\
\midrule
\multirow{3}*{BYOL} & ResNet18  & 0.845 & 0.842  \\
~ &  ResNet34  & 0.850 & 0.847\\
~ &  ResNet50  & 0.857 & 0.855 \\
~ &  DenseNet161 & 0.845 & 0.821\\
~ &  MobileNetV2 & 0.839 & 0.847\\
\midrule
\multirow{3}*{SimSiam} & ResNet18  & 0.856 & 0.835  \\
~ &  ResNet34  & 0.858 & 0.839\\
~ &  ResNet50  & 0.860 & 0.848 \\
~ &  DenseNet161 & 0.791 & 0.783\\
~ &  MobileNetV2 & 0.812 & 0.832\\
\bottomrule
\end{tabular}
}
\label{table:arch}
\end{table}

\mypara{Impact of Surrogate Encoder's Architecture}
Previous experiments are based on the assumption that the adversary knows the target encoder's architecture.
We then investigate whether the attack against the encoder is still effective when the surrogate encoder has different model architectures compared to the target encoder.
Concretely, we perform \ContSteal against the ResNet18 encoder with the surrogate encoder's architecture as ResNet18, ResNet34, ResNet50, DenseNet161, and MobileNetV2, respectively.
As shown in \autoref{table:arch}, we can see that the architecture of the surrogate model only has limited influence on the attack performance.
For instance, the adversary can achieve 0.839 accuracy using the same architecture as the target model, while it can even achieve 0.840 accuracy when using a more complex model architecture (ResNet50) on SimCLR.
The attack performance will drop a little if the adversary uses DenseNet161 and MobileNetV2.
This might be because the architectures of DenseNet161 and MobileNetV2 have larger differences compared to ResNet18.
However, the accuracy with DenseNet161/MobileNetV2 as the surrogate encoder's architecture can still achieve 0.828/0.811.
This demonstrates that the model architectures of the surrogate encoder only have a limited impact on the attack performance, which makes the attack a more realistic threat.

\mypara{Impact of Surrogate Dataset's Size and Surrogate Model's Training Epoch}
We conduct ablation studies here to better illustrate the effectiveness of \ContSteal.
Concretely, we investigate whether conventional attacks and \ContSteal are still effective under limited surrogate dataset size and the number of training epochs.
Ideally, we consider the attack that can reach similar performance but with less surrogate dataset size and fewer training epochs as a better attack as it requires less query and monetary costs.
As shown in \autoref{fig:heatmap}, we observe that both conventional attacks and \ContSteal can have better performance with a larger surrogate dataset size and more training epochs.
For instance, \ContSteal reaches 0.675 agreement when the surrogate encoder is trained with 10\% surrogate dataset for 50 epochs, while the agreement increase to 0.812 with 100\% surrogate dataset and 100 training epochs.
The second observation is that \ContSteal outperforms conventional attacks even with limited data and training epochs.
For instance, even with only 10\% surrogate dataset and 10 training epochs, the surrogate encoder built by \ContSteal can reach 0.562 agreement, while the conventional attack can only achieve 0.479 agreement with the full surrogate dataset and 100 training epochs.
As we mentioned before, this is because \ContSteal can enforce the surrogate embedding of an image close to its target embedding and also push away embeddings of different images irrespective of being generated by the target or the surrogate encoders (see also \autoref{table:half-contrastive} for the necessity of introducing negative pairs from the surrogate encoder).
This makes \ContSteal a more effective model stealing attack against encoders.

\begin{figure*}[t]
\centering
\begin{subfigure}{0.48\columnwidth}
\includegraphics[width=\columnwidth]{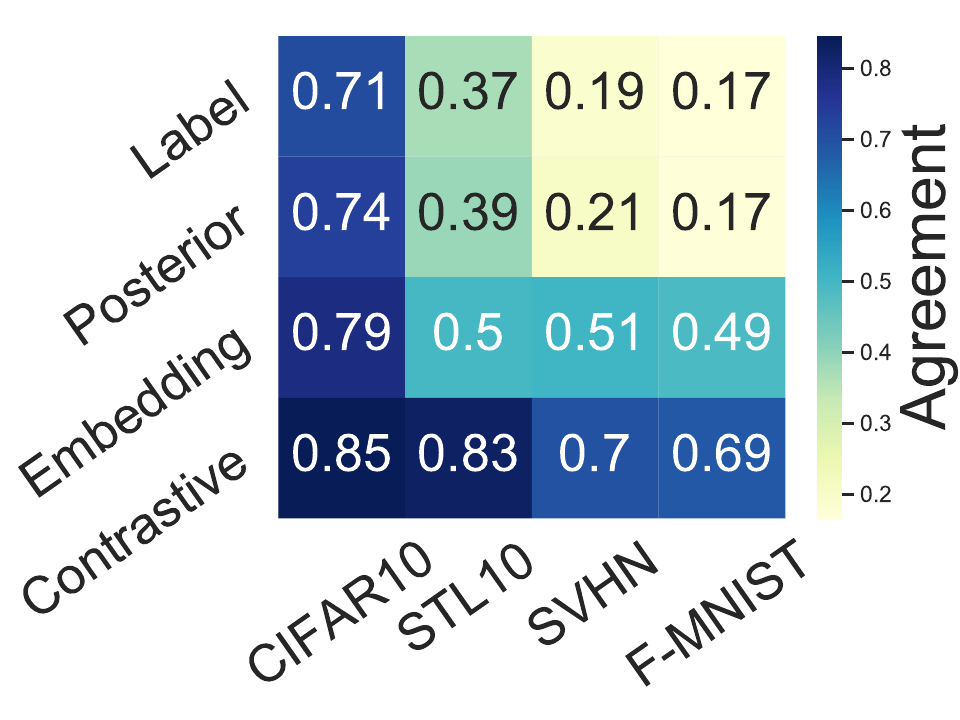}
\caption{SimCLR}
\label{fig:heatmap_simclr}
\end{subfigure}
\begin{subfigure}{0.48\columnwidth}
\includegraphics[width=\columnwidth]{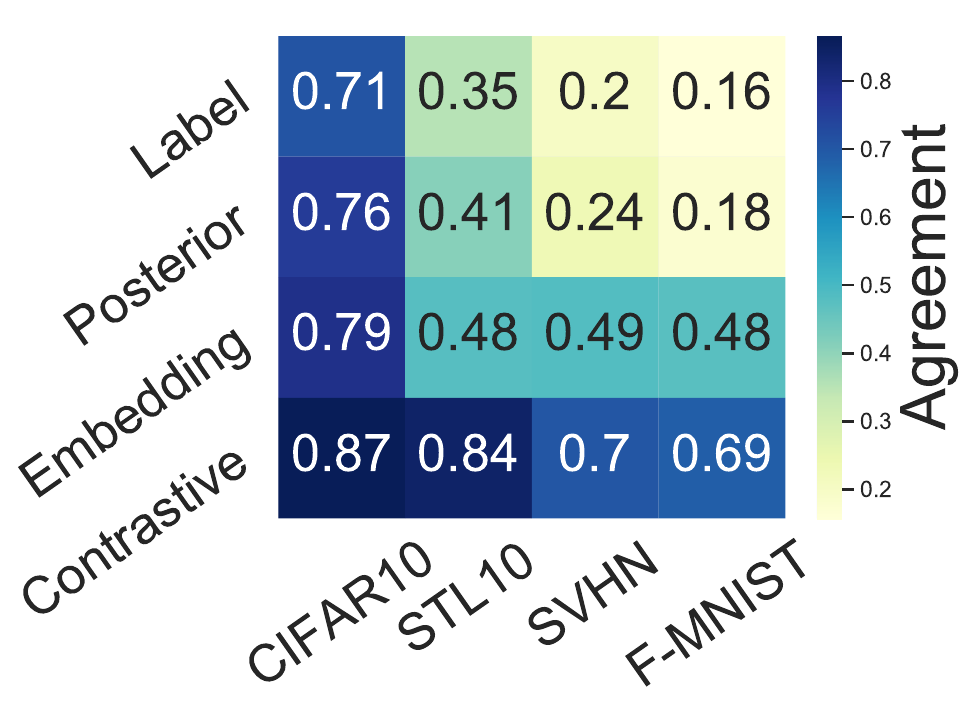}
\caption{MoCo}
\label{fig:heatmap_moco}
\end{subfigure}
\begin{subfigure}{0.48\columnwidth}
\includegraphics[width=\columnwidth]{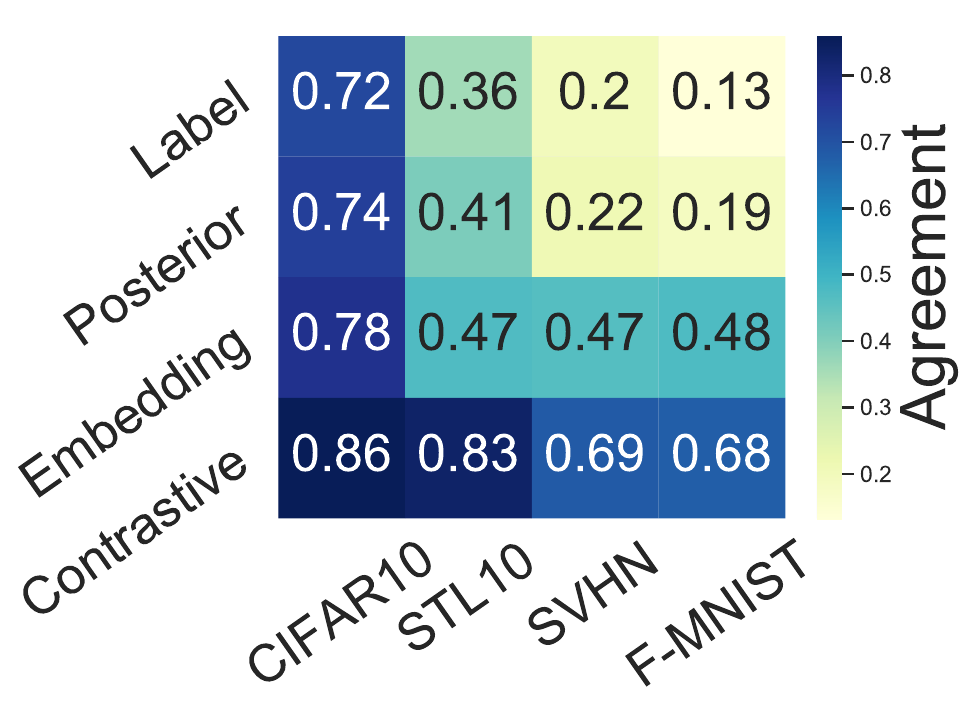}
\caption{BYOL}
\label{fig:heatmap_byol}
\end{subfigure}
\begin{subfigure}{0.48\columnwidth}
\includegraphics[width=\columnwidth]{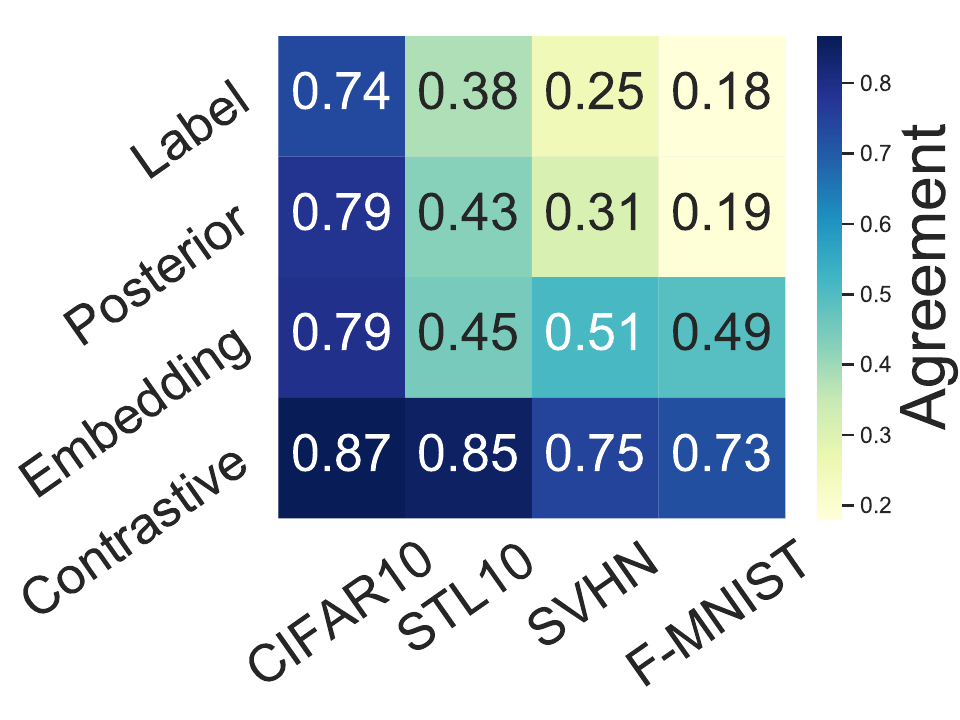}
\caption{SimSiam}
\label{fig:heatmap_simsiam}
\end{subfigure}

\caption{Heatmap of the agreement scores of model stealing attacks.
We show the performance of 16 combinations of different information that the target model outputs and the adversary's knowledge of target training data.
Target models are trained on CIFAR10.
}
\label{fig:heatmap_cifar10}
\end{figure*}

\mypara{Impact of Surrogate Dataset's Correlation With the Target Dataset}
In the meanwhile, since the adversary cannot always have knowledge about the target dataset, the impact of the surrogate dataset's correlation with the target dataset is also worth consideration.
We find that \ContSteal depends less on the surrogate dataset's distribution and can always achieve stable performance.
We plot the attack agreement in \autoref{fig:heatmap_cifar10} where the target encoders and downstream classifiers are trained on CIFAR10.
We can see that when the adversary conducts a conventional attack against the classifier, the adversary's knowledge of target training data is crucial.
For example, when the adversary can only get the predicted label from the target model, he/she can only achieve 0.182 agreement when using F-MNIST to attack the model trained by SimCLR, while it can achieve 0.711 agreement when using CIFAR10 as the surrogate dataset, which is same as target dataset.
However, compared to the predicted label or posterior as the response, embedding depends less on the surrogate dataset distribution, and \ContSteal can better leverage the embedding information, contributing to the less dependent on the surrogate dataset's distribution.
For instance, when the target model is trained by SimCLR, \ContSteal can achieve $0.832$ agreement when the surrogate dataset is STL10, which is even better than the best conventional attack ($0.781$) using the exact same target training dataset as the surrogate dataset and embedding as the response.
Such observation better implies that \ContSteal can always achieve good performance regardless of the surrogate dataset's distribution and can also achieve more generalized performance in practice.

\begin{table}[!htbp]
\centering
\caption{Impact of learning rate and batch size.
The target dataset and downstream dataset are both CIFAR10.
The surrogate dataset is STL10.
Note that for different learning rates, we set the batch size as 128.
For different batch sizes, we set the learning rate as 0.001
}
\scalebox{0.7}{
\begin{tabular}{l c c c}
\toprule
\textbf{Hyperparamter} & \textbf{Different Settings} & \textbf{Agreement}  \\
\midrule
\multirow{3}*{Learning Rate} & 0.001  & 0.813  \\
~ &  0.002  & 0.801  \\
~ &  0.003  & 0.805  \\
~ &  0.004 & 0.819 \\
~ &  0.005 & 0.809 \\
\midrule
\multirow{3}*{Batch Size} & 16  & 0.827  \\
~ &  32  & 0.806 \\
~ &  64  & 0.800 \\
~ &  128 & 0.813 \\
~ &  256 & 0.775 \\
\bottomrule
\end{tabular}
}
\label{table:hyperparameter}
\end{table}

\mypara{Impact of Hyperparameters}
In our experiments, we set batch size as 128 and learning rate as 0.001.
We show in \autoref{table:hyperparameter} that with reasonable batch size and learning rate, our \ContSteal can have stable performance.

\mypara{Impact of Negative Pairs Generated From the Surrogate Encoder}
In \ContSteal's loss functions, besides $D^{-}_{encoder}$, we also consider the distance of negative pairs generated from the surrogate encoder itself, i.e., $D^{-}_{self}$.
To evaluate the necessity of $D^{-}_{self}$, we take the target encoder trained by BYOL on CIFAR10 and the downstream task on STL10 as an example and study the attack performance with and without $D^{-}_{self}$.
The results are summarized in \autoref{table:half-contrastive}.
We find that adding $D^{-}_{self}$ greatly improves the attack performance in both accuracy and agreement.
For instance, when the surrogate dataset is STL10, the surrogate model stolen by \ContSteal with $D^{-}_{self}$ achieves $0.817$ agreement while only $0.314$ if without $D^{-}_{self}$.
The reason behind this is that the negative pairs generated from the surrogate encoder can serve as extra ``anchors'' to better locate the position of the embedding, which leads to higher agreement.
Such observation demonstrates that it is important to introduce $D^{-}_{self}$ in \ContSteal as well.

\begin{table}[!htbp]
\centering
\caption{The agreement and accuracy of different contrastive losses.
We use BYOL trained on STL10 as the target model.
}
\label{table:half-contrastive}
\scalebox{0.7}{
\begin{tabular}{llcc}
\toprule
\multirow{2}*{\textbf{Dataset}} & \multirow{2}*{\textbf{Method}} & \multicolumn{2}{c}{\textbf{BYOL}} \\
~ & ~  & \textbf{Agreement} & \textbf{Accuracy}  \\
\midrule
\multirow{2}*{CIFAR10} & w/o $D^{-}_{encoder}$  & 0.242 & 0.242   \\
~ & w $D^{-}_{encoder}$  &  0.844 & 0.843\\
\midrule
\multirow{2}*{F-MNIST} & w/o $D^{-}_{encoder}$  & 0.215 & 0.217   \\
~ & w $D^{-}_{encoder}$  & 0.647 & 0.641 \\
\midrule
\multirow{2}*{STL10} & w/o $D^{-}_{encoder}$  & 0.314 & 0.320  \\
~ & w $D^{-}_{encoder}$  & 0.817 & 0.811\\
\midrule
\multirow{2}*{SVHN}& w/o $D^{-}_{encoder}$  & 0.176 & 0.175  \\
~ & w $D^{-}_{encoder}$  & 0.655& 0.650\\
\bottomrule
\end{tabular}
}
\end{table}

\subsection{Further Attacks Based on \ContSteal}
\label{appendix:further}

As we have mentioned in the introduction part, model stealing can be used as a stepping stone for further attacks.
In this section, we select adversary sample attacks as a case study to show the importance of model stealing for further attacks on the target model.
Normally, the adversary can not obtain the gradient from the target model.
But to conduct adversary sample attacks, the adversary needs to obtain the gradient in most attack scenarios.
Therefore, the adversary can construct a surrogate model to generate the adversary sample and transfer it to the target model to perform the attack.
We consider three widely used mechanisms to generate adversarial examples, including Fast Gradient Sign Attack (FGSM)~\cite{GSS15}, Basic Iterative Methods (BIM)~\cite{KGB16}, and Projected Gradient Descent (PGD)~\cite{MMSTV18}.
Our target model is SimCLR pre-trained on CIFAR10 and the last layer classifier trained on STL10.
We also use STL10 as the surrogate dataset to conduct \ContSteal and generate adversary samples.
Experiments show that the surrogate model can generate adversary samples that are valid for the target model (\autoref{table:adversary}).
To show the necessity of the surrogate model as a springboard for the attack, we also conduct the baseline attack, which uses another model as the springboard to attack the target model.
We choose the normal ResNet18 model trained on SVHN as our baseline model and then apply the adversary example to attack the target model.
We observe that compared to the adversarial examples generated from the baseline model, those adversarial examples generated from the surrogate model constructed by \ContSteal can better transfer to the target model.
For instance, with PGD, the adversarial examples obtained from the surrogate model can lead to a lower classification accuracy (0.203) on the target model than those generated from the baseline model (0.246).
This implies that the model stealing attack can be a valid stepping stone for more effective further attacks.

\begin{table}[!htbp]
\centering
\caption{The different methods to create adversary sample to attack on surrogate model and target model.
[Lower is better]
}
\label{table:adversary}
\scalebox{0.7}
{
\begin{tabular}{lccc}
\toprule
\textbf{Method} & \textbf{Surrogate model (acc)} & \textbf{Target model (acc)} & \textbf{Baseline (acc)} \\
\midrule
FGSM ~\cite{GSS15} & 0.097 & 0.131 & 0.194 \\
\midrule
BIM ~\cite{KGB16} & 0.054 & 0.192 & 0.235\\
\midrule
PGD ~\cite{MMSTV18} & 0.092 & 0.203 & 0.246\\
\bottomrule
\end{tabular}
}
\end{table}

\subsection{More Results on Conventional Attacks}
\label{subsection:appendix_conventional}

\begin{figure*}[!t]
\centering
\begin{subfigure}{0.44\columnwidth}
\includegraphics[width=\columnwidth]{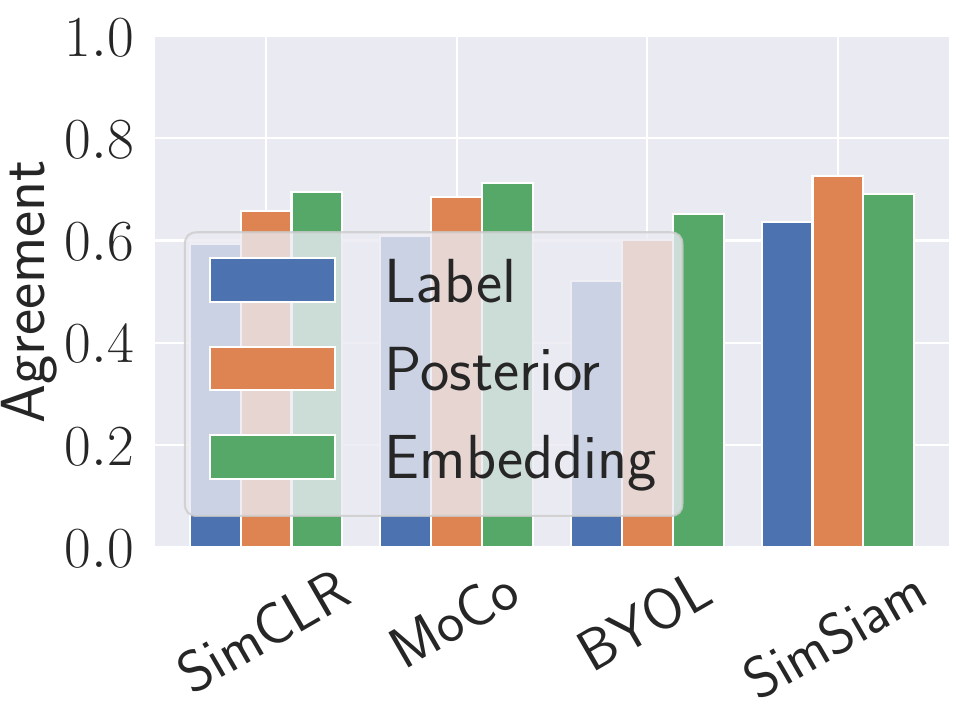}
\caption{CIFAR10}
\label{fig:agreement_normal_target_cifar10_stl10_cifar10}
\end{subfigure}
\begin{subfigure}{0.44\columnwidth}
\includegraphics[width=\columnwidth]{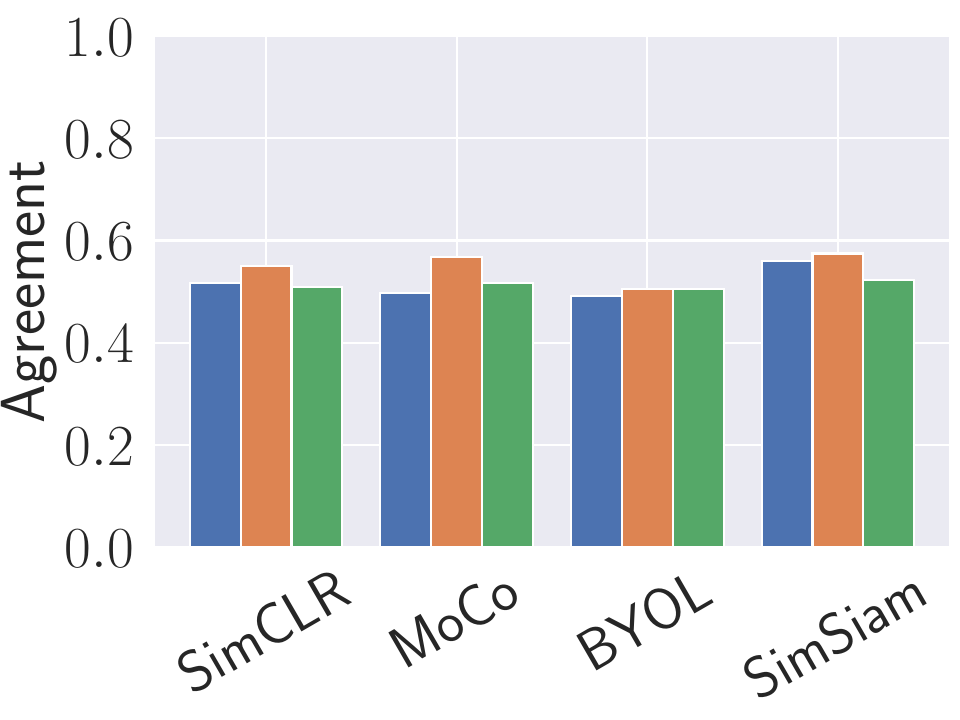}
\caption{STL10}
\label{fig:agreement_normal_target_cifar10_stl10_stl10}
\end{subfigure}
\begin{subfigure}{0.44\columnwidth}
\includegraphics[width=\columnwidth]{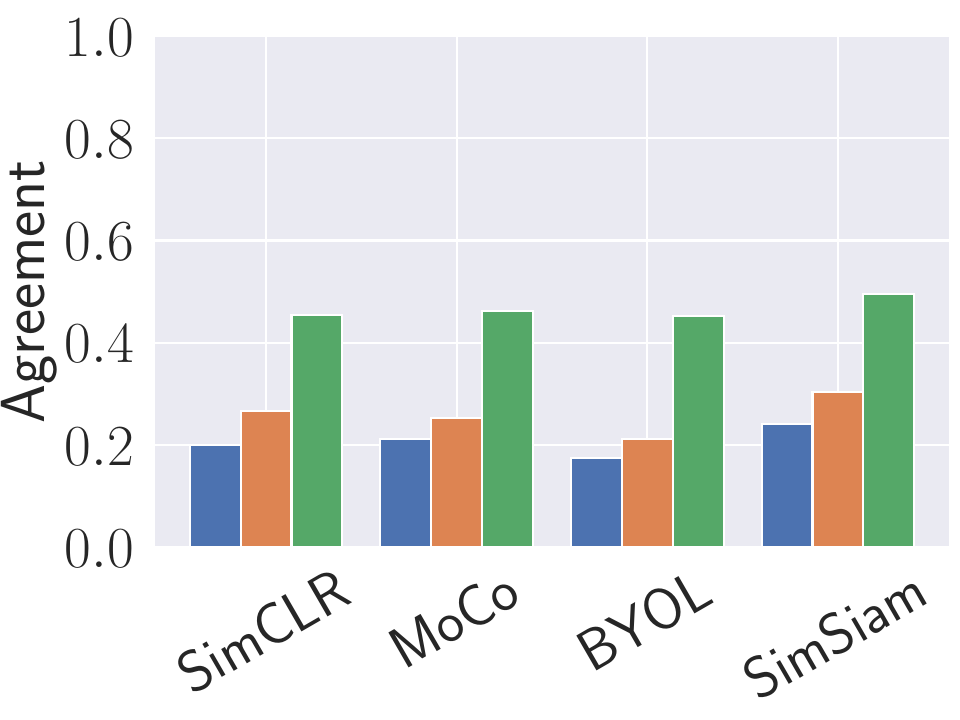}
\caption{F-MNIST}
\label{fig:agreement_normal_target_cifar10_stl10_mnist}
\end{subfigure}
\begin{subfigure}{0.44\columnwidth}
\includegraphics[width=\columnwidth]{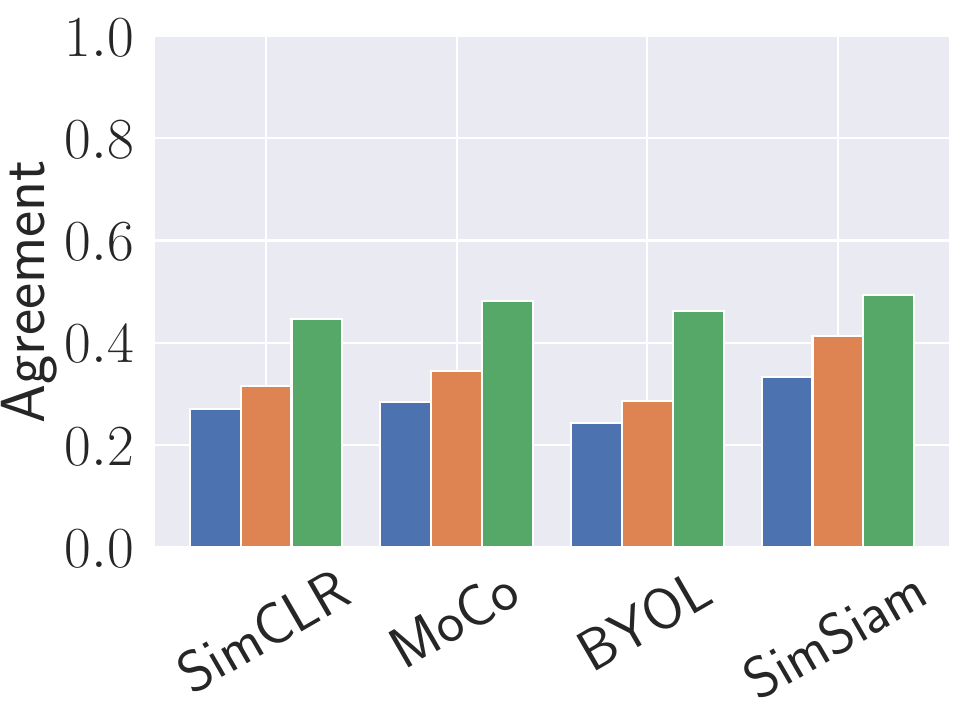}
\caption{SVHN}
\label{fig:agreement_normal_target_cifar10_stl10_svhn}
\end{subfigure}
\begin{subfigure}{0.44\columnwidth}
\includegraphics[width=\columnwidth]{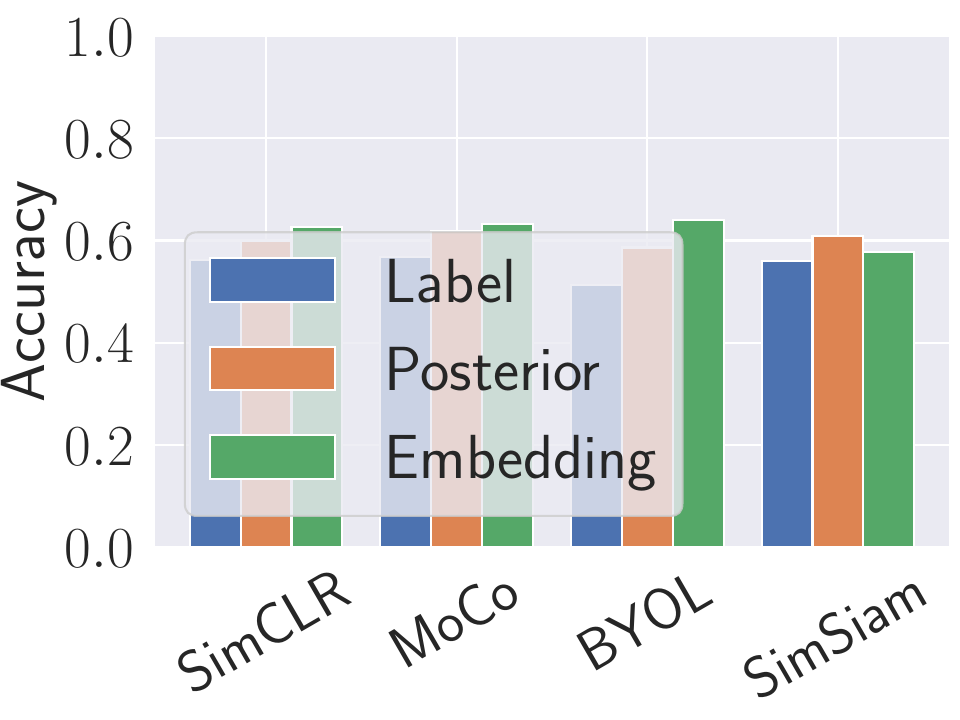}
\caption{CIFAR10}
\label{fig:accuracy_normal_target_cifar10_stl10_cifar10}
\end{subfigure}
\begin{subfigure}{0.44\columnwidth}
\includegraphics[width=\columnwidth]{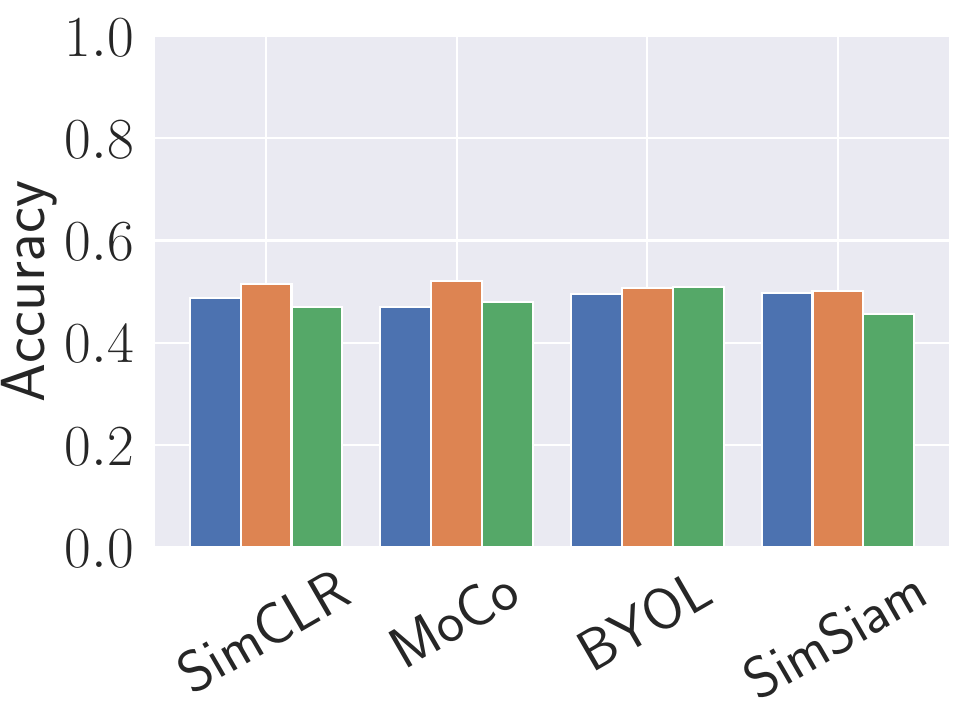}
\caption{STL10}
\label{fig:accuracy_normal_target_cifar10_stl10_stl10}
\end{subfigure}
\begin{subfigure}{0.44\columnwidth}
\includegraphics[width=\columnwidth]{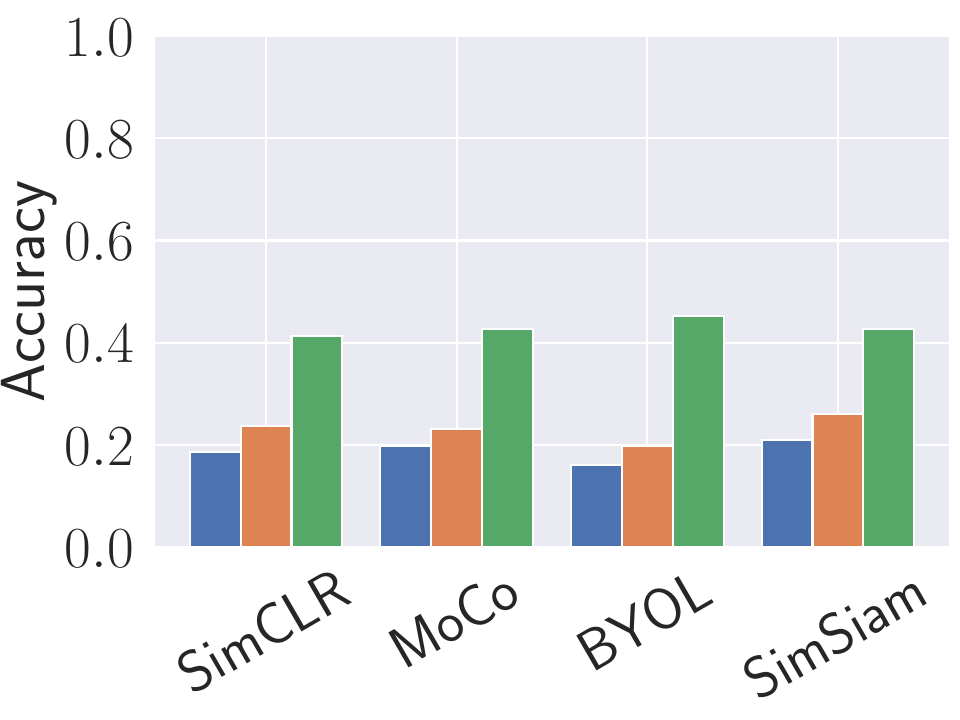}
\caption{F-MNIST}
\label{fig:accuracy_normal_target_cifar10_stl10_mnist}
\end{subfigure}
\begin{subfigure}{0.44\columnwidth}
\includegraphics[width=\columnwidth]{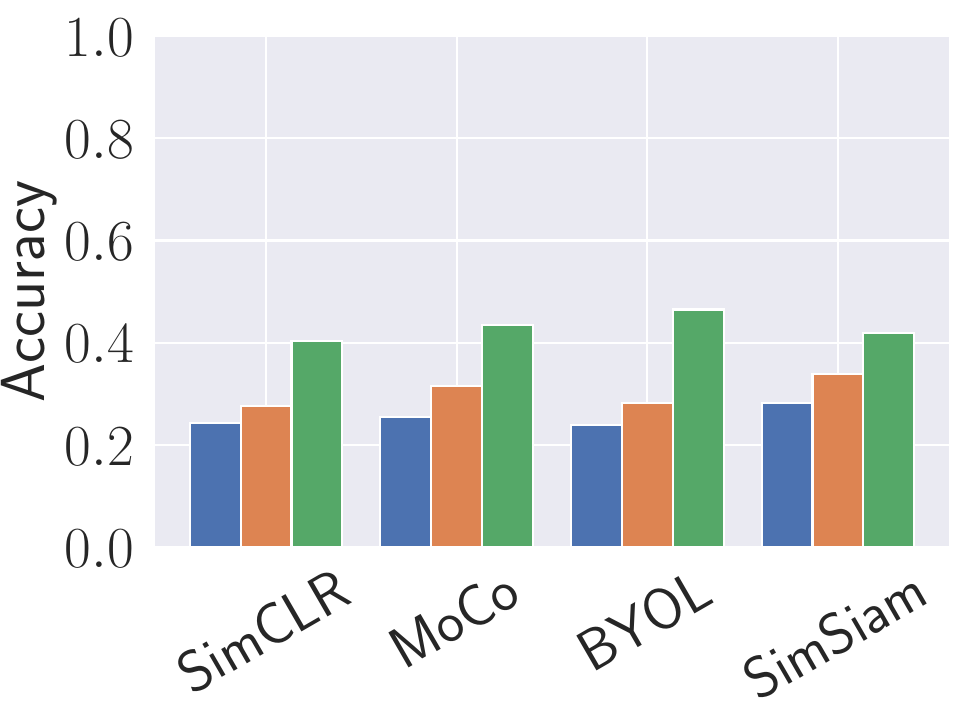}
\caption{SVHN}
\label{fig:accuracy_normal_target_cifar10_stl10_svhn}
\end{subfigure}
\caption{The performance of model stealing attack against target encodes and downstream classifiers trained on CIFAR10 and STL10.
Target models can output predicted labels, posteriors, or embeddings.
The adversary uses CIFAR10, STL10, Fashion-MNIST (F-MNIST), SVHN to conduct model stealing attacks.
The x-axis represents different kinds of target models.
The first line's y-axis represents the agreement of the model stealing attack. 
The second line's y-axis represents the accuracy of the model stealing attack.}
\label{fig:encoder_cifar10_stl10}
\end{figure*}

\begin{figure*}[!t]
\centering
\begin{subfigure}{0.44\columnwidth}
\includegraphics[width=\columnwidth]{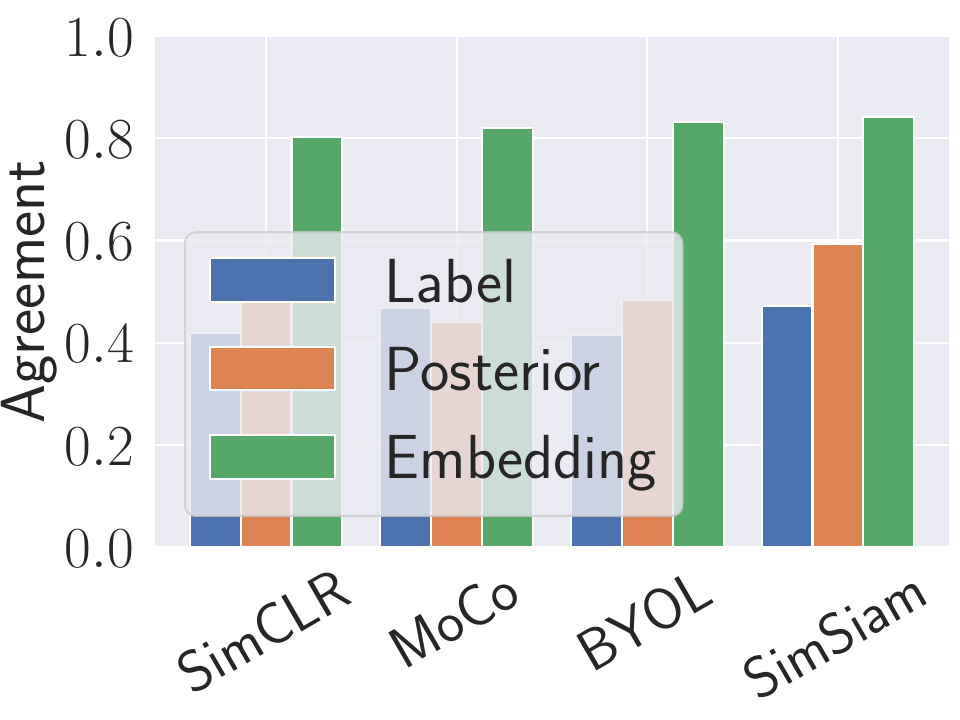}
\caption{CIFAR10}
\label{fig:agreement_normal_target_cifar10_mnist_cifar10}
\end{subfigure}
\begin{subfigure}{0.44\columnwidth}
\includegraphics[width=\columnwidth]{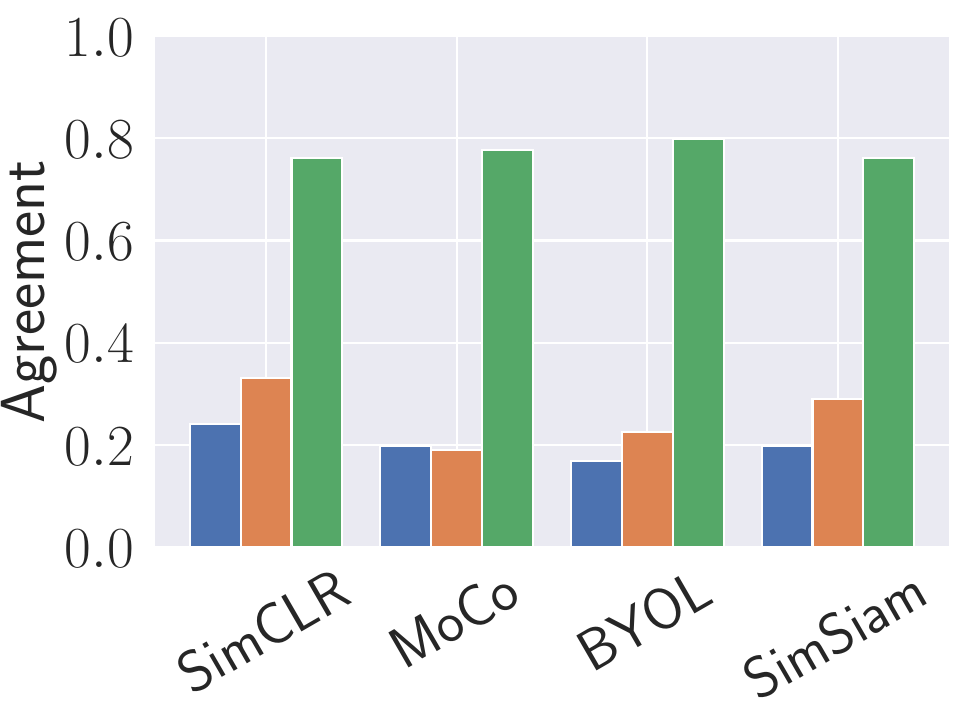}
\caption{STL10}
\label{fig:agreement_normal_target_cifar10_mnist_stl10}
\end{subfigure}
\begin{subfigure}{0.44\columnwidth}
\includegraphics[width=\columnwidth]{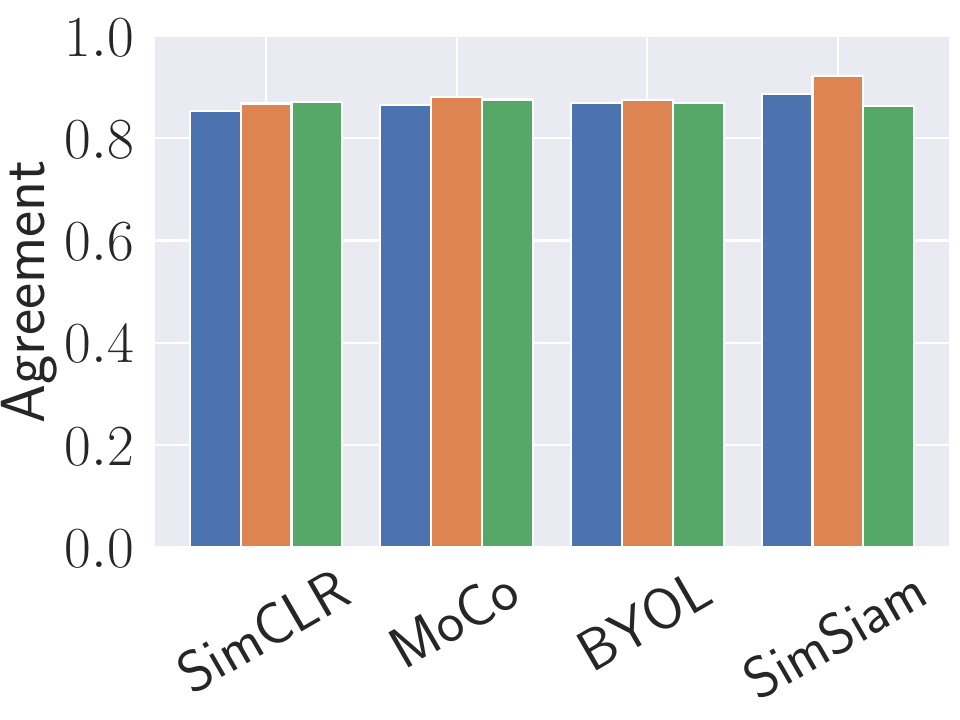}
\caption{F-MNIST}
\label{fig:agreement_normal_target_cifar10_mnist_mnist}
\end{subfigure}
\begin{subfigure}{0.44\columnwidth}
\includegraphics[width=\columnwidth]{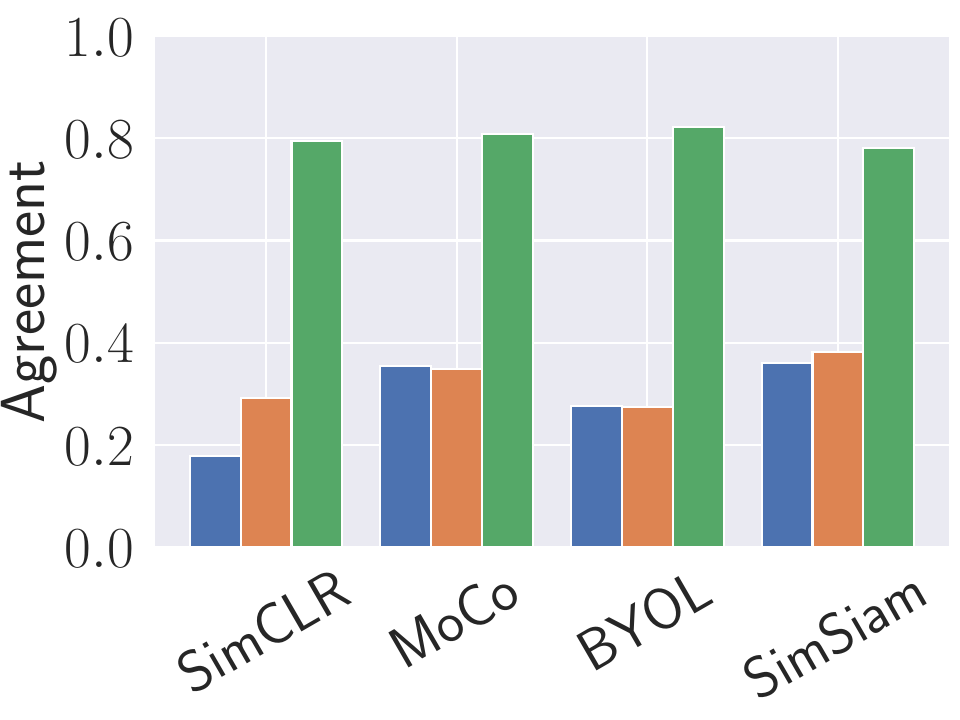}
\caption{SVHN}
\label{fig:agreement_normal_target_cifar10_mnist_svhn}
\end{subfigure}
\begin{subfigure}{0.44\columnwidth}
\includegraphics[width=\columnwidth]{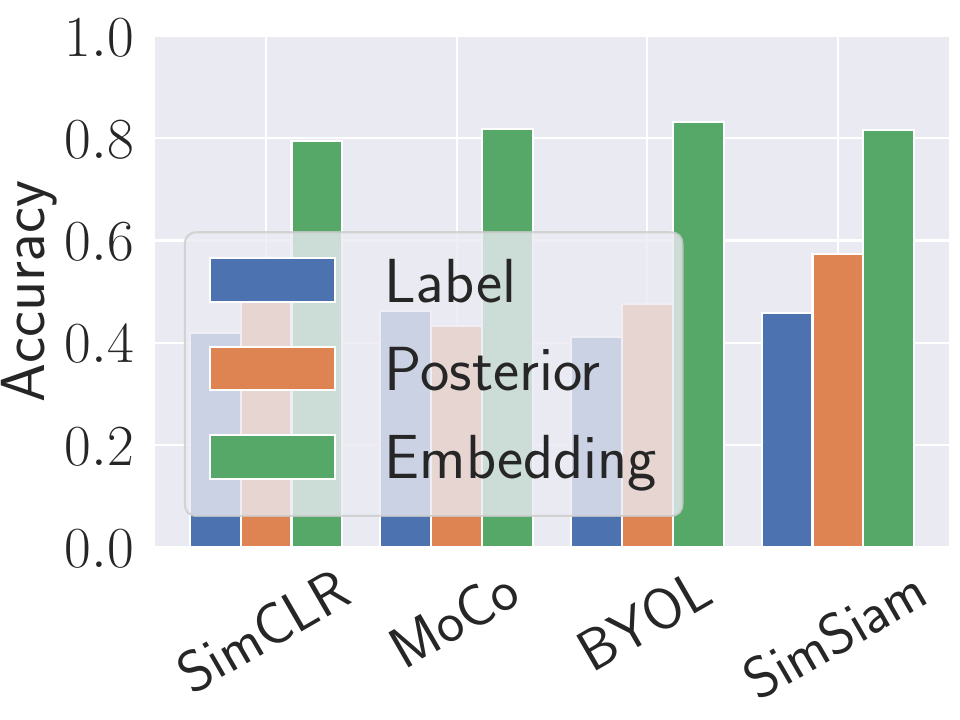}
\caption{CIFAR10}
\label{fig:accuracy_normal_target_cifar10_mnist_cifar10}
\end{subfigure}
\begin{subfigure}{0.44\columnwidth}
\includegraphics[width=\columnwidth]{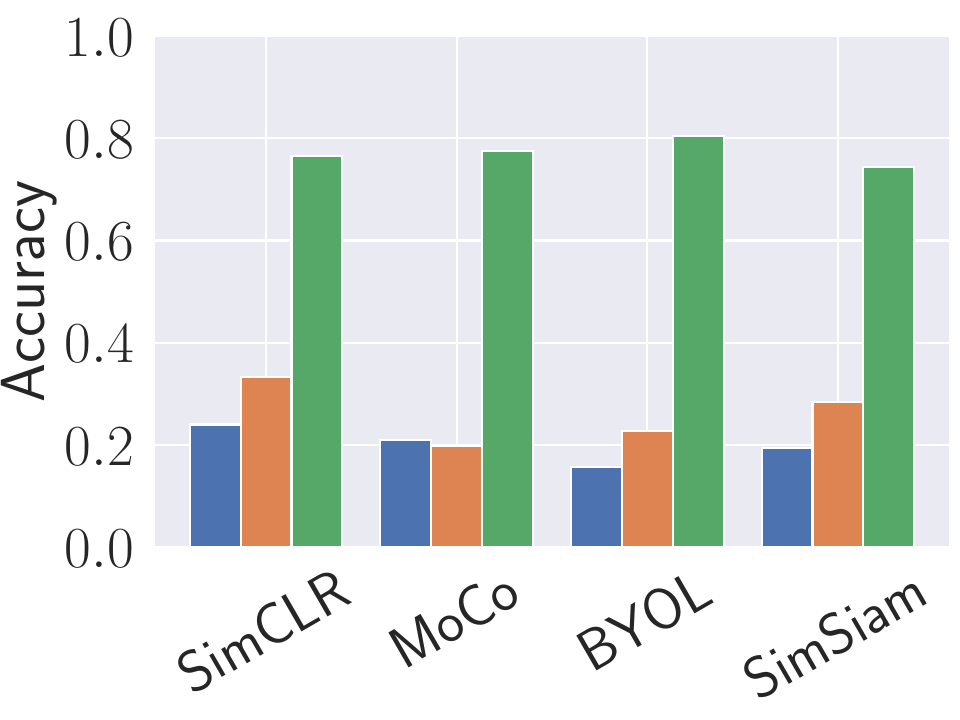}
\caption{STL10}
\label{fig:accuracy_normal_target_cifar10_mnist_stl10}
\end{subfigure}
\begin{subfigure}{0.44\columnwidth}
\includegraphics[width=\columnwidth]{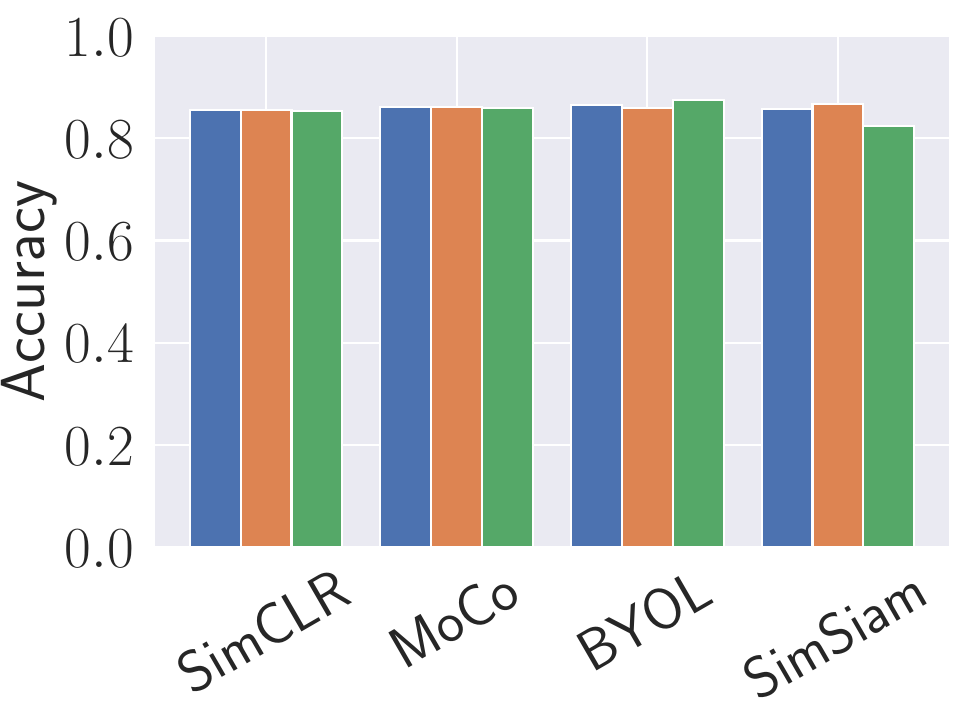}
\caption{F-MNIST}
\label{fig:accuracy_normal_target_cifar10_mnist_mnist}
\end{subfigure}
\begin{subfigure}{0.44\columnwidth}
\includegraphics[width=\columnwidth]{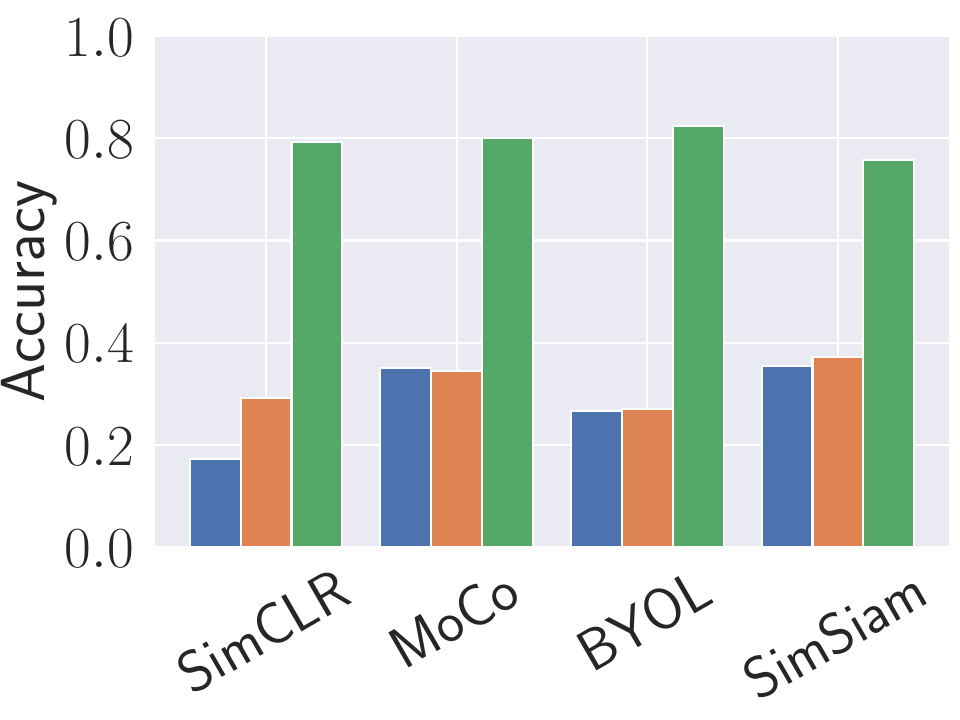}
\caption{SVHN}
\label{fig:accuracy_normal_target_cifar10_mnist_svhn}
\end{subfigure}
\caption{The performance of model stealing attack against target encodes and downstream classifiers trained on CIFAR10 and Fashon-MNIST.
Target models can output predicted labels, posteriors, or embeddings.
The adversary uses CIFAR10, STL10, Fashion-MNIST (F-MNIST), SVHN to conduct model stealing attacks.
The x-axis represents different kinds of target models.
The first line's y-axis represents the agreement of the model stealing attack. 
The second line's y-axis represents the accuracy of the model stealing attack.}
\label{fig:encoder_cifar10_mnist}
\end{figure*}

\begin{figure*}[!t]
\centering
\begin{subfigure}{0.44\columnwidth}
\includegraphics[width=\columnwidth]{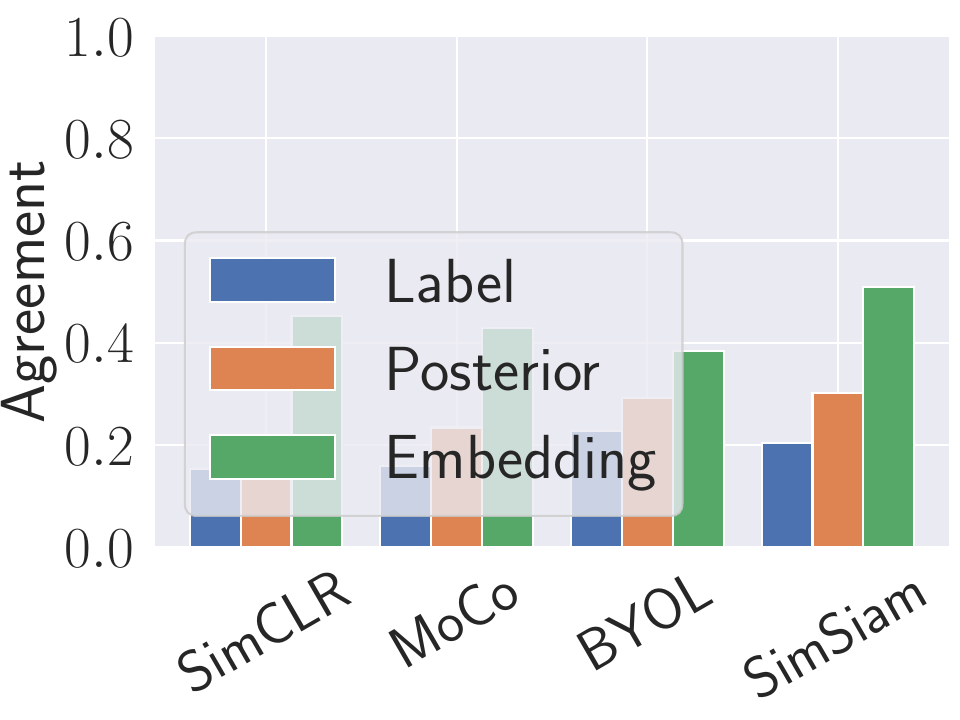}
\caption{CIFAR10}
\label{fig:agreement_normal_target_cifar10_svhn_cifar10}
\end{subfigure}
\begin{subfigure}{0.44\columnwidth}
\includegraphics[width=\columnwidth]{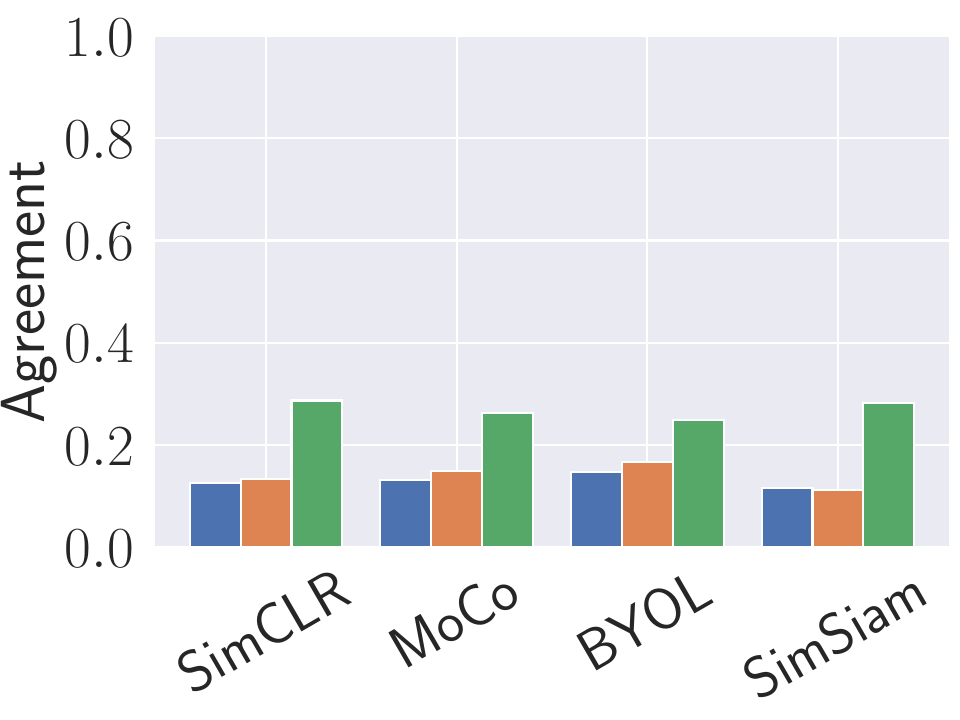}
\caption{STL10}
\label{fig:agreement_normal_target_cifar10_svhn_stl10}
\end{subfigure}
\begin{subfigure}{0.44\columnwidth}
\includegraphics[width=\columnwidth]{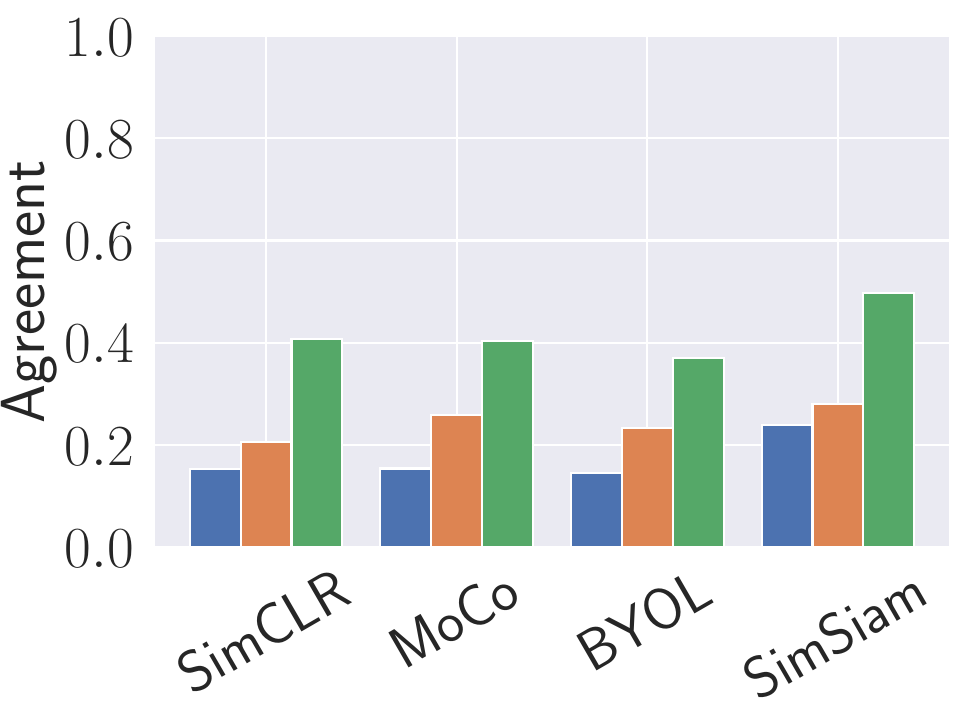}
\caption{F-MNIST}
\label{fig:agreement_normal_target_cifar10_svhn_mnist}
\end{subfigure}
\begin{subfigure}{0.44\columnwidth}
\includegraphics[width=\columnwidth]{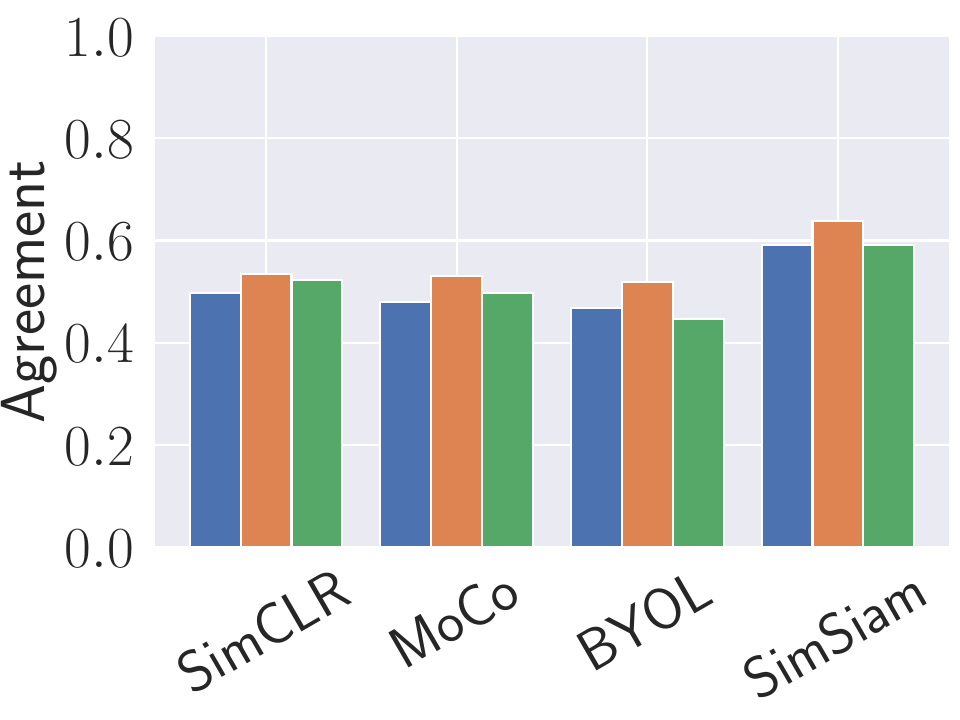}
\caption{SVHN}
\label{fig:agreement_normal_target_cifar10_svhn_svhn}
\end{subfigure}
\begin{subfigure}{0.44\columnwidth}
\includegraphics[width=\columnwidth]{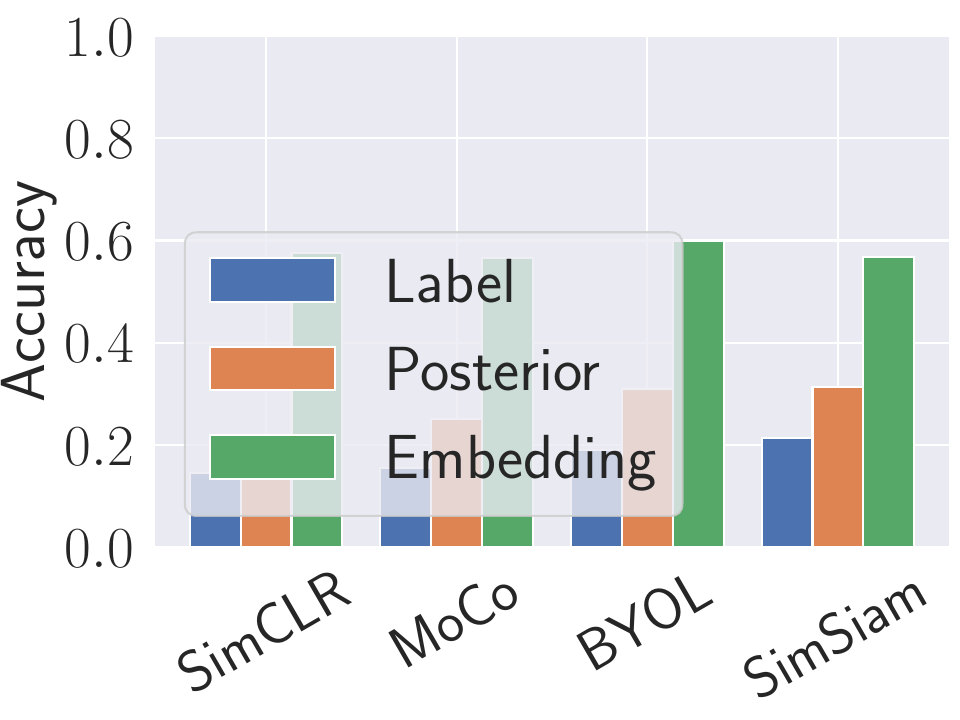}
\caption{CIFAR10}
\label{fig:accuracy_normal_target_cifar10_svhn_cifar10}
\end{subfigure}
\begin{subfigure}{0.44\columnwidth}
\includegraphics[width=\columnwidth]{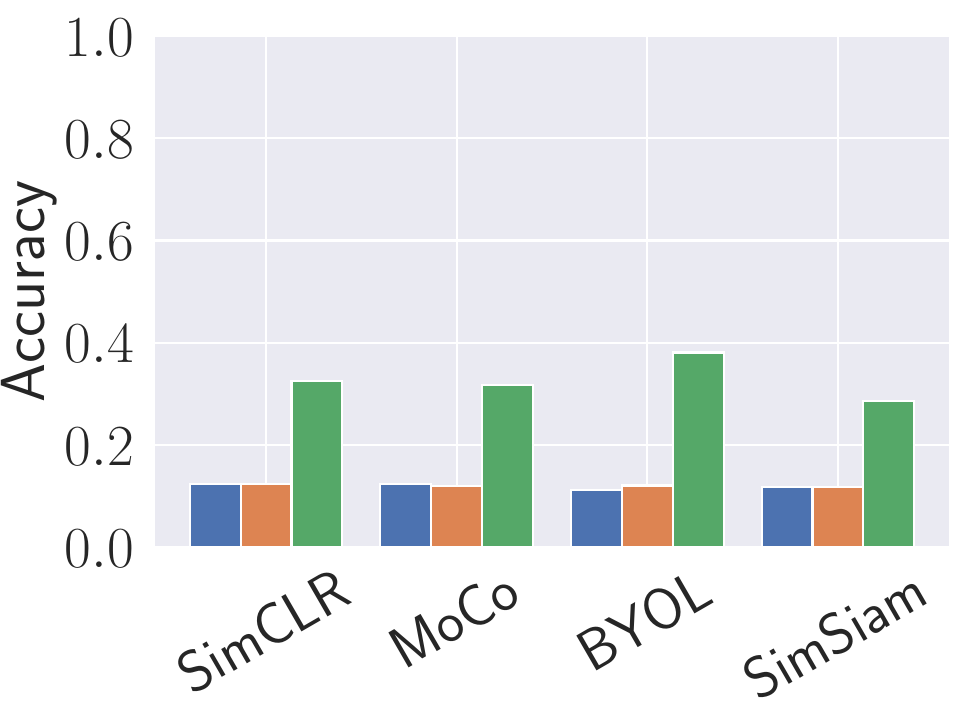}
\caption{STL10}
\label{fig:accuracy_normal_target_cifar10_svhn_stl10}
\end{subfigure}
\begin{subfigure}{0.44\columnwidth}
\includegraphics[width=\columnwidth]{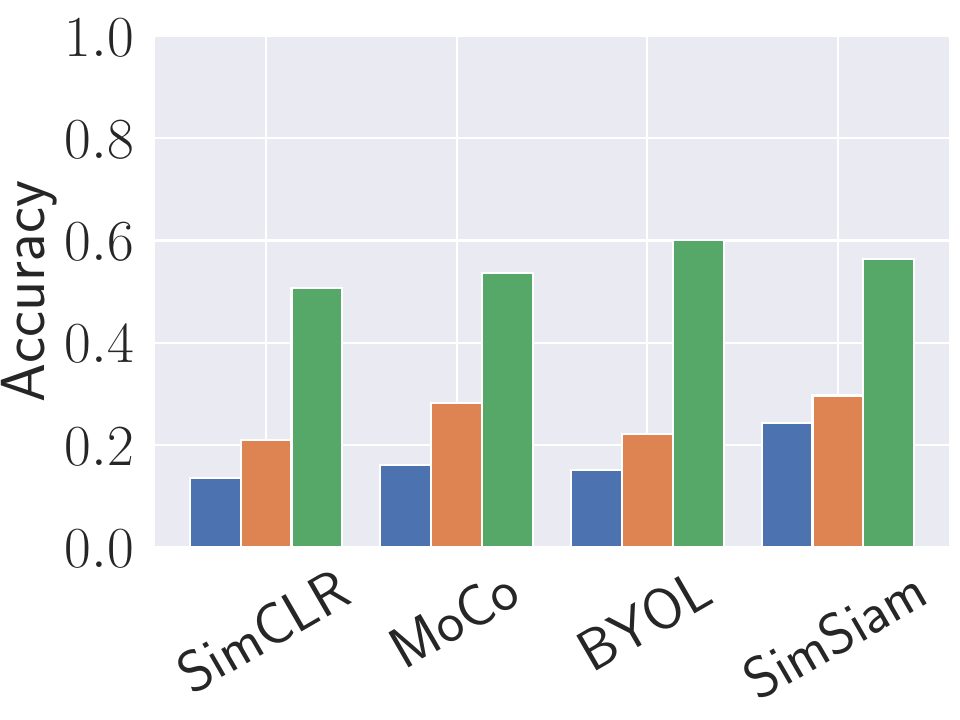}
\caption{F-MNIST}
\label{fig:accuracy_normal_target_cifar10_svhn_mnist}
\end{subfigure}
\begin{subfigure}{0.44\columnwidth}
\includegraphics[width=\columnwidth]{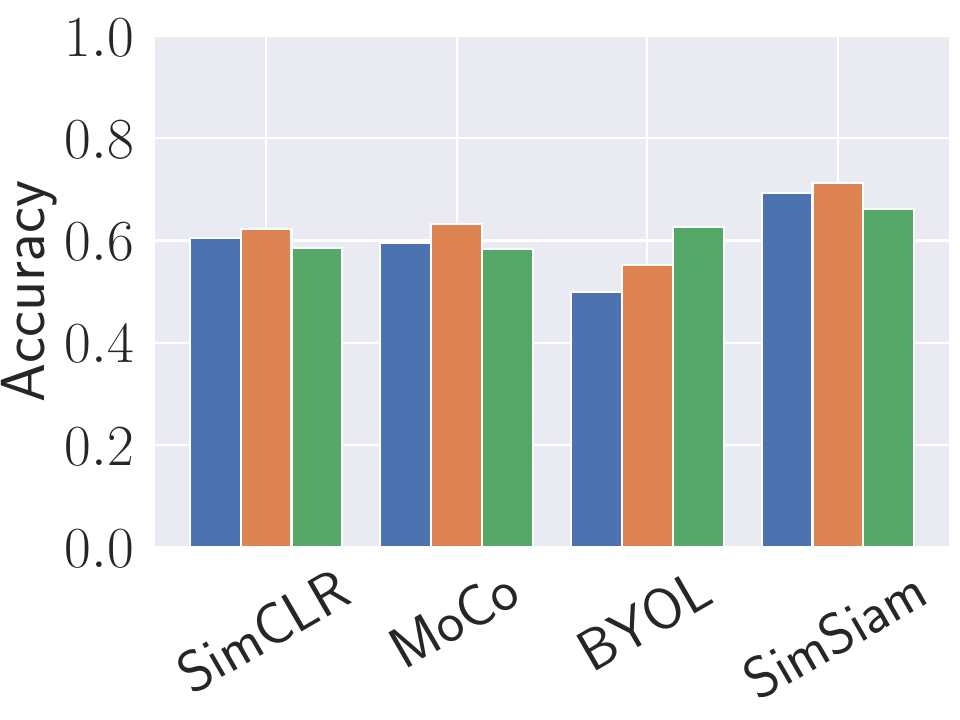}
\caption{SVHN}
\label{fig:accuracy_normal_target_cifar10_svhn_svhn}
\end{subfigure}
\caption{The performance of model stealing attack against target encodes and downstream classifiers trained on CIFAR10 and SVHN.
Target models can output predicted labels, posteriors, or embeddings.
The adversary uses CIFAR10, STL10, Fashion-MNIST (F-MNIST), SVHN to conduct model stealing attacks.
The x-axis represents different kinds of target models.
The first line's y-axis represents the agreement of the model stealing attack. 
The second line's y-axis represents the accuracy of the model stealing attack.}
\label{fig:encoder_cifar10_svhn}
\end{figure*}

\begin{figure*}[!t]
\centering
\begin{subfigure}{0.44\columnwidth}
\includegraphics[width=\columnwidth]{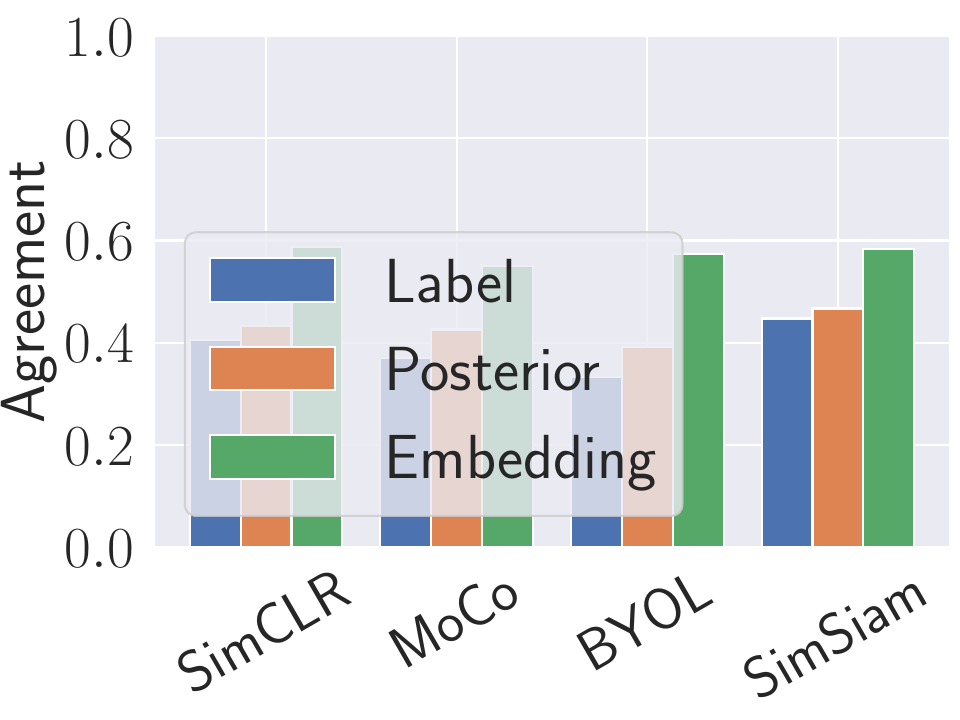}
\caption{CIFAR10}
\label{fig:agreement_normal_target_imagenet_stl10_cifar10}
\end{subfigure}
\begin{subfigure}{0.44\columnwidth}
\includegraphics[width=\columnwidth]{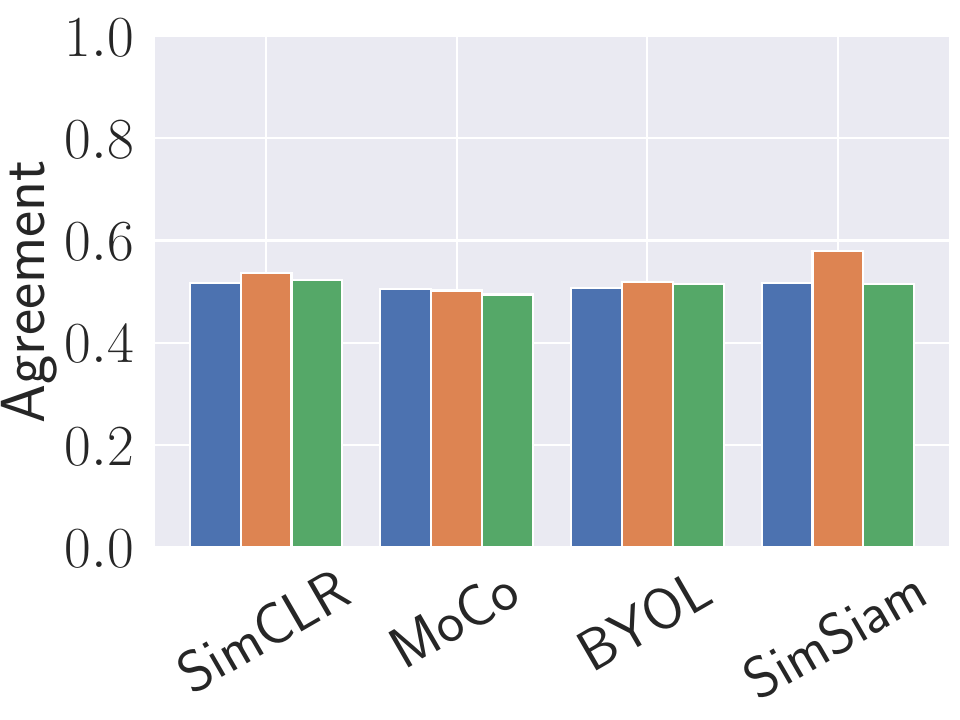}
\caption{STL10}
\label{fig:agreement_normal_target_imagenet_stl10_stl10}
\end{subfigure}
\begin{subfigure}{0.44\columnwidth}
\includegraphics[width=\columnwidth]{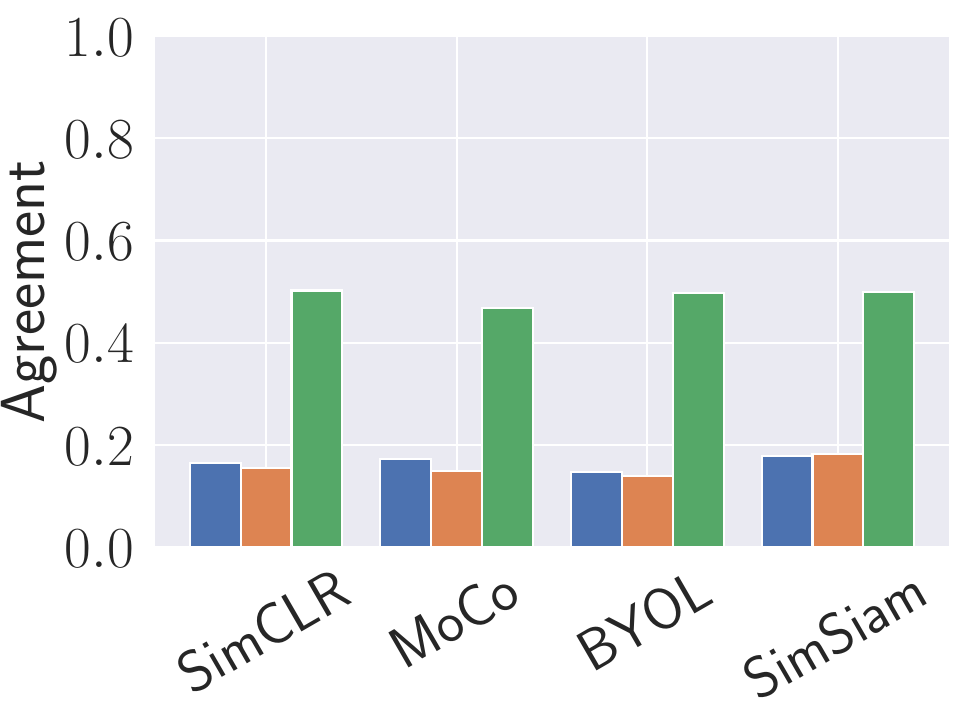}
\caption{F-MNIST}
\label{fig:agreement_normal_target_imagenet_stl10_mnist}
\end{subfigure}
\begin{subfigure}{0.44\columnwidth}
\includegraphics[width=\columnwidth]{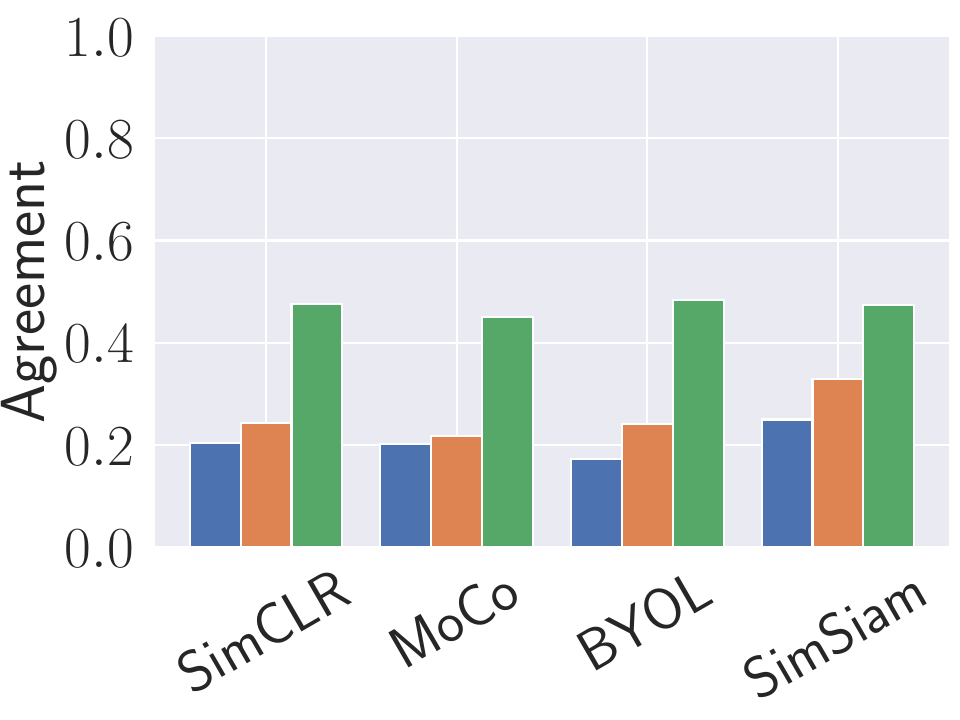}
\caption{SVHN}
\label{fig:agreement_normal_target_imagenet_stl10_svhn}
\end{subfigure}
\begin{subfigure}{0.44\columnwidth}
\includegraphics[width=\columnwidth]{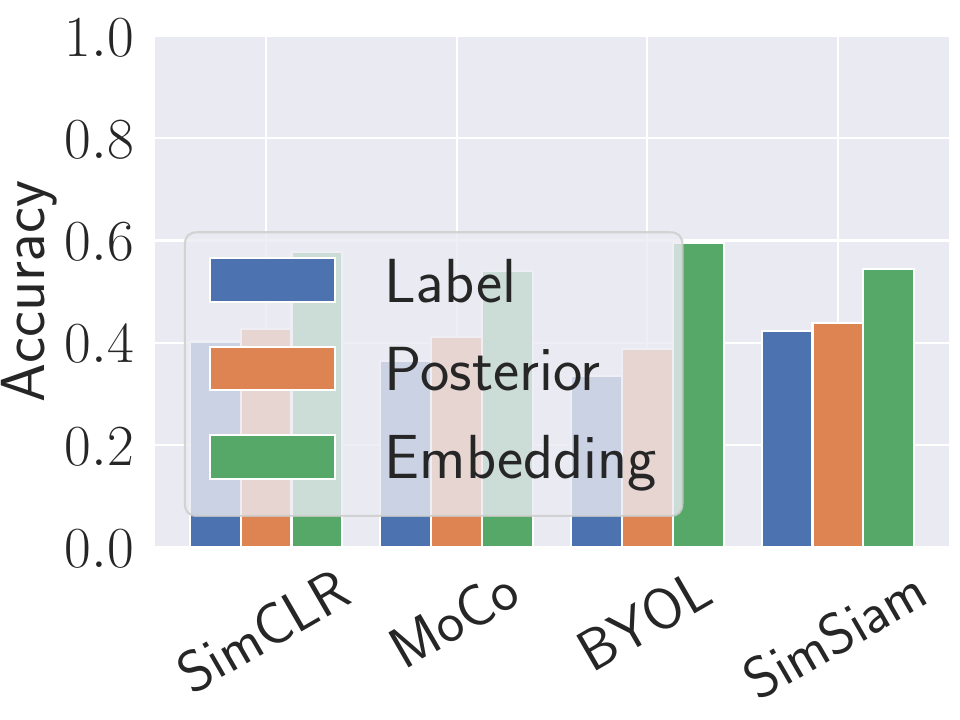}
\caption{CIFAR10}
\label{fig:accuracy_normal_target_imagenet_stl10_cifar10}
\end{subfigure}
\begin{subfigure}{0.44\columnwidth}
\includegraphics[width=\columnwidth]{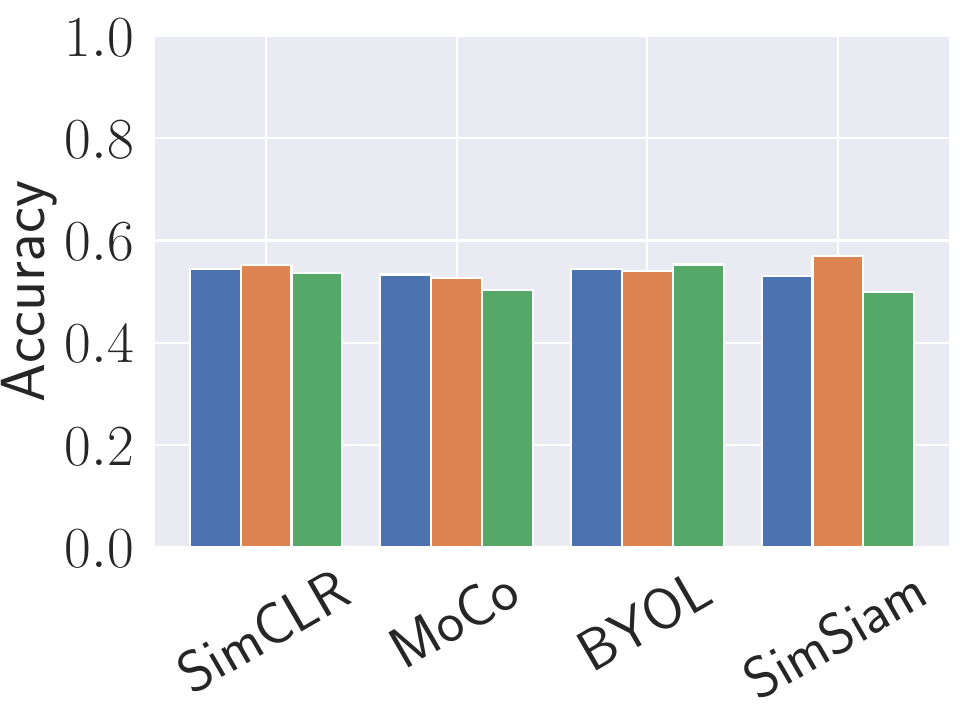}
\caption{STL10}
\label{fig:accuracy_normal_target_imagenet_stl10_stl10}
\end{subfigure}
\begin{subfigure}{0.44\columnwidth}
\includegraphics[width=\columnwidth]{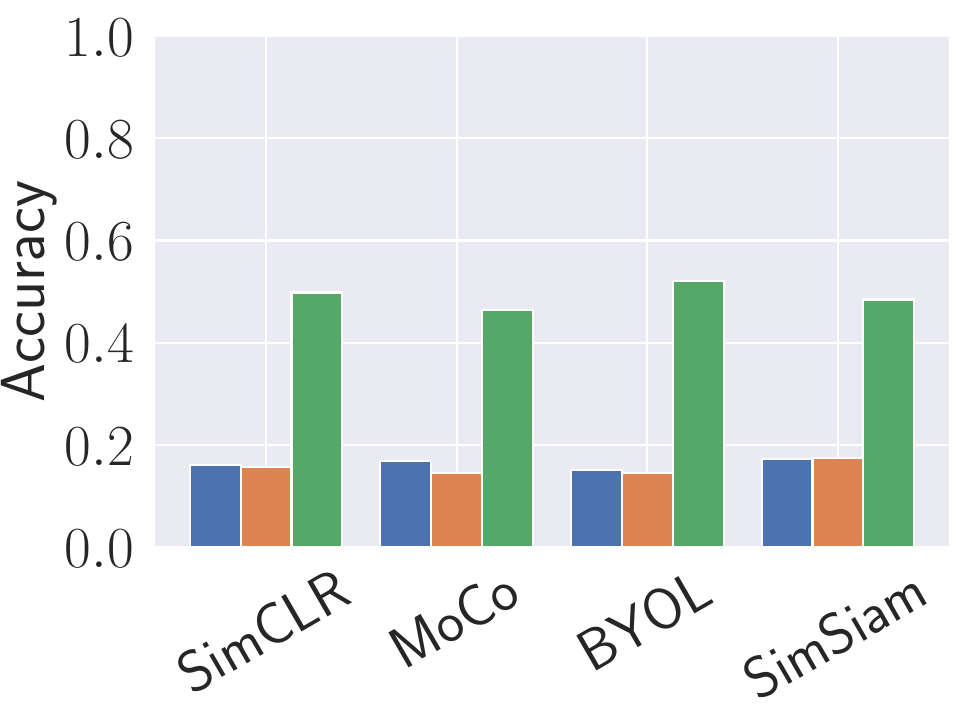}
\caption{F-MNIST}
\label{fig:accuracy_normal_target_imagenet_stl10_mnist}
\end{subfigure}
\begin{subfigure}{0.44\columnwidth}
\includegraphics[width=\columnwidth]{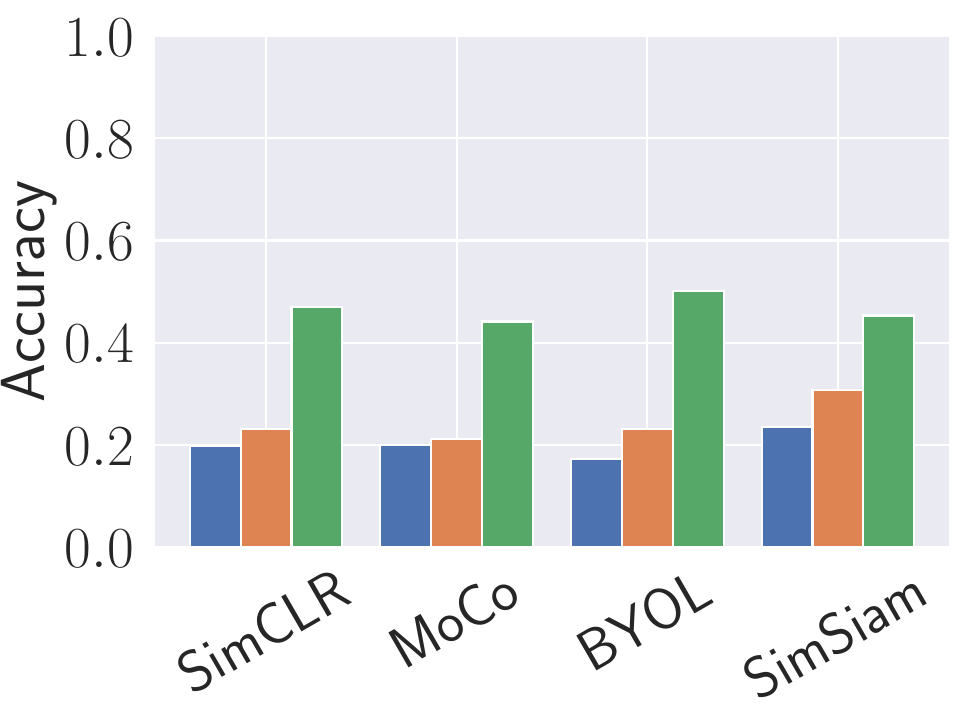}
\caption{SVHN}
\label{fig:accuracy_normal_target_imagenet_stl10_svhn}
\end{subfigure}
\caption{The performance of model stealing attack against target encodes and downstream classifiers trained on ImageNet and STL10.
Target models can output predicted labels, posteriors, or embeddings.
The adversary uses CIFAR10, STL10, Fashion-MNIST (F-MNIST), SVHN to conduct model stealing attacks.
The x-axis represents different kinds of target models.
The first line's y-axis represents the agreement of the model stealing attack. 
The second line's y-axis represents the accuracy of the model stealing attack.}
\label{fig:encoder_imagenet_stl10}
\end{figure*}

\begin{figure*}[!t]
\centering
\begin{subfigure}{0.44\columnwidth}
\includegraphics[width=\columnwidth]{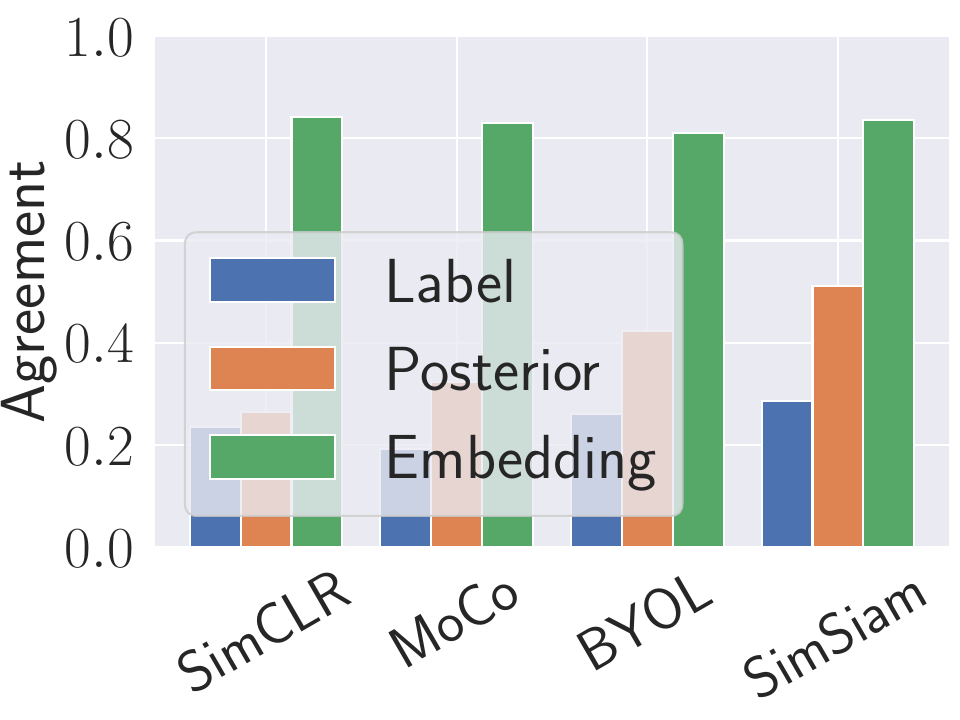}
\caption{CIFAR10}
\label{fig:agreement_normal_target_imagnet_mnist_cifar10}
\end{subfigure}
\begin{subfigure}{0.44\columnwidth}
\includegraphics[width=\columnwidth]{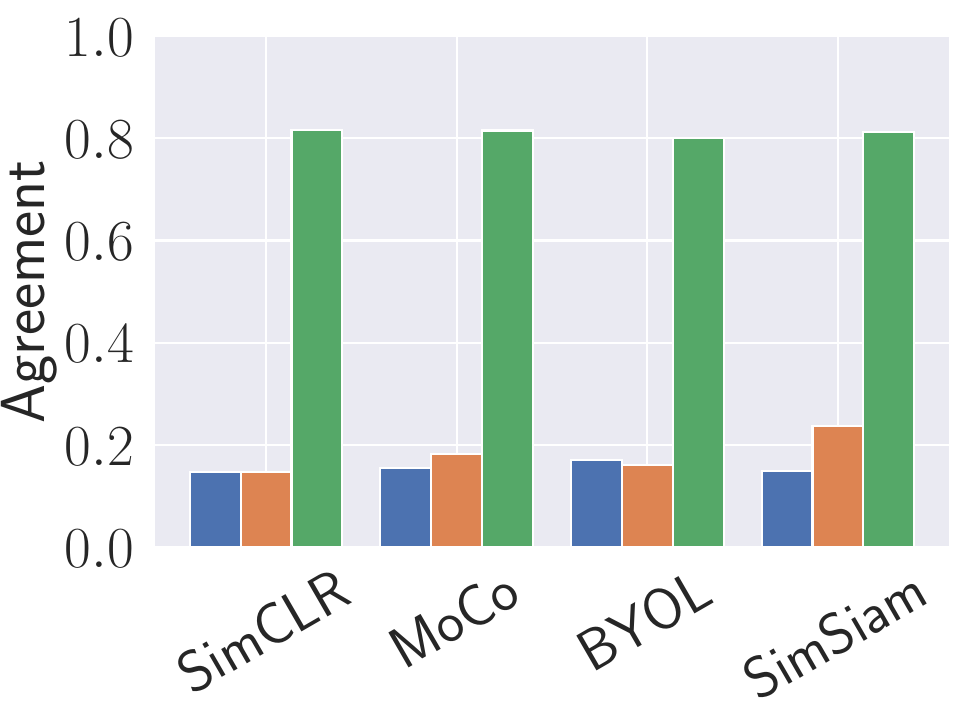}
\caption{STL10}
\label{fig:agreement_normal_target_imagenet_mnist_stl10}
\end{subfigure}
\begin{subfigure}{0.44\columnwidth}
\includegraphics[width=\columnwidth]{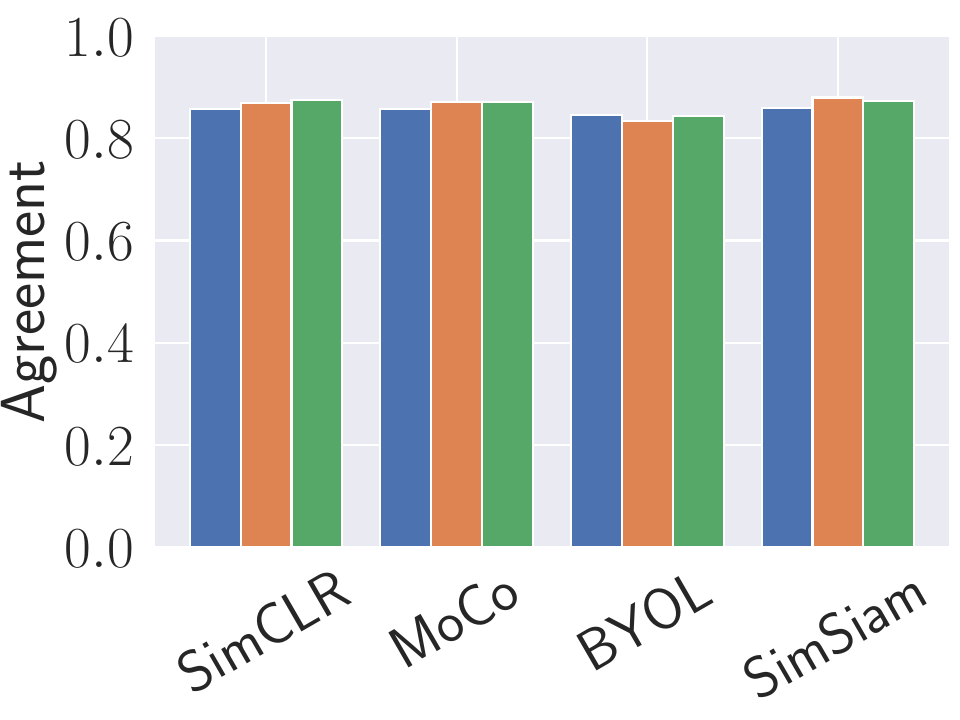}
\caption{F-MNIST}
\label{fig:agreement_normal_target_imagenet_mnist_mnist}
\end{subfigure}
\begin{subfigure}{0.44\columnwidth}
\includegraphics[width=\columnwidth]{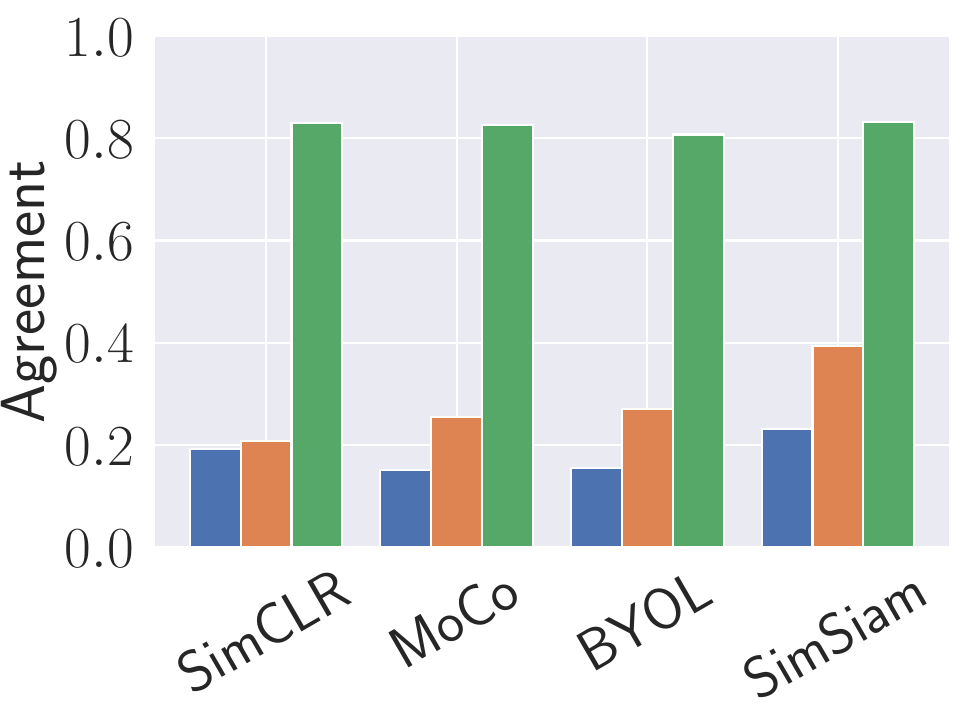}
\caption{SVHN}
\label{fig:agreement_normal_target_imagenet_mnist_svhn}
\end{subfigure}
\begin{subfigure}{0.44\columnwidth}
\includegraphics[width=\columnwidth]{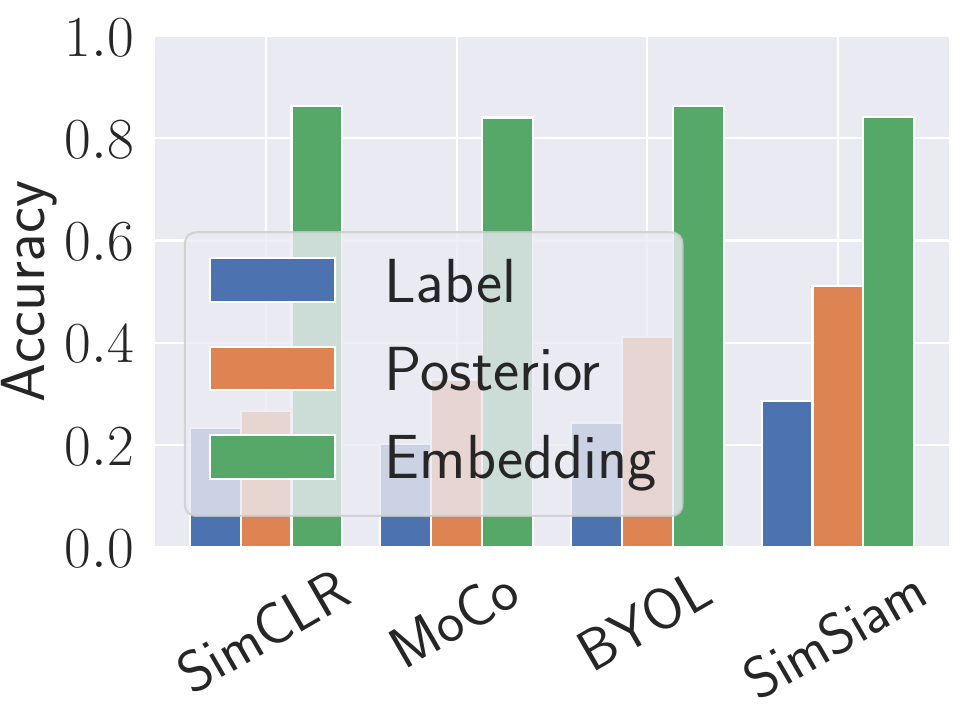}
\caption{CIFAR10}
\label{fig:accuracy_normal_target_imagenet_mnist_cifar10}
\end{subfigure}
\begin{subfigure}{0.44\columnwidth}
\includegraphics[width=\columnwidth]{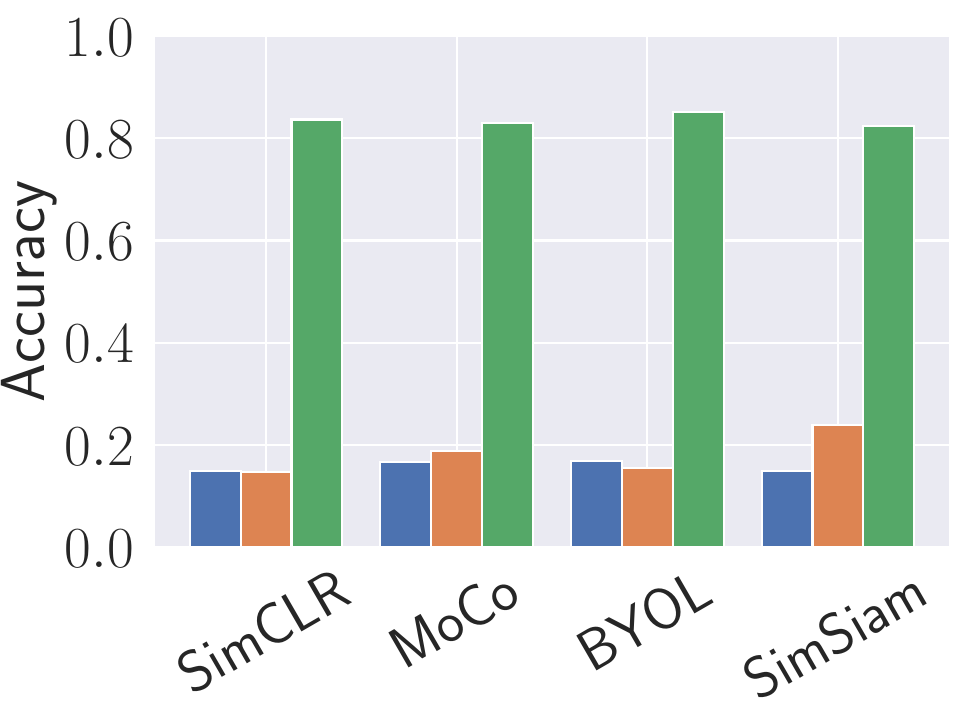}
\caption{STL10}
\label{fig:accuracy_normal_target_imagenet_mnist_stl10}
\end{subfigure}
\begin{subfigure}{0.44\columnwidth}
\includegraphics[width=\columnwidth]{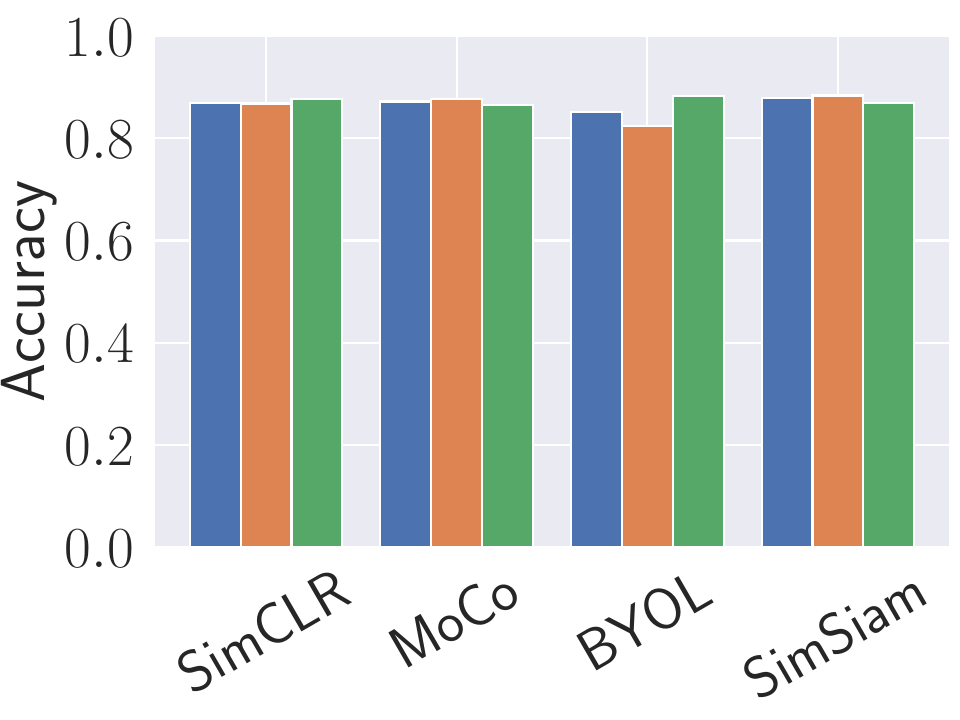}
\caption{F-MNIST}
\label{fig:accuracy_normal_target_imagenet_mnist_mnist}
\end{subfigure}
\begin{subfigure}{0.44\columnwidth}
\includegraphics[width=\columnwidth]{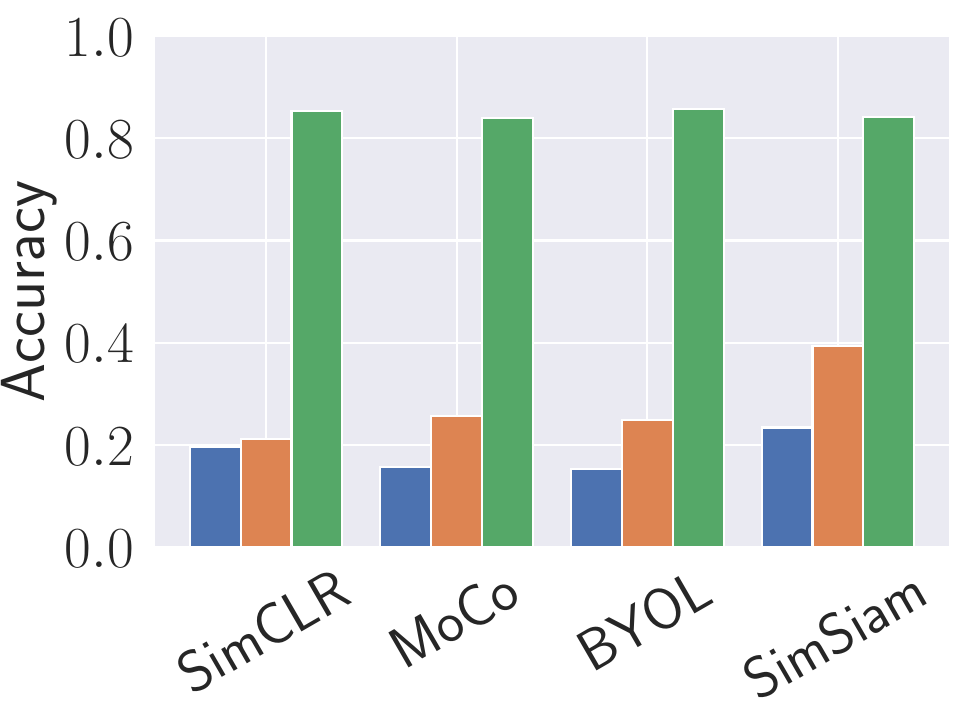}
\caption{SVHN}
\label{fig:accuracy_normal_target_imagenet_mnist_svhn}
\end{subfigure}
\caption{The performance of model stealing attack against target encodes and downstream classifiers trained on ImageNet and Fashion-MNIST.
Target models can output predicted labels, posteriors, or embeddings.
The adversary uses CIFAR10, STL10, Fashion-MNIST (F-MNIST), SVHN to conduct model stealing attacks.
The x-axis represents different kinds of target models.
The first line's y-axis represents the agreement of the model stealing attack. 
The second line's y-axis represents the accuracy of the model stealing attack.}
\label{fig:encoder_imagenet_mnist}
\end{figure*}

\begin{figure*}[!t]
\centering
\begin{subfigure}{0.44\columnwidth}
\includegraphics[width=\columnwidth]{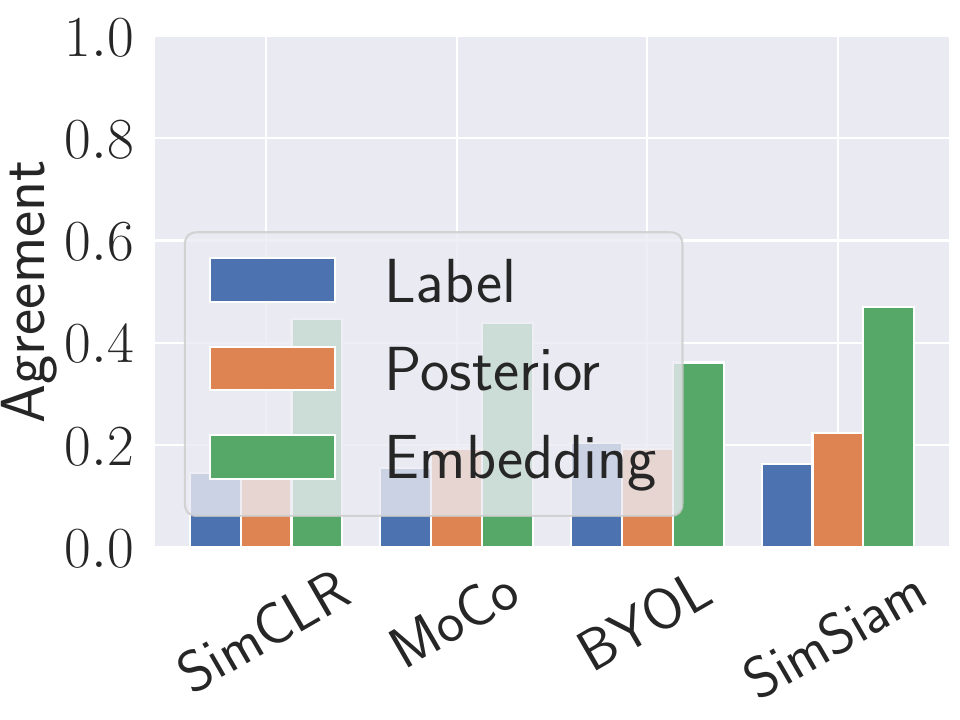}
\caption{CIFAR10}
\label{fig:agreement_normal_target_imagnet_cifar10_cifar10}
\end{subfigure}
\begin{subfigure}{0.44\columnwidth}
\includegraphics[width=\columnwidth]{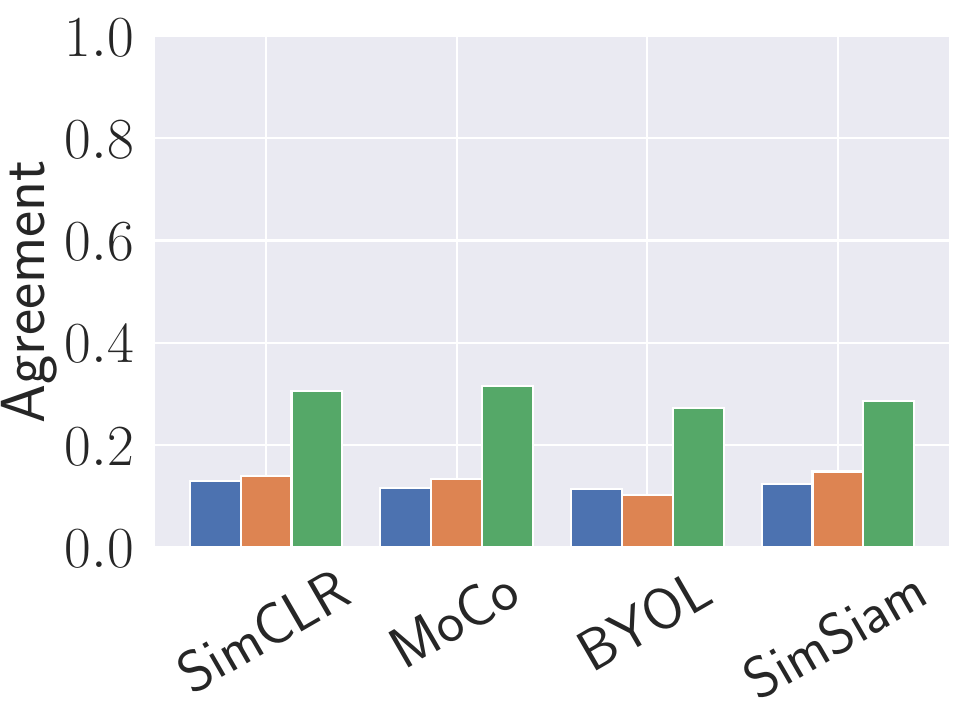}
\caption{STL10}
\label{fig:agreement_normal_target_imagenet_svhn_stl10}
\end{subfigure}
\begin{subfigure}{0.44\columnwidth}
\includegraphics[width=\columnwidth]{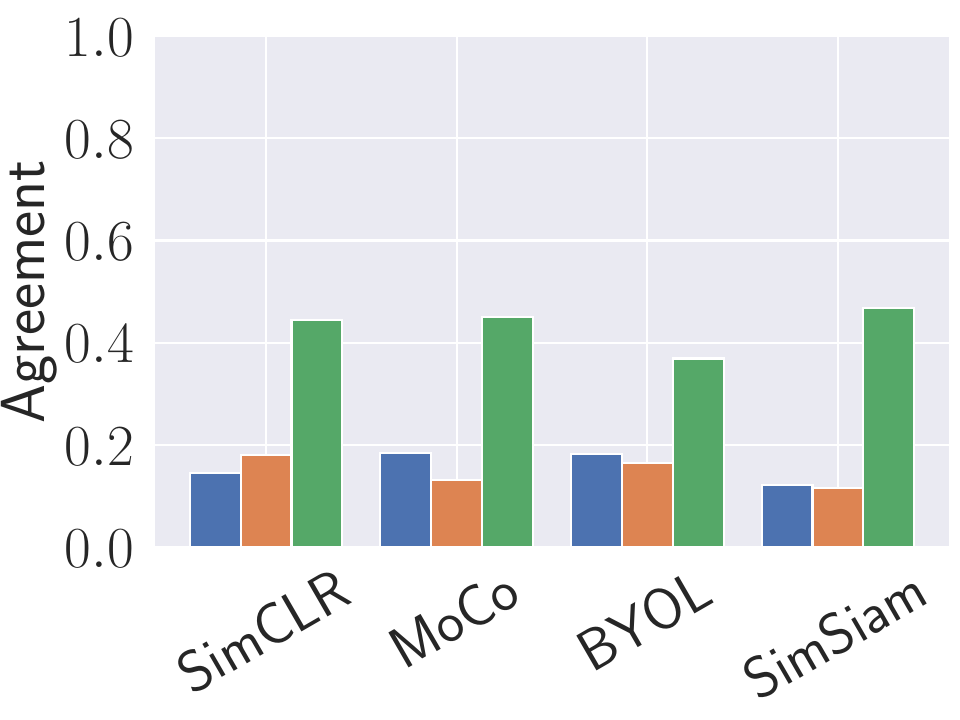}
\caption{F-MNIST}
\label{fig:agreement_normal_target_imagenet_svhn_mnist}
\end{subfigure}
\begin{subfigure}{0.44\columnwidth}
\includegraphics[width=\columnwidth]{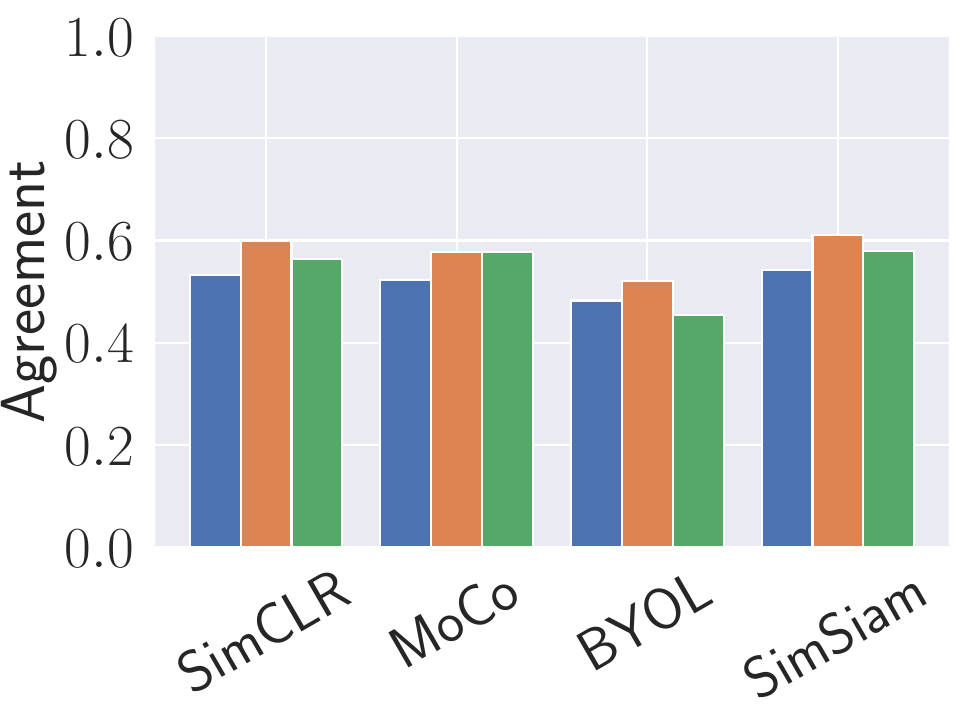}
\caption{SVHN}
\label{fig:agreement_normal_target_imagenet_svhn_svhn}
\end{subfigure}
\begin{subfigure}{0.44\columnwidth}
\includegraphics[width=\columnwidth]{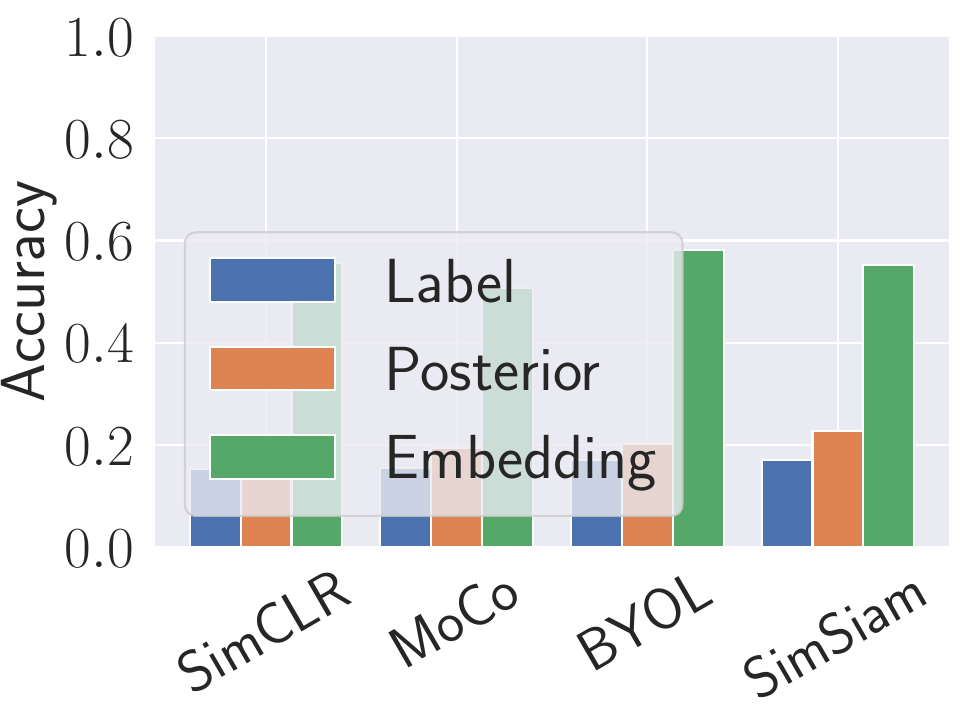}
\caption{CIFAR10}
\label{fig:accuracy_normal_target_imagenet_svhn_cifar10}
\end{subfigure}
\begin{subfigure}{0.44\columnwidth}
\includegraphics[width=\columnwidth]{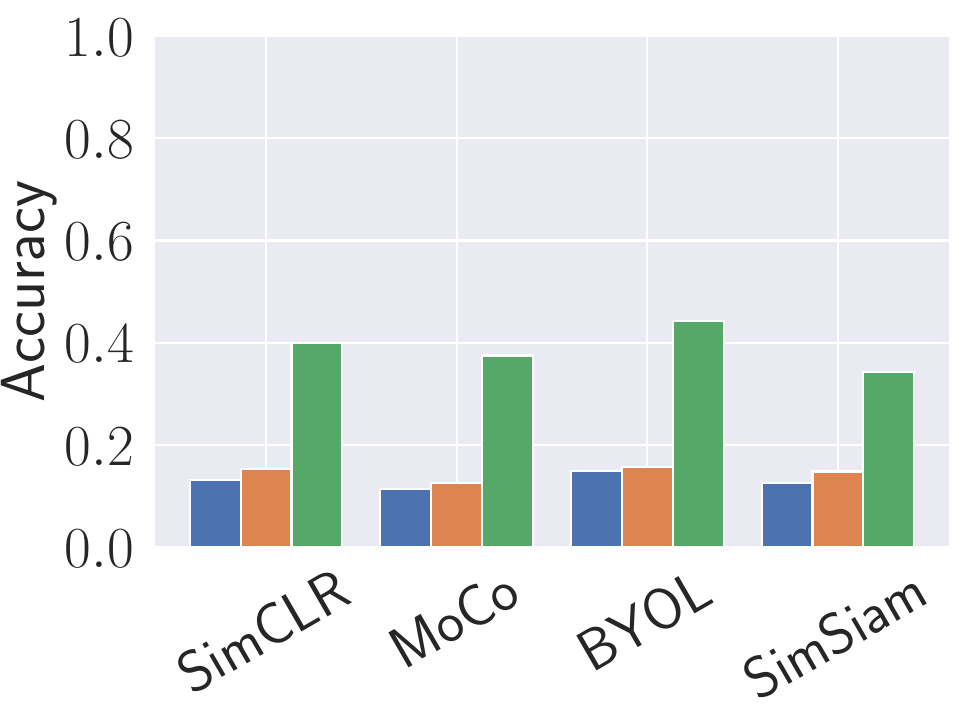}
\caption{STL10}
\label{fig:accuracy_normal_target_imagenet_svhn_stl10}
\end{subfigure}
\begin{subfigure}{0.44\columnwidth}
\includegraphics[width=\columnwidth]{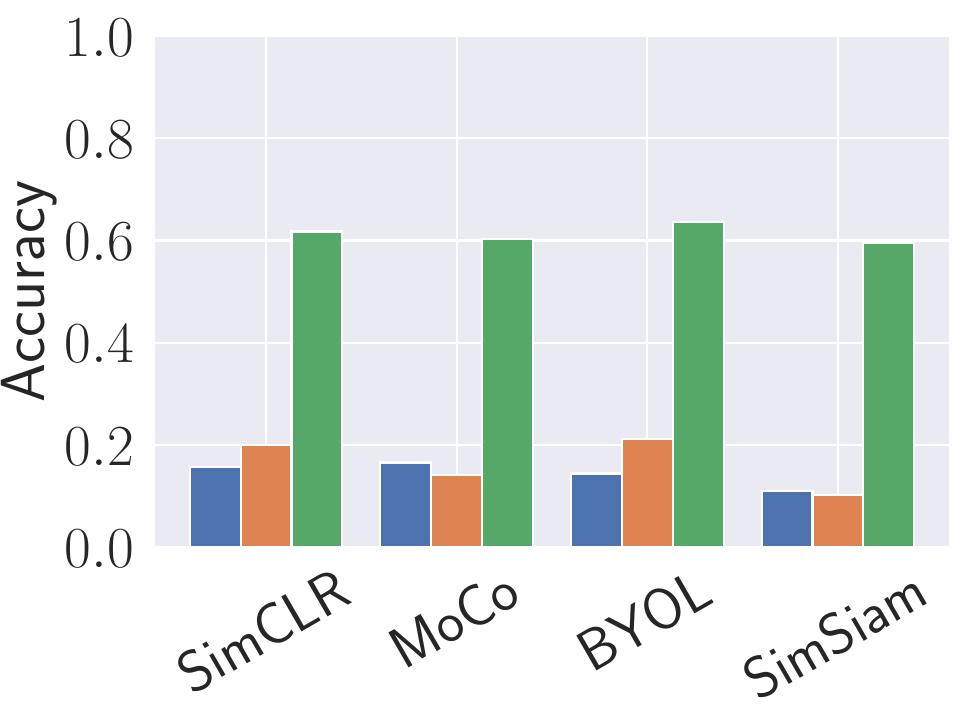}
\caption{F-MNIST}
\label{fig:accuracy_normal_target_imagenet_svhn_mnist}
\end{subfigure}
\begin{subfigure}{0.44\columnwidth}
\includegraphics[width=\columnwidth]{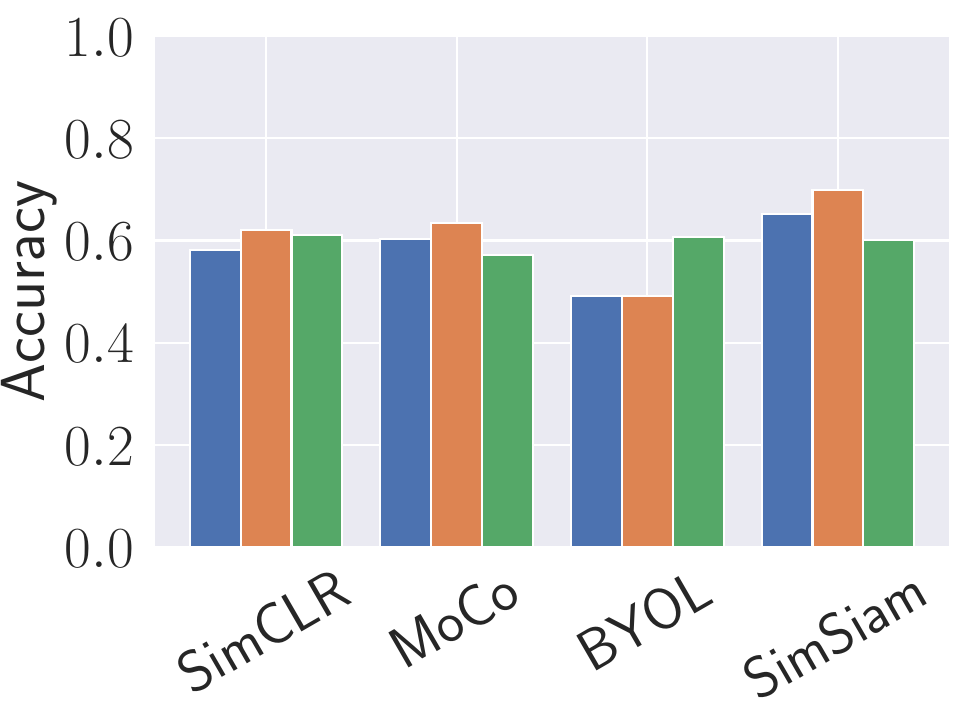}
\caption{SVHN}
\label{fig:accuracy_normal_imagenet_svhn_svhn}
\end{subfigure}
\caption{The performance of model stealing attack against target encodes and downstream classifiers trained on ImageNet and SVHN.
Target models can output predicted labels, posteriors, or embeddings.
The adversary uses CIFAR10, STL10, Fashion-MNIST (F-MNIST), SVHN to conduct model stealing attacks.
The x-axis represents different kinds of target models.
The first line's y-axis represents the agreement of the model stealing attack. 
The second line's y-axis represents the accuracy of the model stealing attack.}
\label{fig:encoder_imagenet_svhn}
\end{figure*}

\autoref{fig:encoder_cifar10_stl10}, \autoref{fig:encoder_cifar10_mnist}, and \autoref{fig:encoder_cifar10_svhn} show the results of the conventional attacks on target models whose encoders are pre-trained on CIFAR10 and downstream classifiers are trained on STL10, F-MNIST, and SVHN, respectively.
\autoref{fig:encoder_imagenet_stl10}, \autoref{fig:encoder_imagenet_mnist}, and \autoref{fig:encoder_imagenet_svhn} show the results of the conventional attacks on classifiers whose encoders are pre-trained on ImageNet100 and downstream classifiers are trained on STL10, F-MNIST, and SVHN, respectively.


\subsection{More Results on \ContSteal}
\label{subsection:appendix_contsteal}

\begin{figure*}[!t]
\centering
\begin{subfigure}{0.44\columnwidth}
\includegraphics[width=\columnwidth]{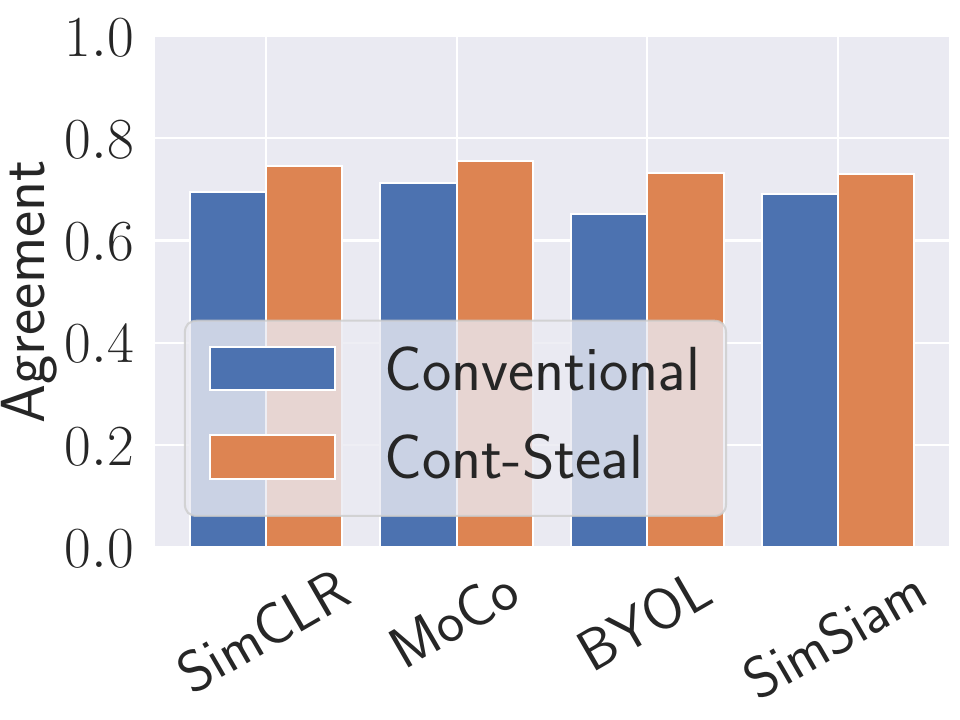}
\caption{CIFAR10}
\label{fig:contrastive_agreement_stl10_cifar10}
\end{subfigure}
\begin{subfigure}{0.44\columnwidth}
\includegraphics[width=\columnwidth]{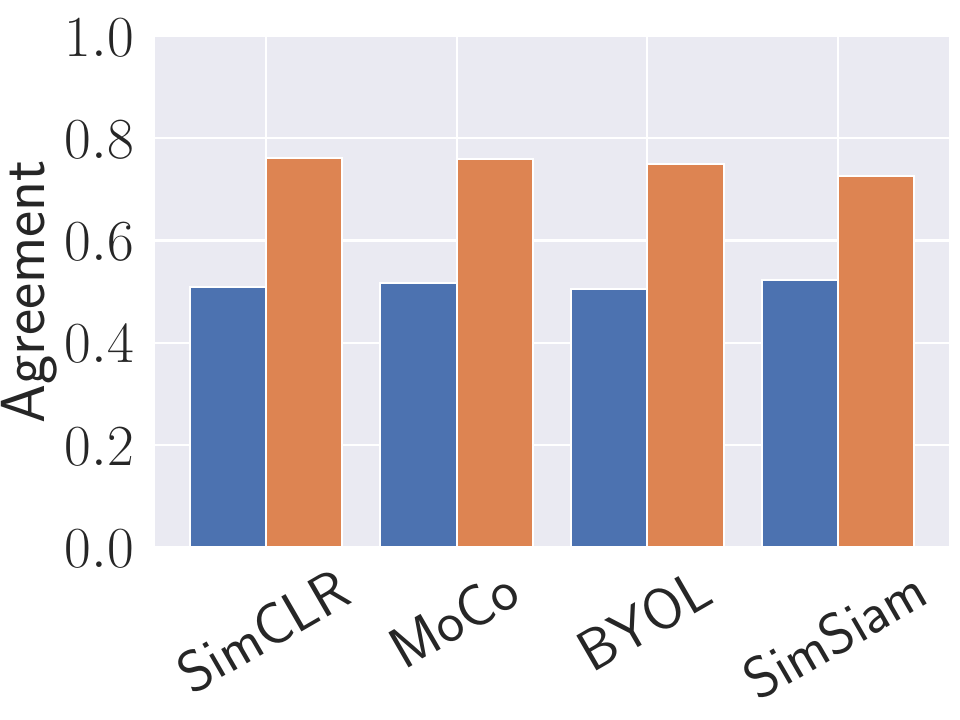}
\caption{STL10}
\label{fig:contrastive_agreement_stl10_stl10}
\end{subfigure}
\begin{subfigure}{0.44\columnwidth}
\includegraphics[width=\columnwidth]{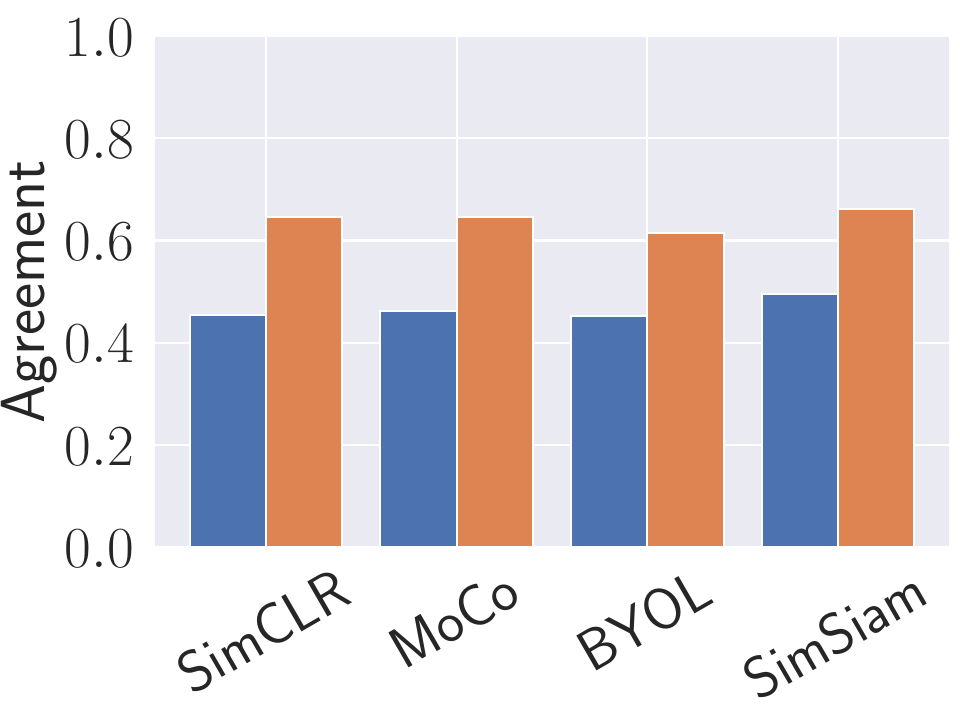}
\caption{F-MNIST}
\label{fig:contrastive_agreement_stl10_mnist}
\end{subfigure}
\begin{subfigure}{0.44\columnwidth}
\includegraphics[width=\columnwidth]{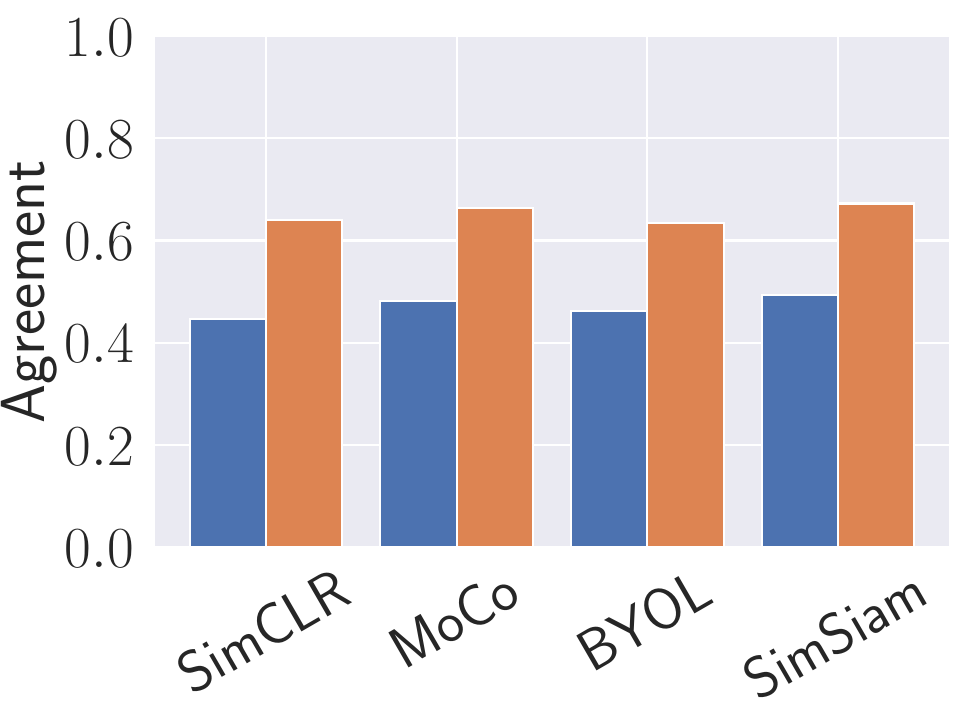}
\caption{SVHN}
\label{fig:contrastive_agreement_stl10_svhn}
\end{subfigure}
\begin{subfigure}{0.44\columnwidth}
\includegraphics[width=\columnwidth]{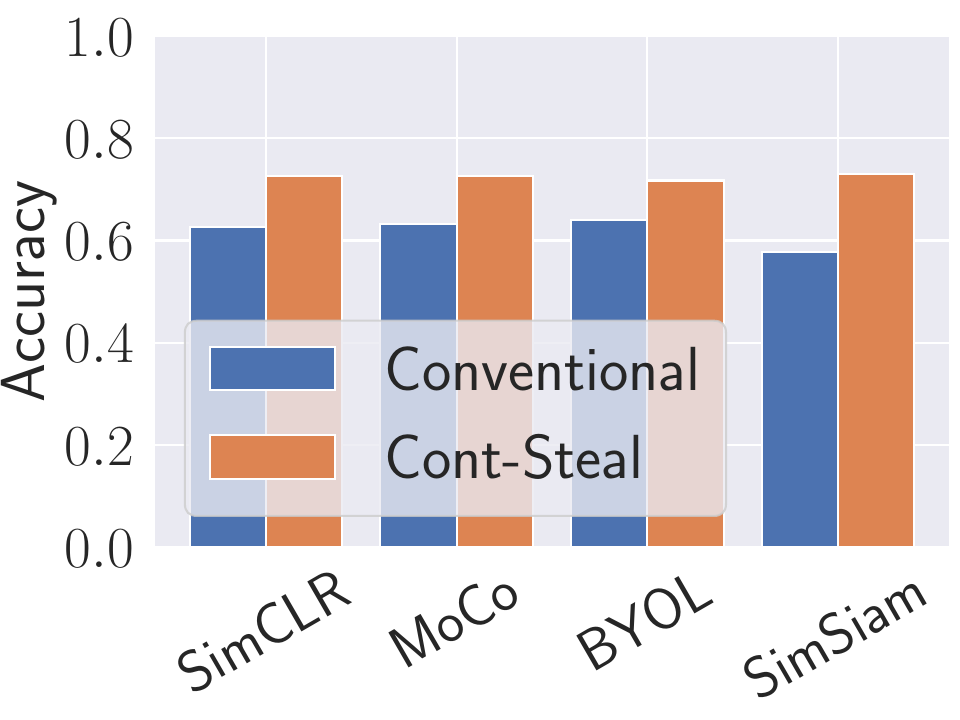}
\caption{CIFAR10}
\label{fig:contrastive_accuracy_stl10_cifar10}
\end{subfigure}
\begin{subfigure}{0.44\columnwidth}
\includegraphics[width=\columnwidth]{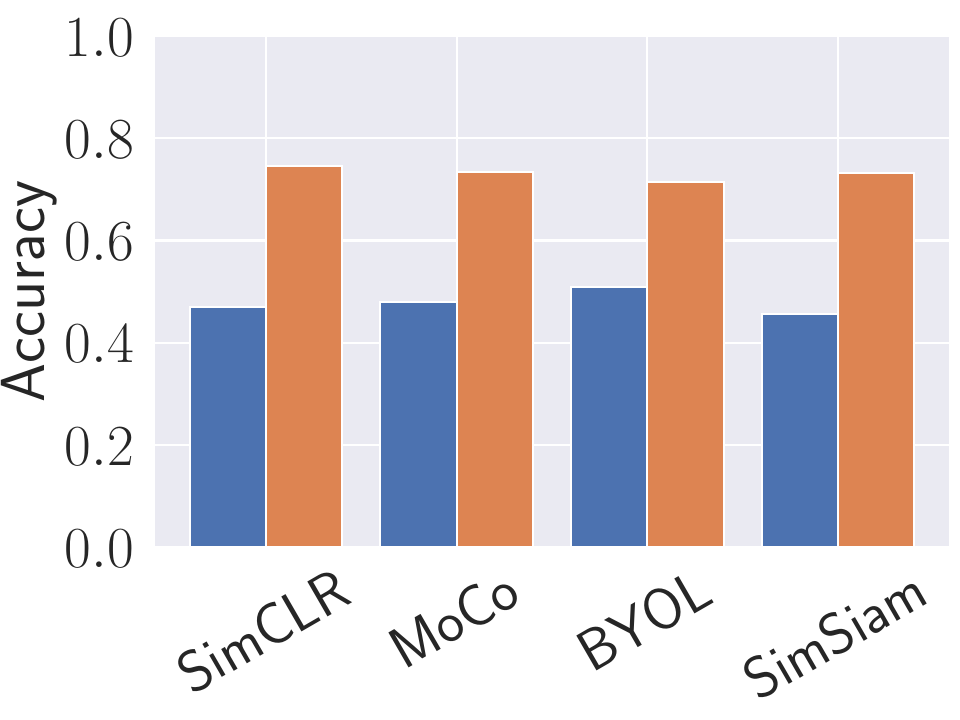}
\caption{STL10}
\label{fig:contrastive_accuracy_stl10_stl10}
\end{subfigure}
\begin{subfigure}{0.44\columnwidth}
\includegraphics[width=\columnwidth]{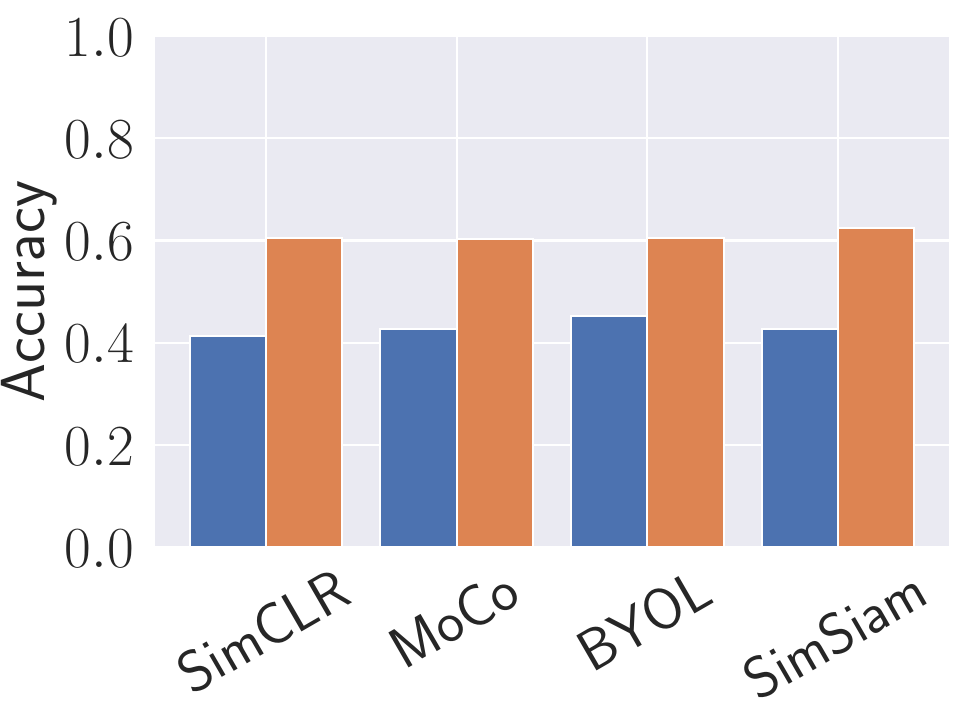}
\caption{F-MNIST}
\label{fig:contrastive_accuracy_stl10_mnist}
\end{subfigure}
\begin{subfigure}{0.44\columnwidth}
\includegraphics[width=\columnwidth]{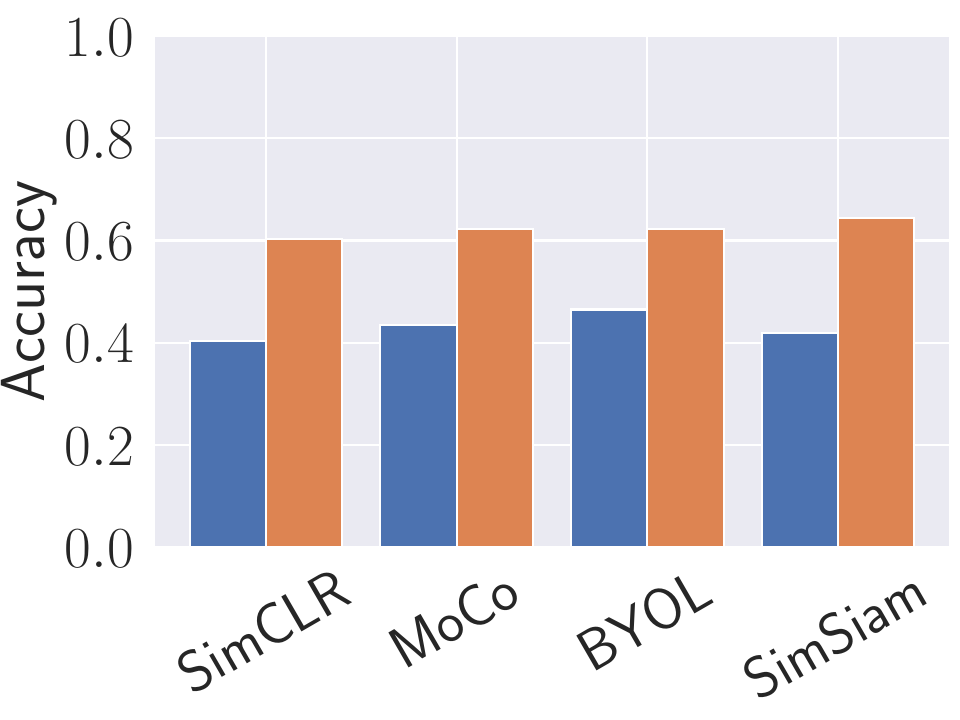}
\caption{SVHN}
\label{fig:contrastive_accuracy_stl10_svhn}
\end{subfigure}
\caption{The performance of Cont-Steal and conventional attack against target encoders trained on CIFAR10.
The adversary uses CIFAR10, STL10, F-MNIST, and SVHN to conduct model stealing attacks.
The adversary uses STL10 as the downstream task to evaluate the attack performance.
The x-axis represents different kinds of the target model.
The first line's y-axis represents the agreement of the model stealing attack. 
The second line's y-axis represents the accuracy of the model stealing attack.}
\label{fig:contrastive_cifar10_stl10}
\end{figure*}

\begin{figure*}[!t]
\centering
\begin{subfigure}{0.44\columnwidth}
\includegraphics[width=\columnwidth]{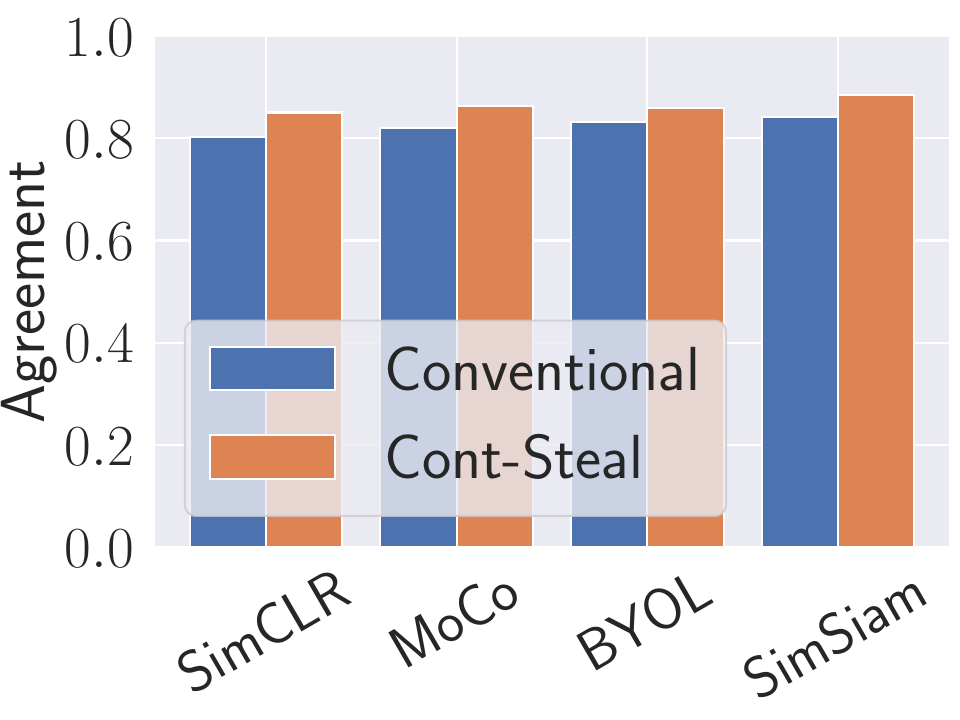}
\caption{CIFAR10}
\label{fig:contrastive_agreement_mnist_cifar10}
\end{subfigure}
\begin{subfigure}{0.44\columnwidth}
\includegraphics[width=\columnwidth]{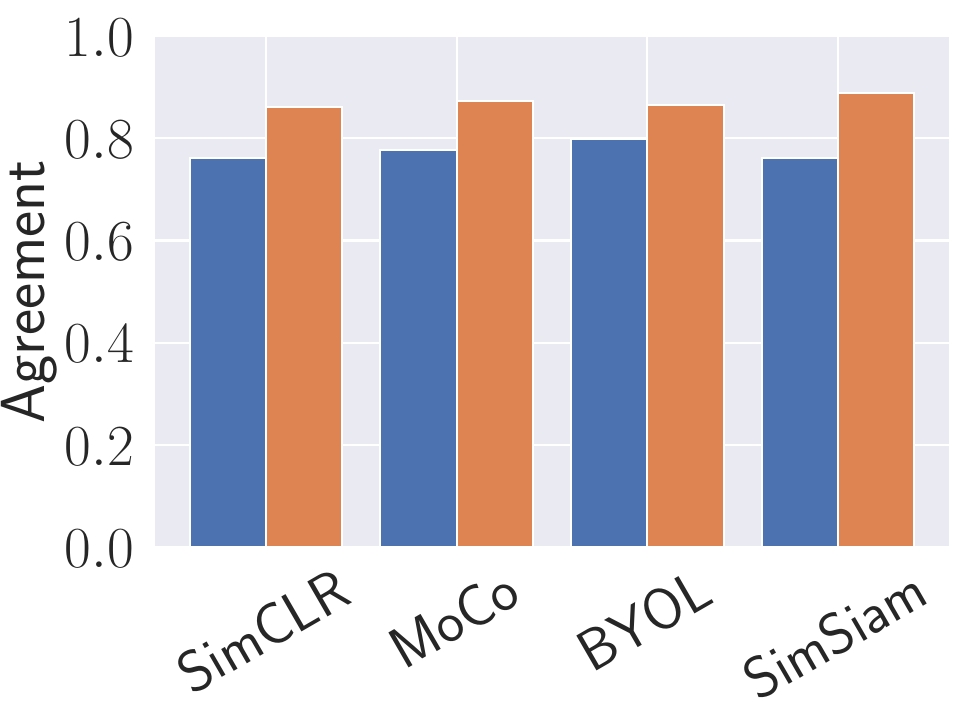}
\caption{STL10}
\label{fig:contrastive_agreement_mnist_stl10}
\end{subfigure}
\begin{subfigure}{0.44\columnwidth}
\includegraphics[width=\columnwidth]{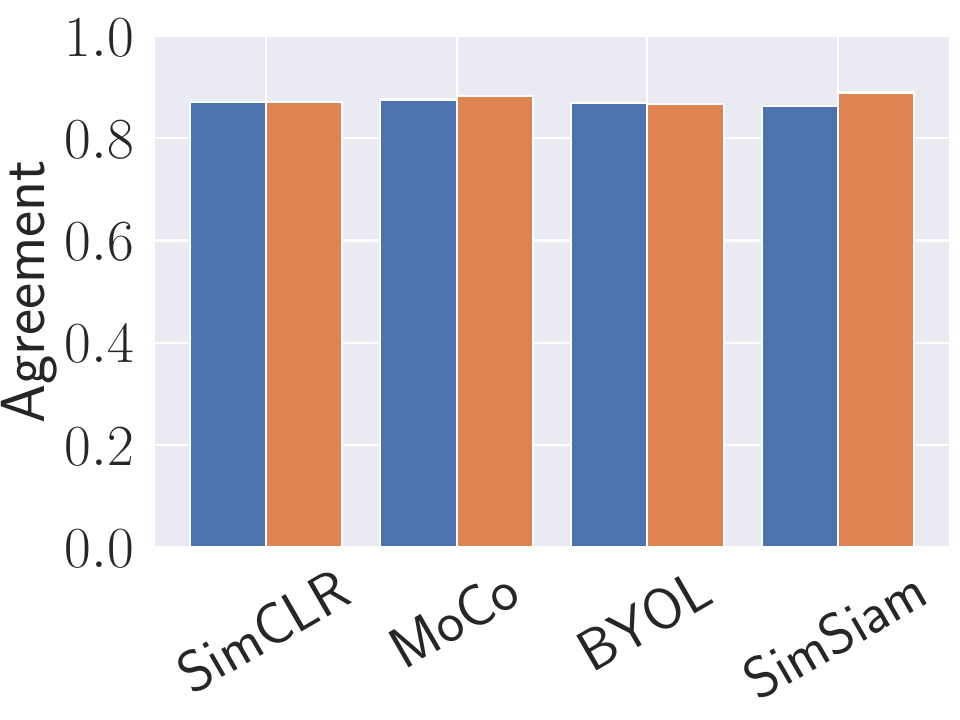}
\caption{F-MNIST}
\label{fig:contrastive_agreement_mnist_mnist}
\end{subfigure}
\begin{subfigure}{0.44\columnwidth}
\includegraphics[width=\columnwidth]{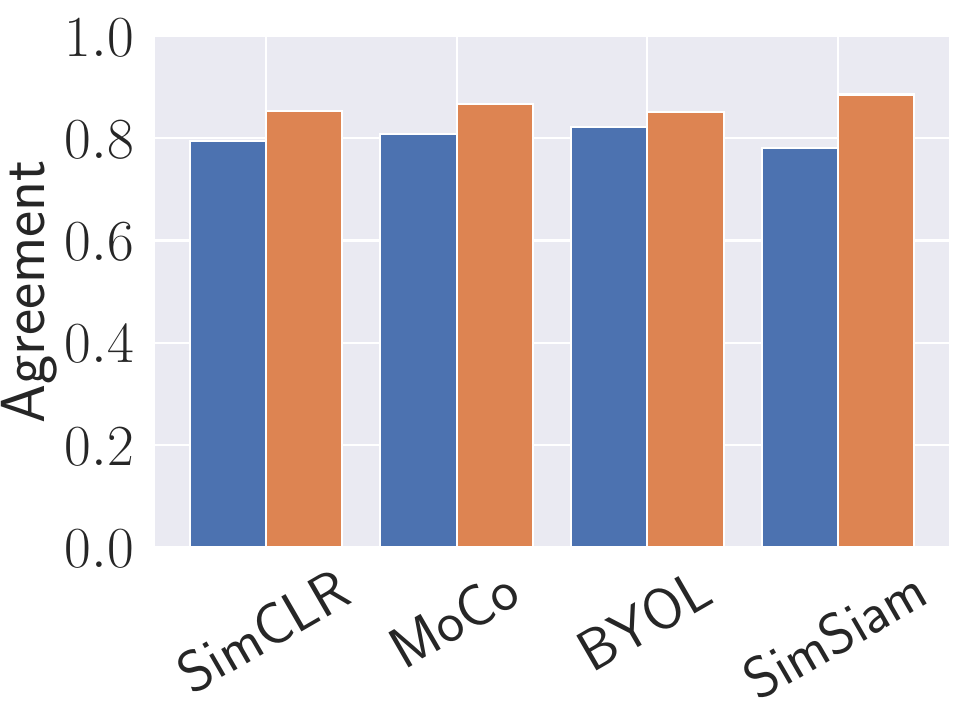}
\caption{SVHN}
\label{fig:contrastive_agreement_mnist_svhn}
\end{subfigure}
\begin{subfigure}{0.44\columnwidth}
\includegraphics[width=\columnwidth]{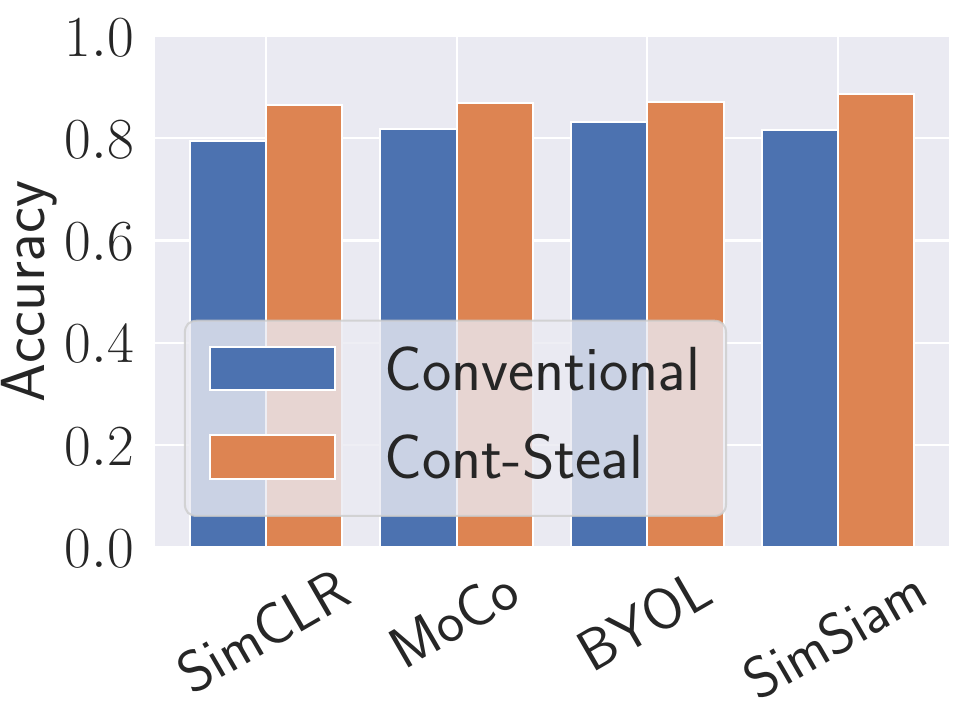}
\caption{CIFAR10}
\label{fig:contrastive_accuracy_mnist_cifar10}
\end{subfigure}
\begin{subfigure}{0.44\columnwidth}
\includegraphics[width=\columnwidth]{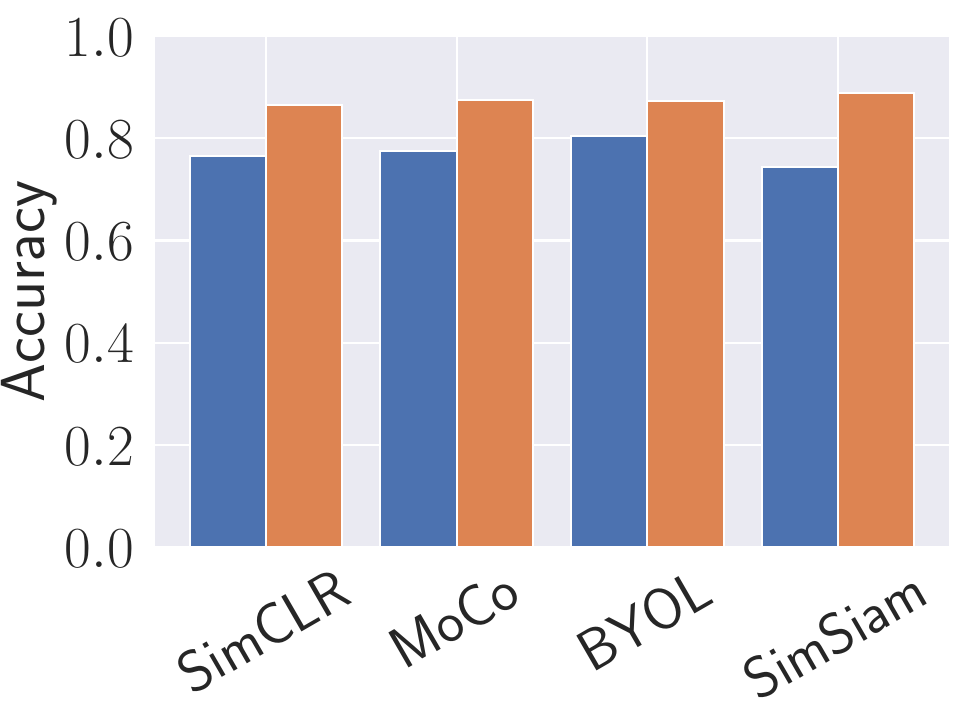}
\caption{STL10}
\label{fig:contrastive_accuracy_mnist_stl10}
\end{subfigure}
\begin{subfigure}{0.44\columnwidth}
\includegraphics[width=\columnwidth]{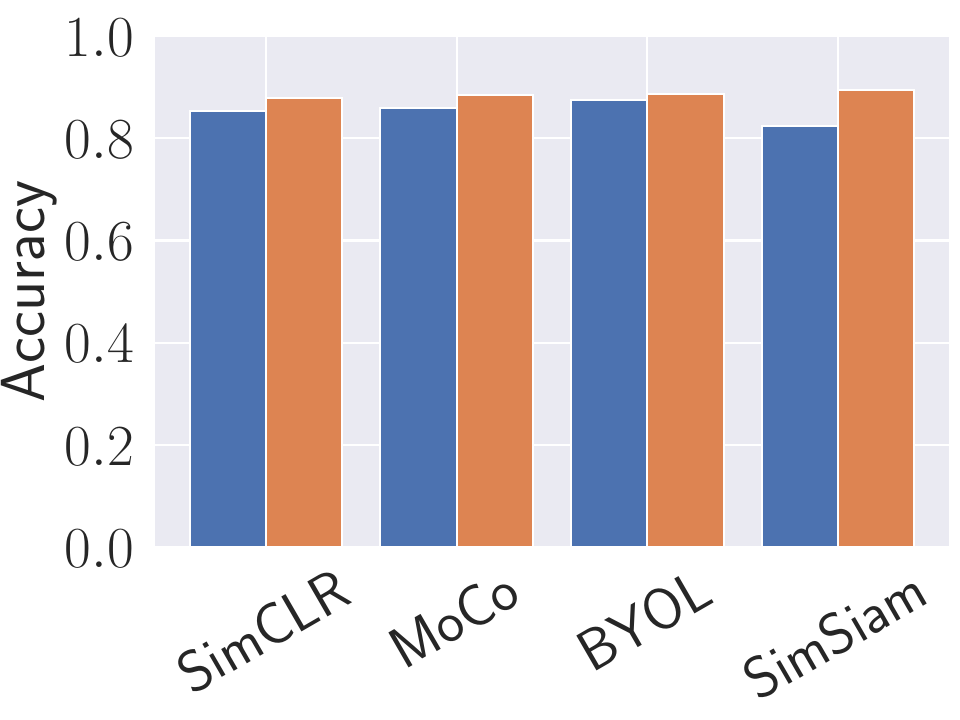}
\caption{F-MNIST}
\label{fig:contrastive_accuracy_mnist_mnist}
\end{subfigure}
\begin{subfigure}{0.44\columnwidth}
\includegraphics[width=\columnwidth]{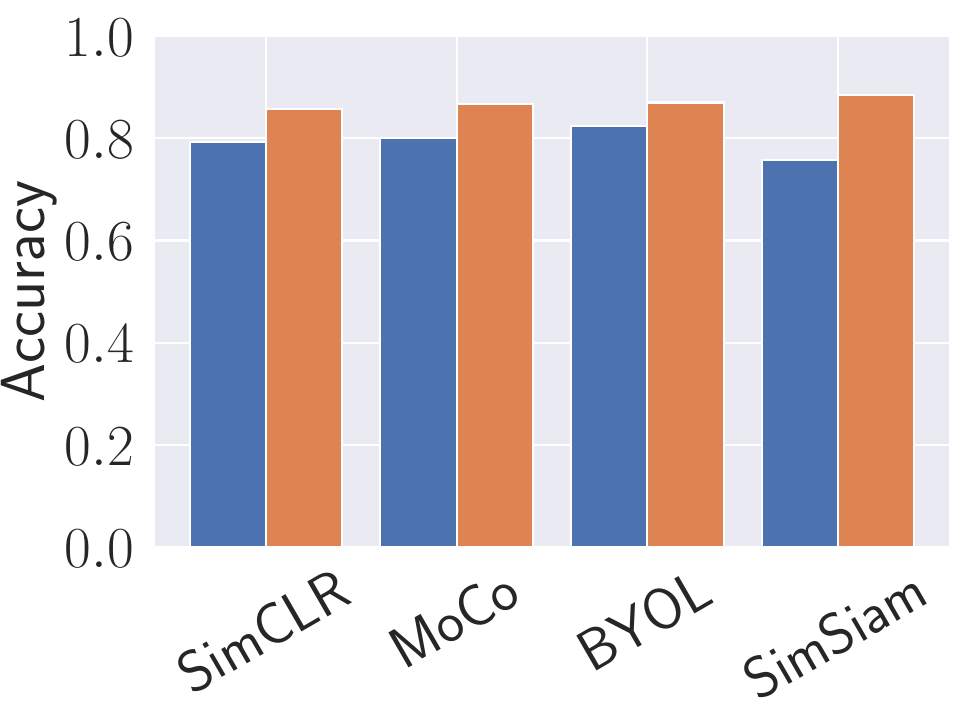}
\caption{SVHN}
\label{fig:contrastive_accuracy_mnist_svhn}
\end{subfigure}
\caption{The performance of \ContSteal and conventional attack against target encoders trained on CIFAR10.
The adversary uses CIFAR10, STL10, F-MNIST, and SVHN to conduct model stealing attacks.
The adversary uses F-MNIST as the downstream task to evaluate the attack performance.
The x-axis represents different kinds of the target model.
The first line's y-axis represents the agreement of the model stealing attack. 
The second line's y-axis represents the accuracy of the model stealing attack.}
\label{fig:contrastive_cifar10_mnist}
\end{figure*}

\begin{figure*}[!t]
\centering
\begin{subfigure}{0.44\columnwidth}
\includegraphics[width=\columnwidth]{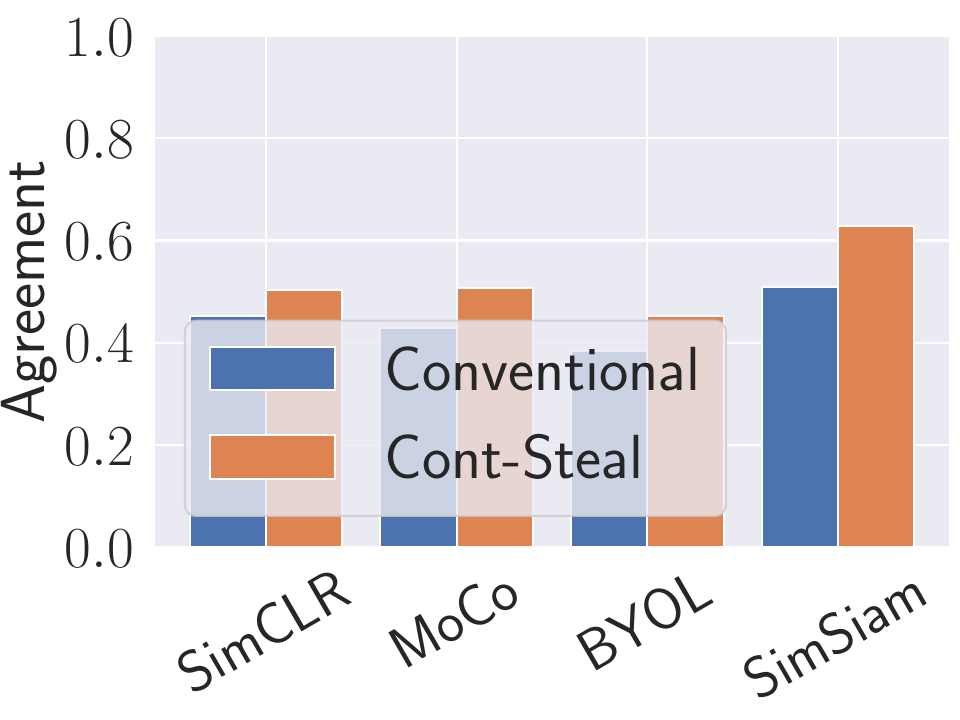}
\caption{CIFAR10}
\label{fig:contrastive_agreement_svhn_cifar10}
\end{subfigure}
\begin{subfigure}{0.44\columnwidth}
\includegraphics[width=\columnwidth]{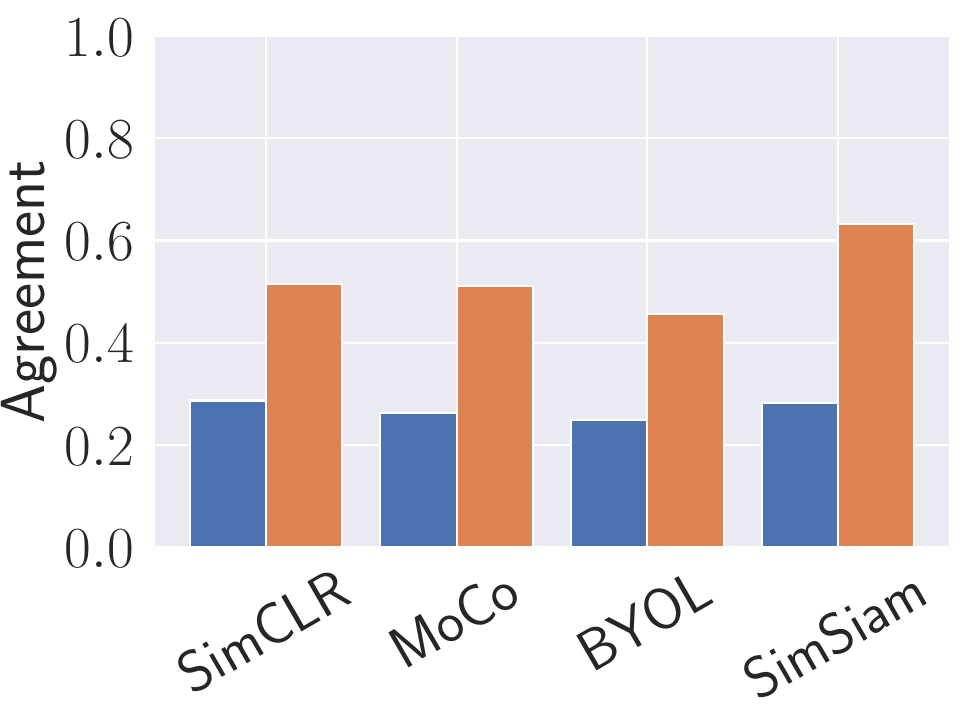}
\caption{STL10}
\label{fig:contrastive_agreement_svhn_stl10}
\end{subfigure}
\begin{subfigure}{0.44\columnwidth}
\includegraphics[width=\columnwidth]{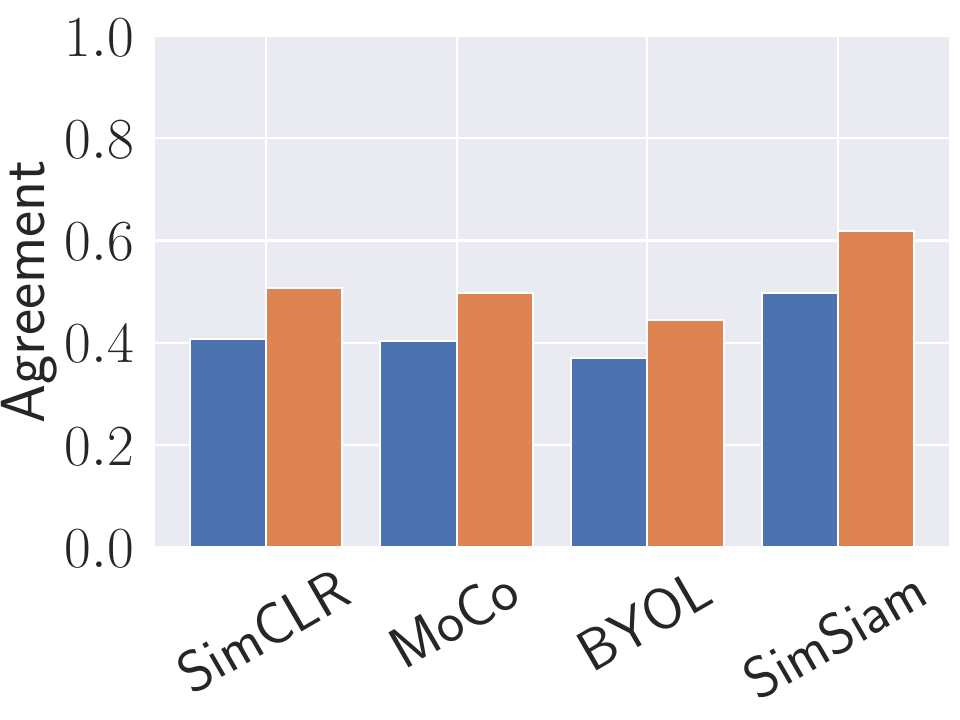}
\caption{F-MNIST}
\label{fig:contrastive_agreement_svhn_mnist}
\end{subfigure}
\begin{subfigure}{0.44\columnwidth}
\includegraphics[width=\columnwidth]{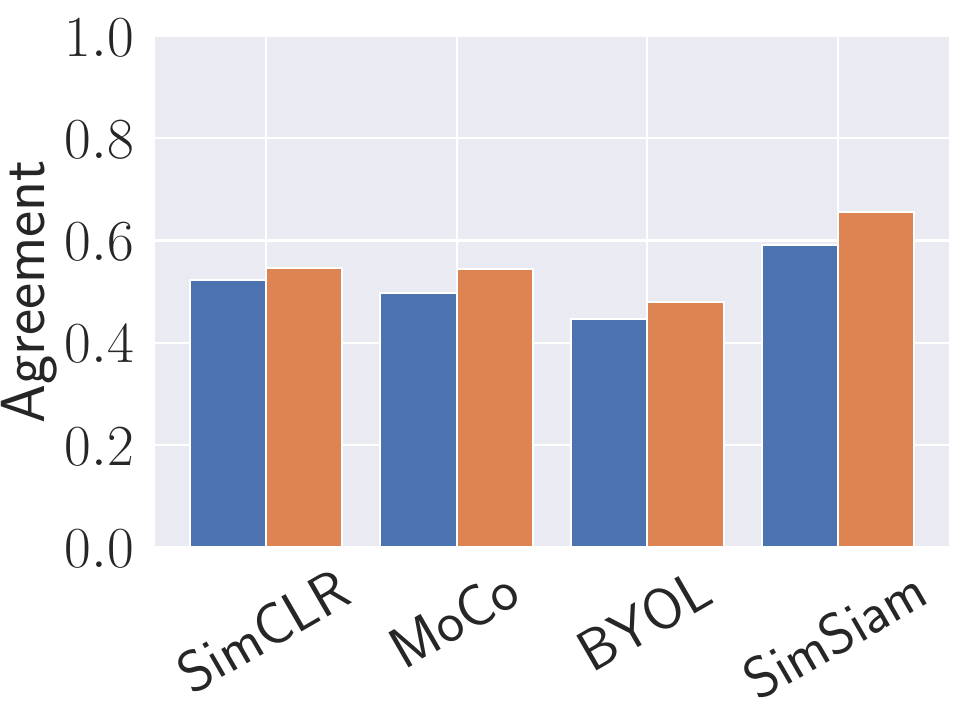}
\caption{SVHN}
\label{fig:contrastive_agreement_svhn_svhn}
\end{subfigure}
\begin{subfigure}{0.44\columnwidth}
\includegraphics[width=\columnwidth]{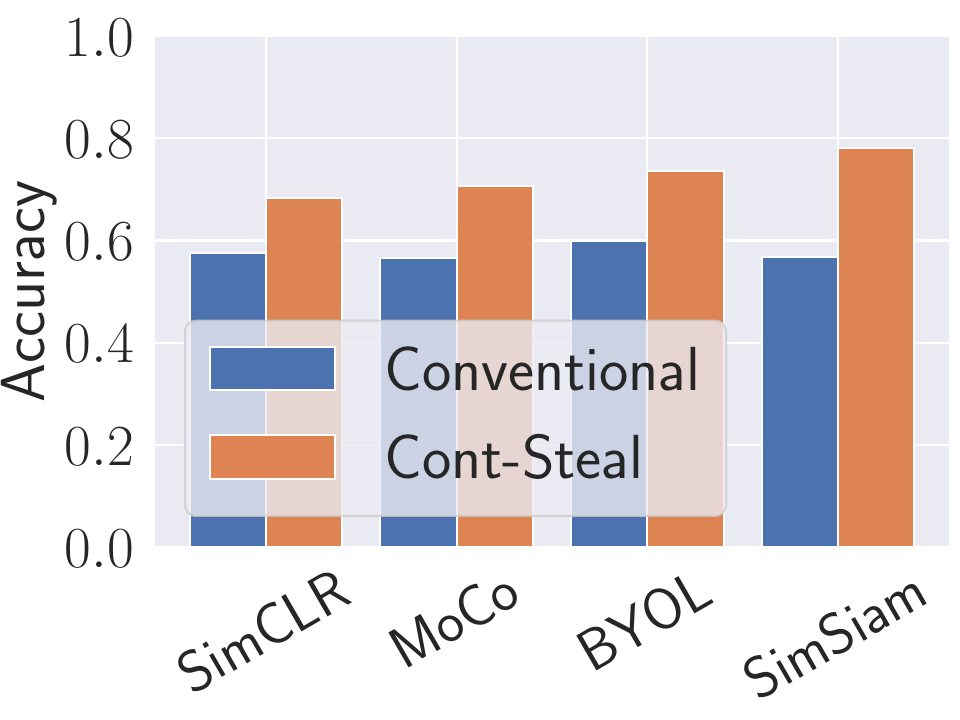}
\caption{CIFAR10}
\label{fig:contrastive_accuracy_svhn_cifar10}
\end{subfigure}
\begin{subfigure}{0.44\columnwidth}
\includegraphics[width=\columnwidth]{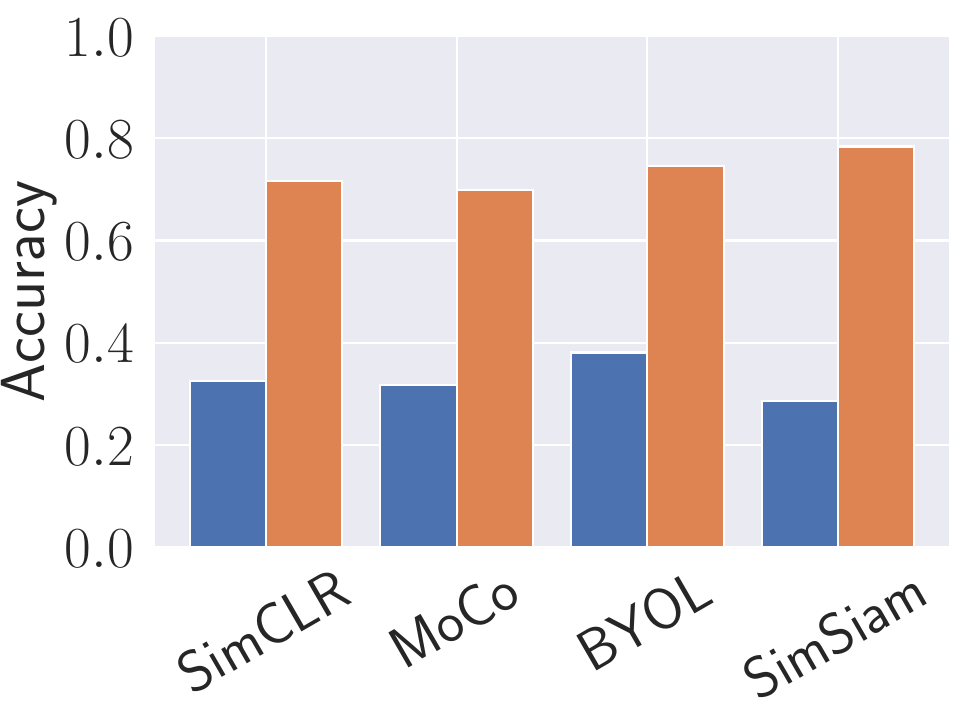}
\caption{STL10}
\label{fig:contrastive_accuracy_svhn_stl10}
\end{subfigure}
\begin{subfigure}{0.44\columnwidth}
\includegraphics[width=\columnwidth]{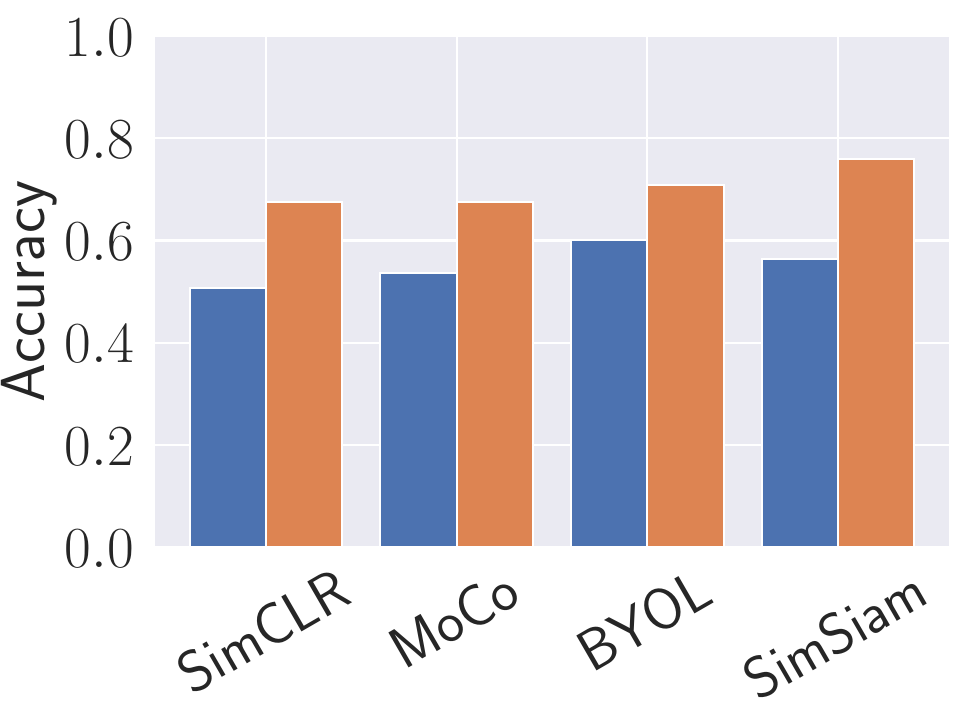}
\caption{F-MNIST}
\label{fig:contrastive_accuracy_svhn_mnist}
\end{subfigure}
\begin{subfigure}{0.44\columnwidth}
\includegraphics[width=\columnwidth]{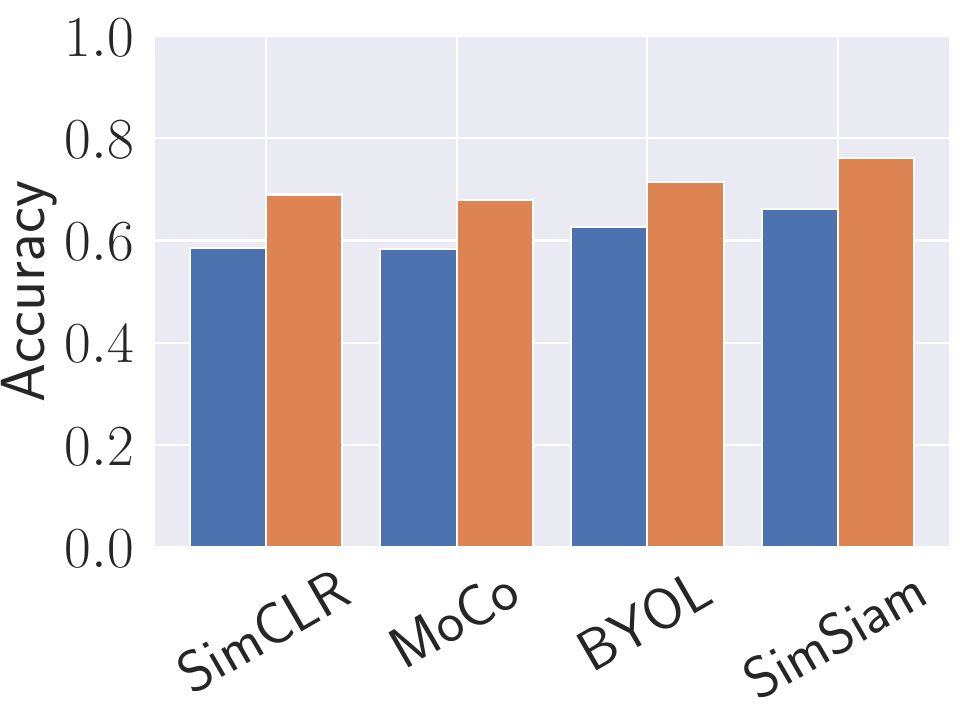}
\caption{SVHN}
\label{fig:contrastive_accuracy_svhn_svhn}
\end{subfigure}
\caption{The performance of Cont-Steal and conventional attack against target encoders trained on CIFAR10.
The adversary uses CIFAR10, STL10, F-MNIST, and SVHN to conduct model stealing attacks.
The adversary uses SVHN as the downstream task to evaluate the attack performance.
The x-axis represents different kinds of the target model.
The first line's y-axis represents the agreement of the model stealing attack. 
The second line's y-axis represents the accuracy of the model stealing attack.}
\label{fig:contrastive_cifar10_svhn}
\end{figure*}

\begin{figure*}[!t]
\centering
\begin{subfigure}{0.44\columnwidth}
\includegraphics[width=\columnwidth]{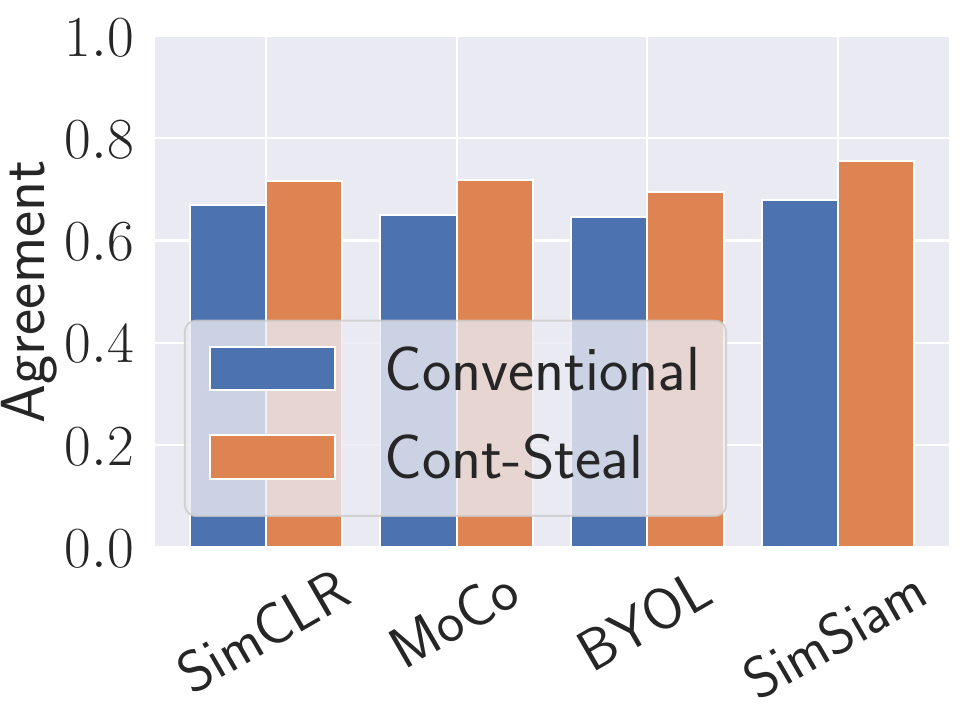}
\caption{CIFAR10}
\label{fig:contrastive_agreement_imagenet_cifar10_cifar10}
\end{subfigure}
\begin{subfigure}{0.44\columnwidth}
\includegraphics[width=\columnwidth]{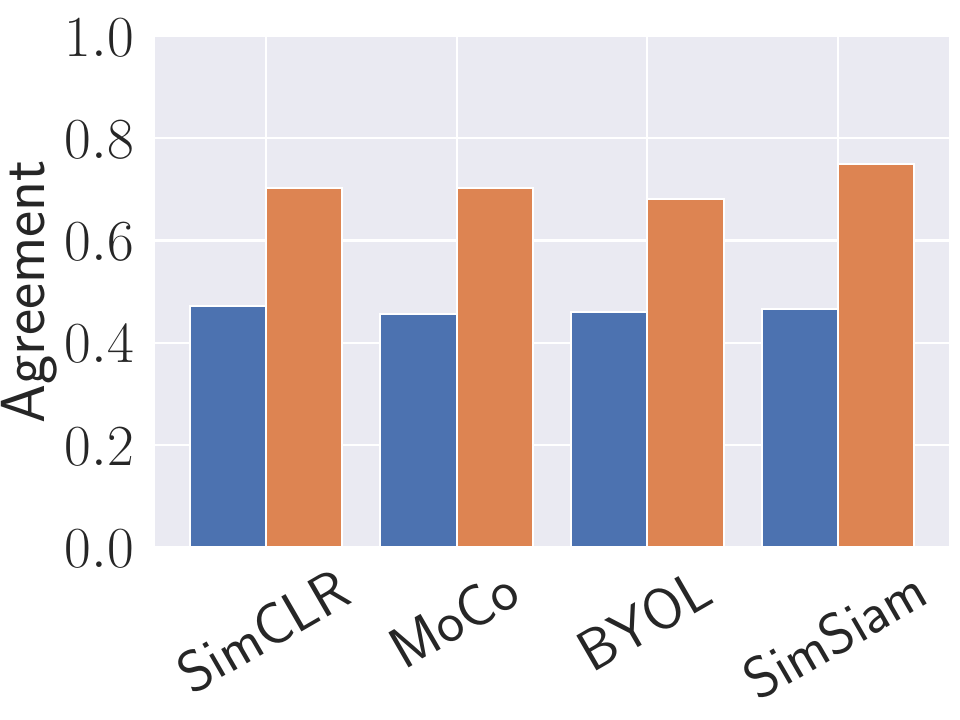}
\caption{STL10}
\label{fig:contrastive_agreement_imagenet_cifar10_wild}
\end{subfigure}
\begin{subfigure}{0.44\columnwidth}
\includegraphics[width=\columnwidth]{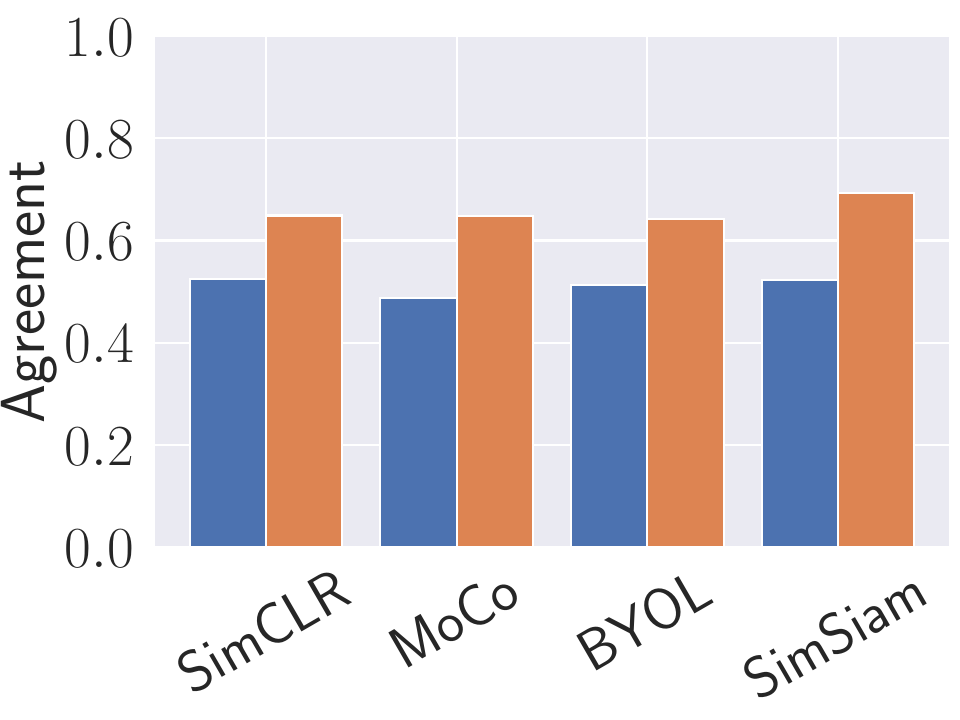}
\caption{F-MNIST}
\label{fig:contrastive_agreement_imagenet_cifar10_cifar100}
\end{subfigure}
\begin{subfigure}{0.44\columnwidth}
\includegraphics[width=\columnwidth]{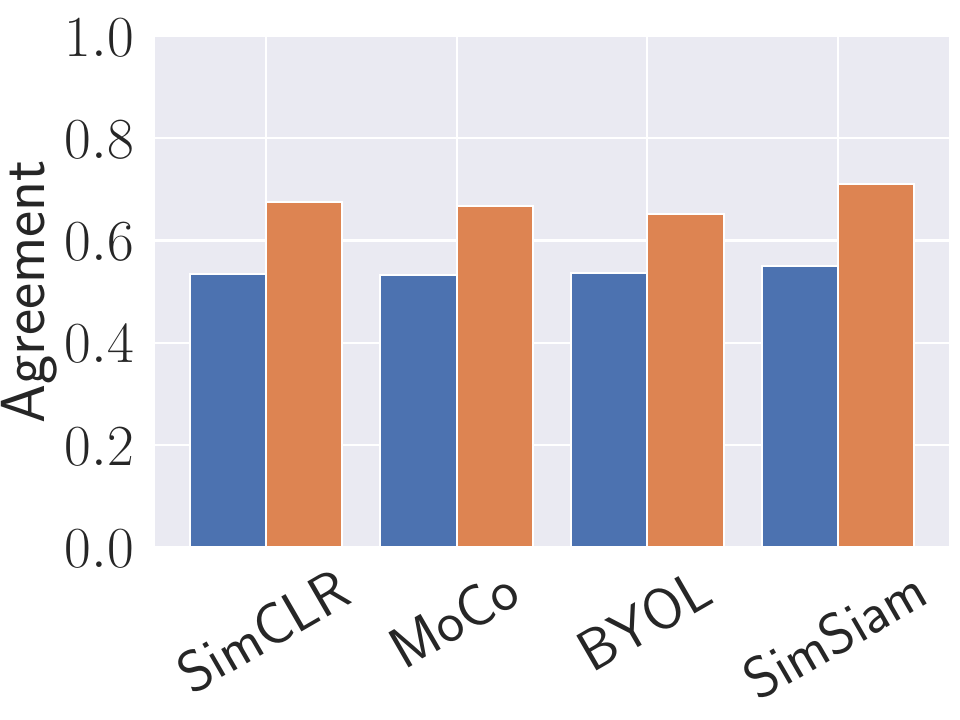}
\caption{SVHN}
\label{fig:contrastive_agreement_imagenet_cifar10_stl10}
\end{subfigure}
\begin{subfigure}{0.44\columnwidth}
\includegraphics[width=\columnwidth]{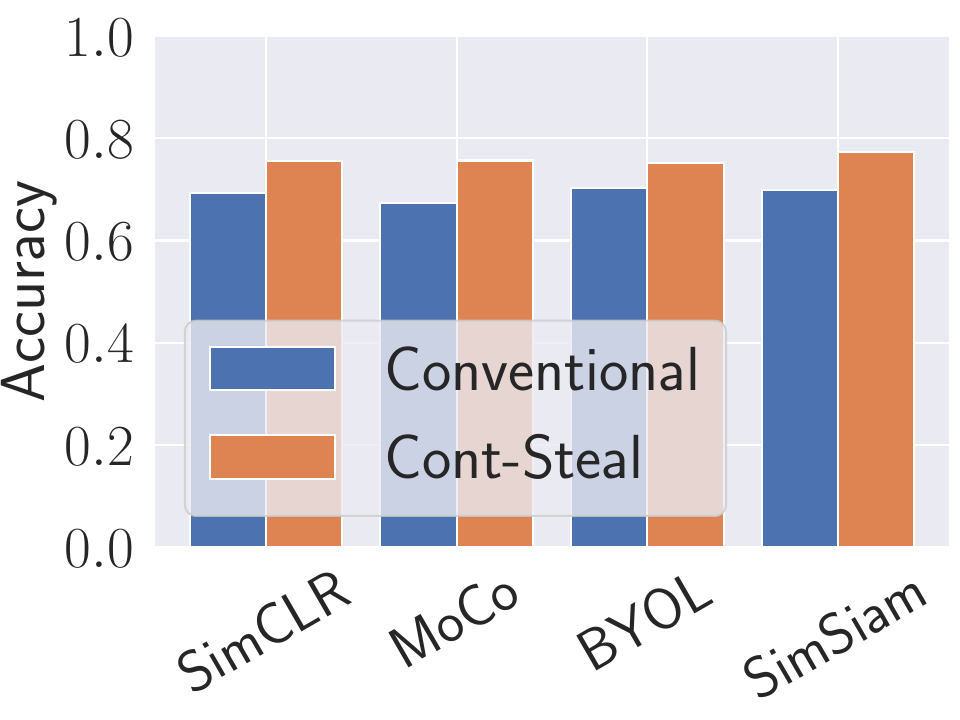}
\caption{CIFAR10}
\label{fig:contrastive_accuracy_imagenet_cifar10_cifar10}
\end{subfigure}
\begin{subfigure}{0.44\columnwidth}
\includegraphics[width=\columnwidth]{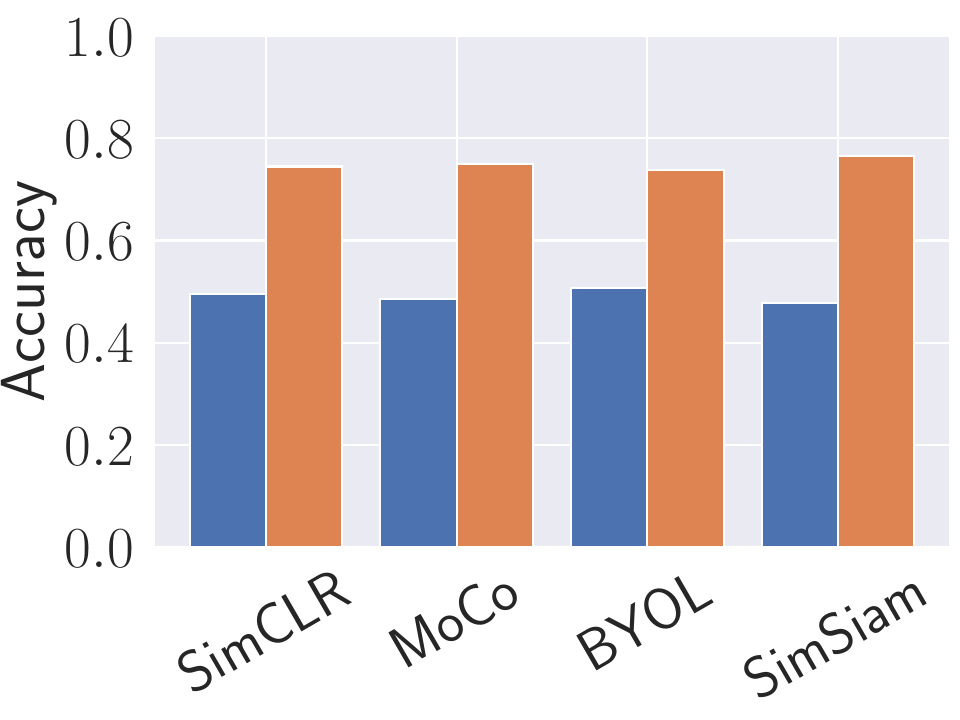}
\caption{STL10}
\label{fig:contrastive_accuracy_imagenet_cifar10_wild}
\end{subfigure}
\begin{subfigure}{0.44\columnwidth}
\includegraphics[width=\columnwidth]{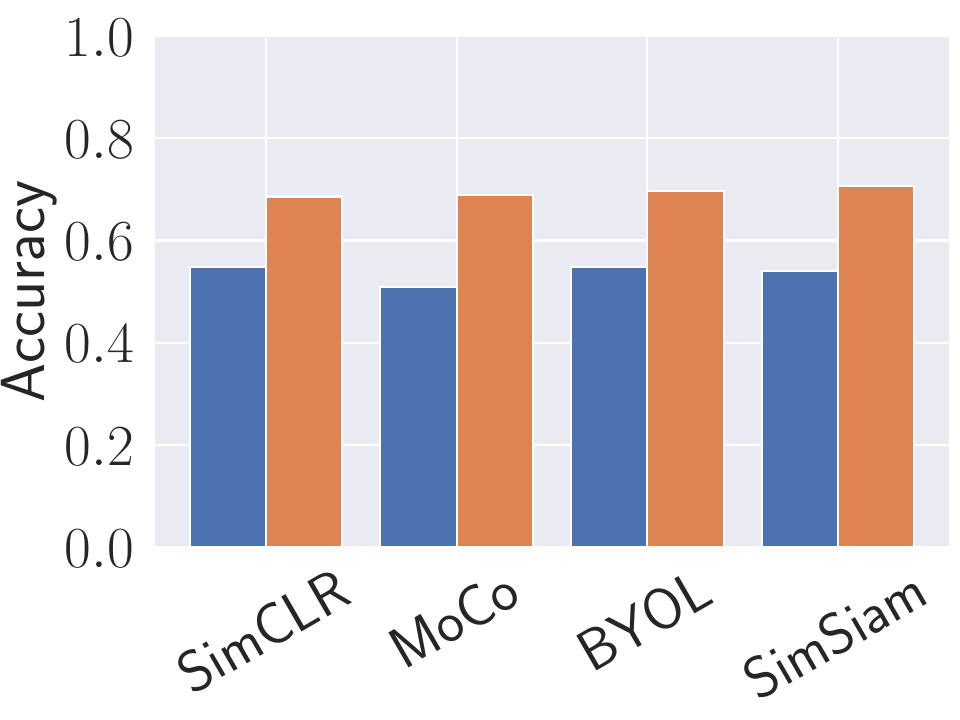}
\caption{F-MNIST}
\label{fig:contrastive_accuracy_imagenet_cifar10_cifar100}
\end{subfigure}
\begin{subfigure}{0.44\columnwidth}
\includegraphics[width=\columnwidth]{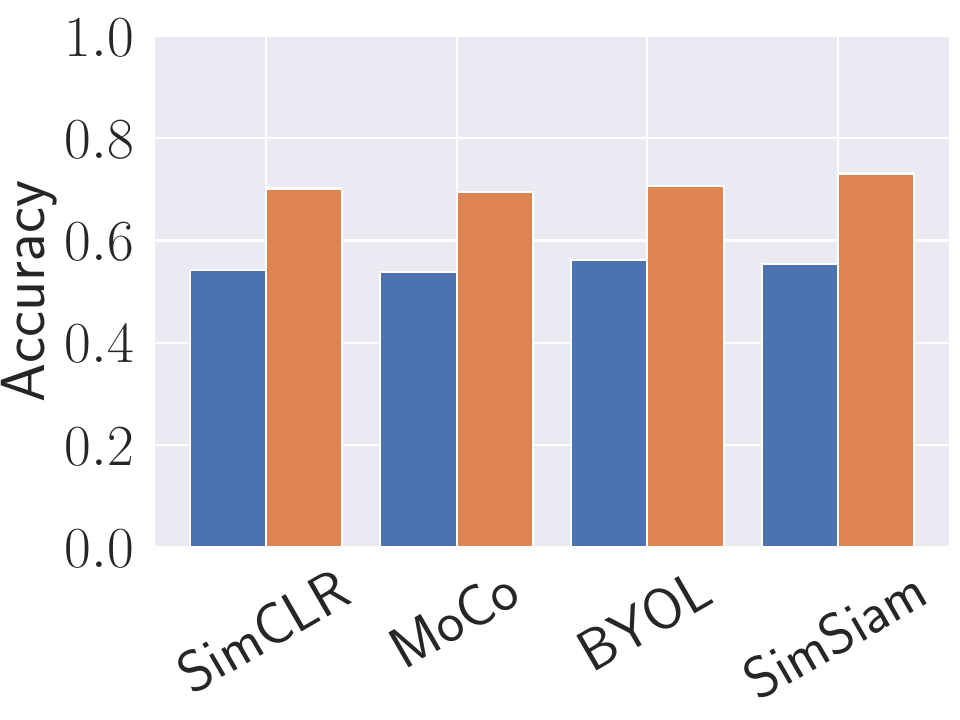}
\caption{SVHN}
\label{fig:contrastive_accuracy_imagenet_cifar10_stl10}
\end{subfigure}
\caption{The performance of Cont-Steal and conventional attack against target encoders trained on ImageNet100.
The adversary uses CIFAR10, STL10, F-MNIST, and SVHN to conduct model stealing attacks.
The adversary uses CIFAR10 as the downstream task to evaluate the attack performance.
The x-axis represents different kinds of the target model.
The first line's y-axis represents the agreement of the model stealing attack. 
The second line's y-axis represents the accuracy of the model stealing attack.}
\label{fig:contrastive_imagenet_cifar10}
\end{figure*}

\begin{figure*}[!t]
\centering
\begin{subfigure}{0.44\columnwidth}
\includegraphics[width=\columnwidth]{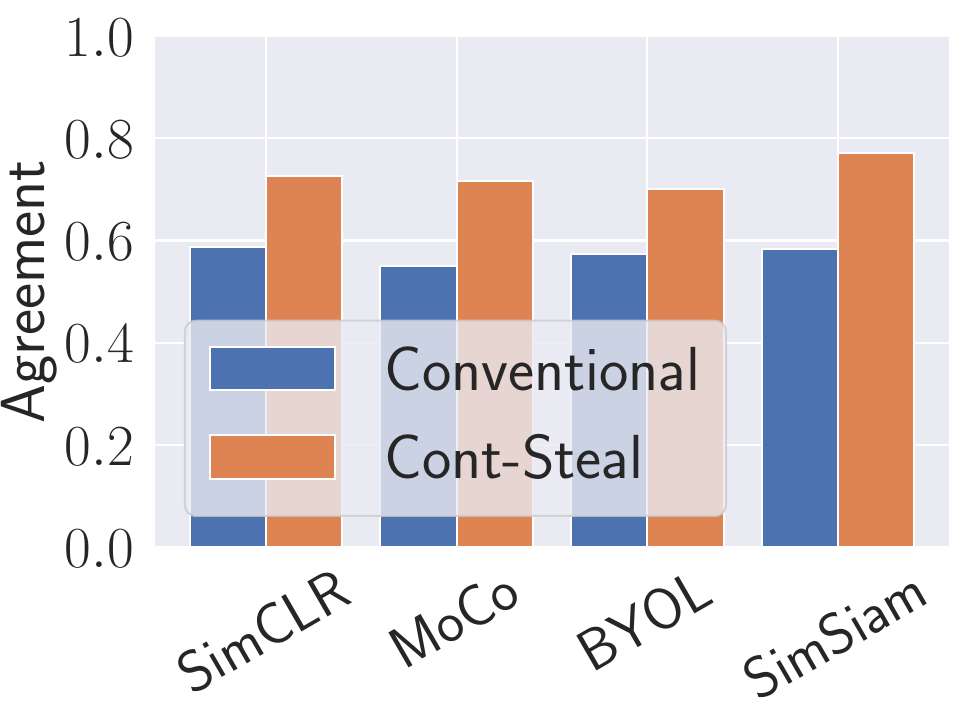}
\caption{CIFAR10}
\label{fig:contrastive_agreement_imagenet_stl10_cifar10}
\end{subfigure}
\begin{subfigure}{0.44\columnwidth}
\includegraphics[width=\columnwidth]{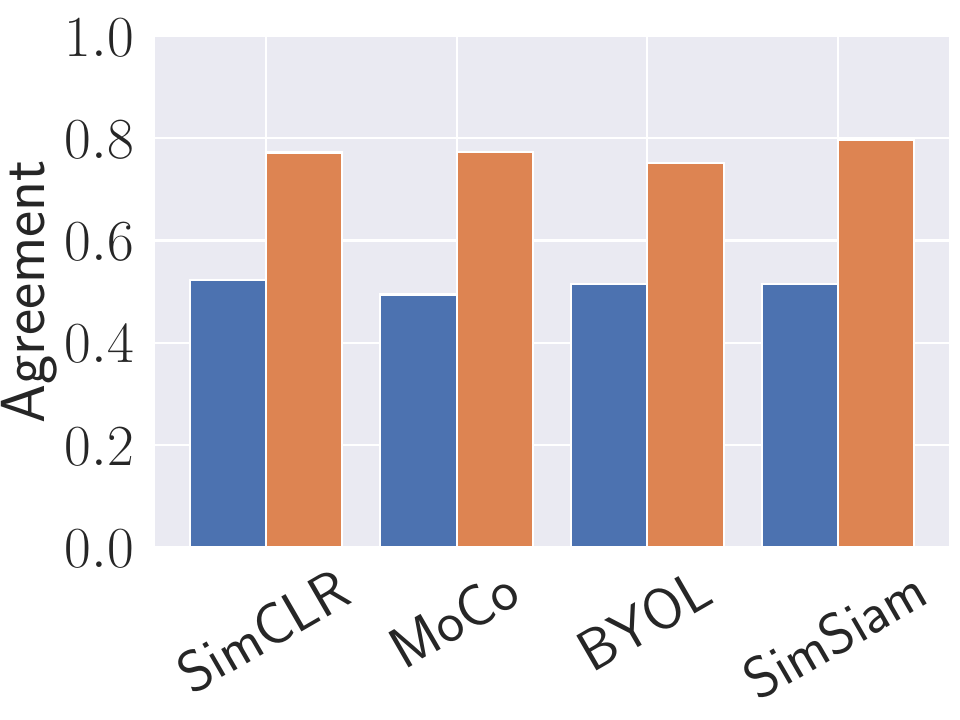}
\caption{STL10}
\label{fig:contrastive_agreement_imagenet_stl10_stl10}
\end{subfigure}
\begin{subfigure}{0.44\columnwidth}
\includegraphics[width=\columnwidth]{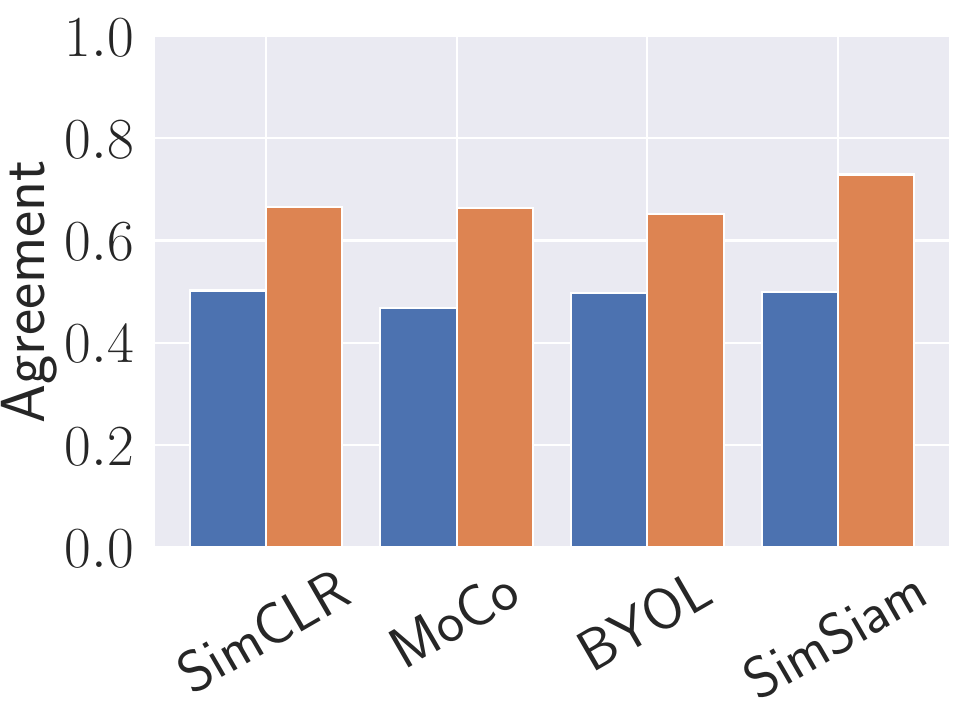}
\caption{F-MNIST}
\label{fig:contrastive_agreement_imagenet_stl10_mnist}
\end{subfigure}
\begin{subfigure}{0.44\columnwidth}
\includegraphics[width=\columnwidth]{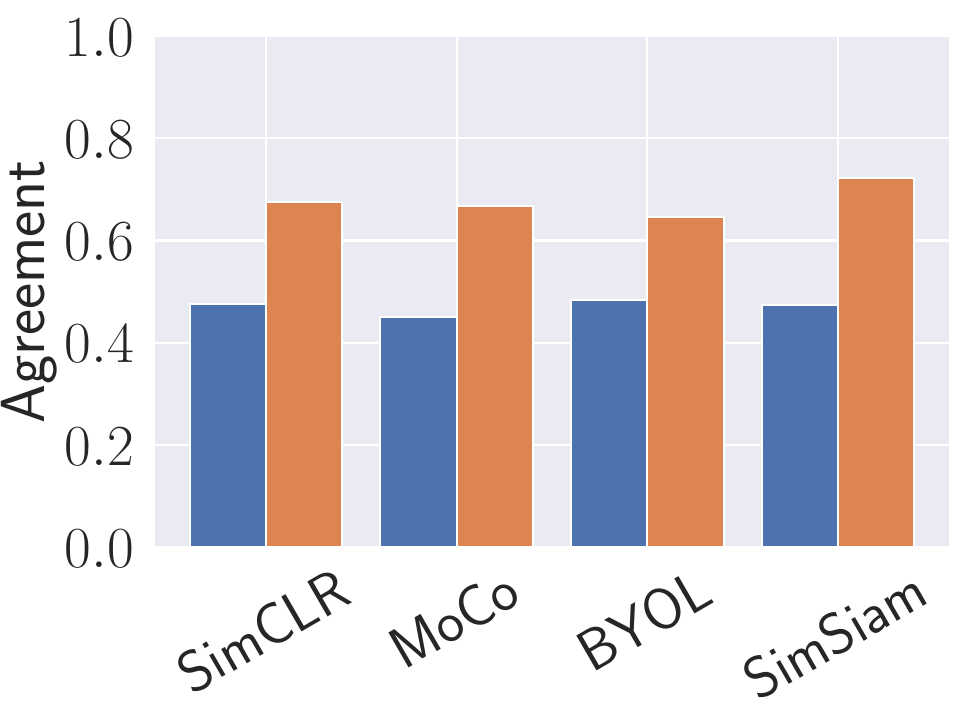}
\caption{SVHN}
\label{fig:contrastive_agreement_imagenet_stl10_svhn}
\end{subfigure}
\begin{subfigure}{0.44\columnwidth}
\includegraphics[width=\columnwidth]{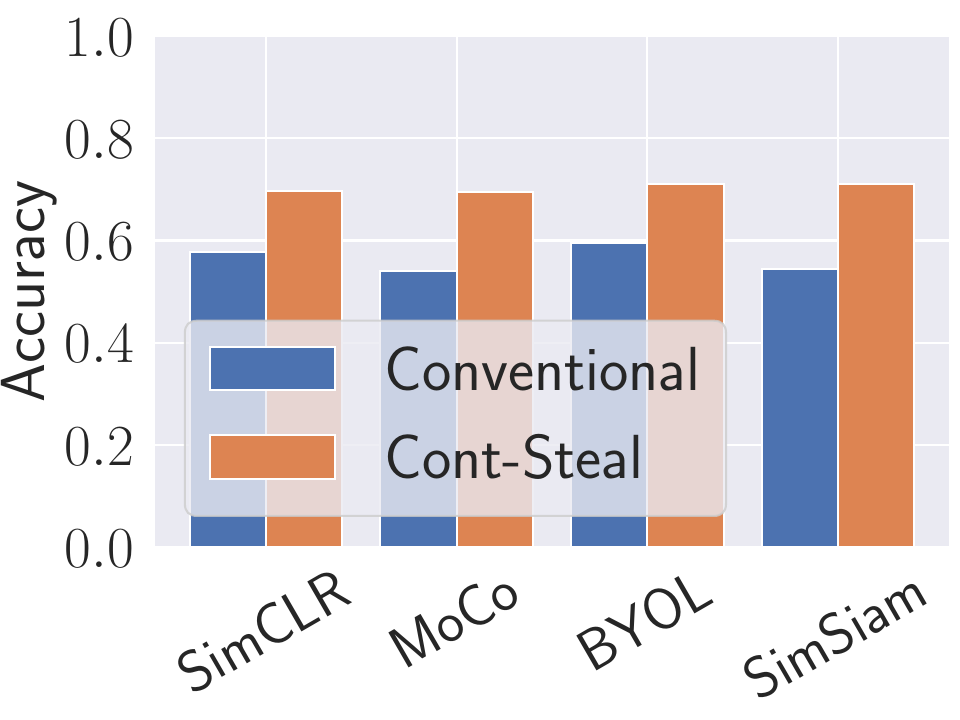}
\caption{CIFAR10}
\label{fig:contrastive_accuracy_imagenet_stl10_cifar10}
\end{subfigure}
\begin{subfigure}{0.44\columnwidth}
\includegraphics[width=\columnwidth]{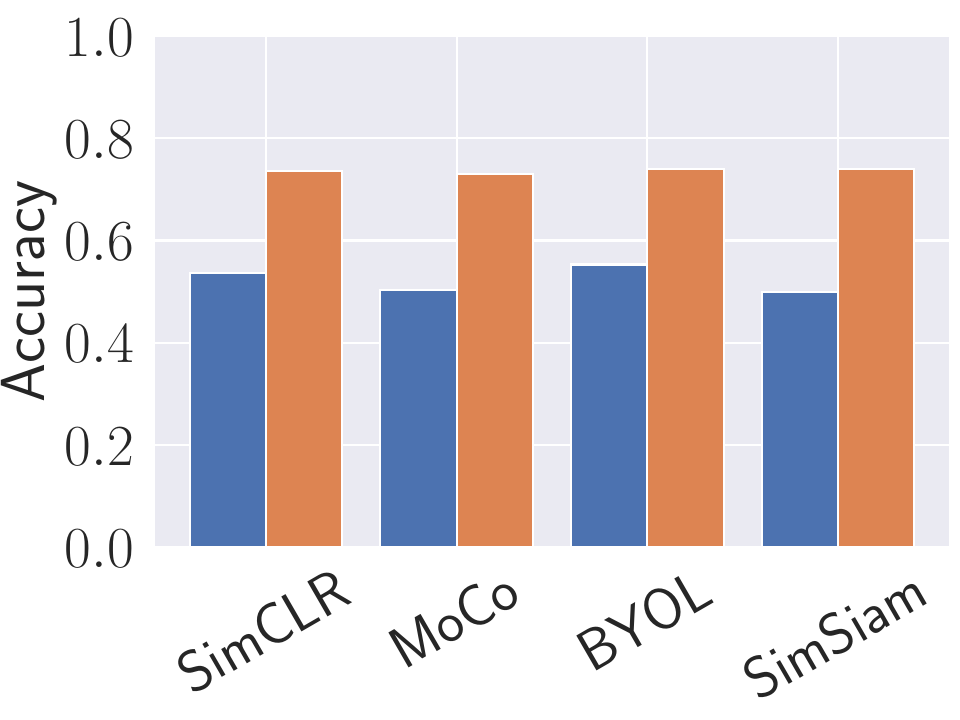}
\caption{STL10}
\label{fig:contrastive_accuracy_imagenet_stl10_stl10}
\end{subfigure}
\begin{subfigure}{0.44\columnwidth}
\includegraphics[width=\columnwidth]{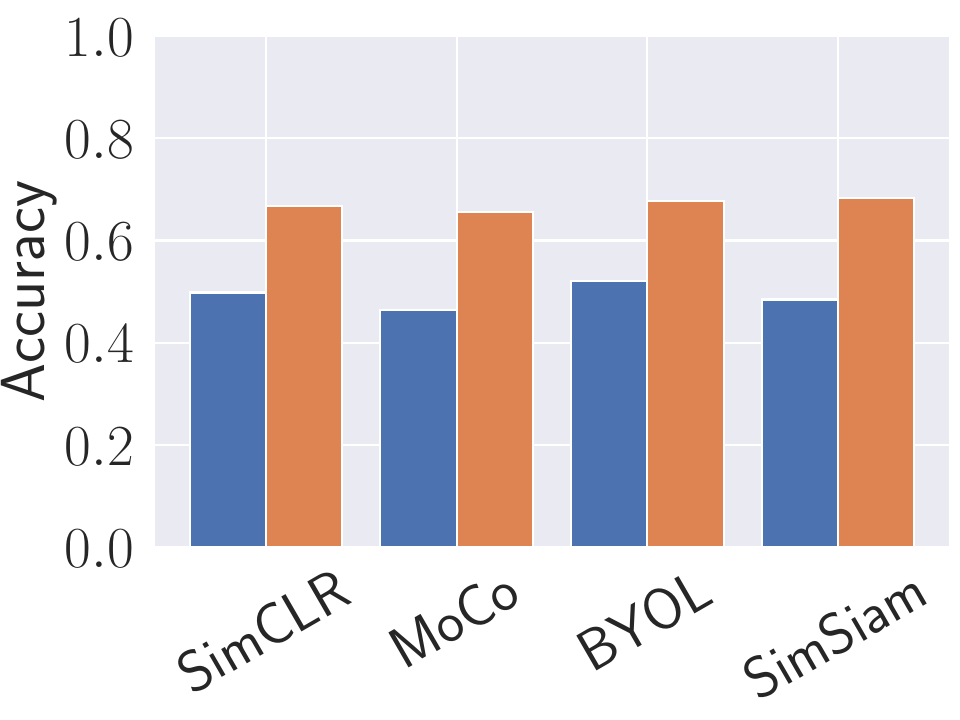}
\caption{F-MNIST}
\label{fig:contrastive_accuracy_imagenet_stl10_mnist}
\end{subfigure}
\begin{subfigure}{0.44\columnwidth}
\includegraphics[width=\columnwidth]{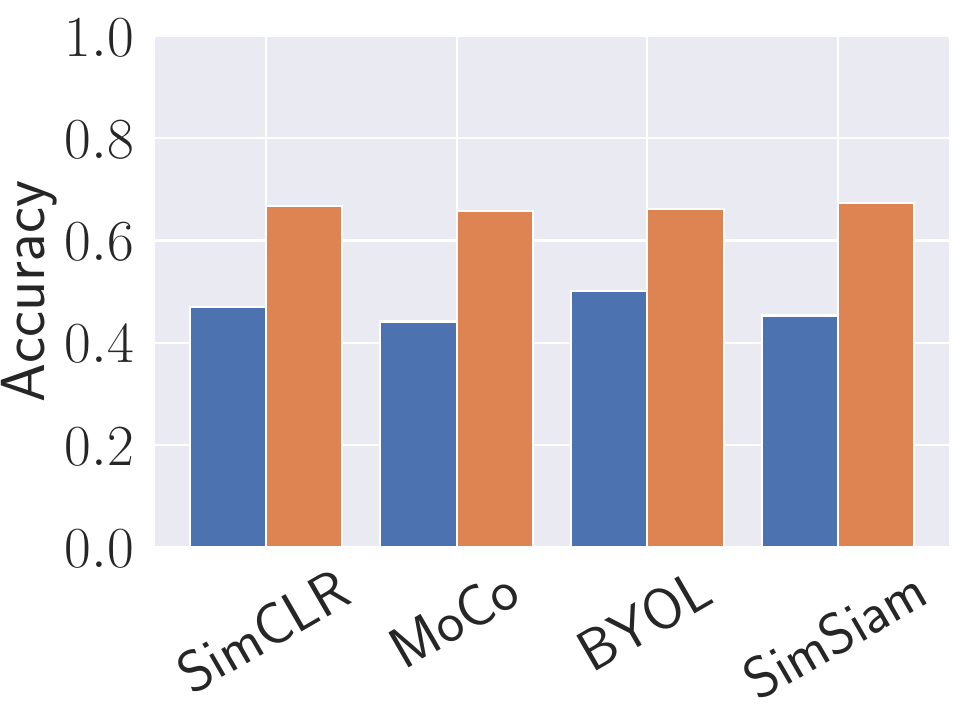}
\caption{SVHN}
\label{fig:contrastive_accuracy_imagenet_stl10_svhn}
\end{subfigure}
\caption{The performance of Cont-Steal and conventional attack against target encoders trained on ImageNet100.
The adversary uses CIFAR10, STL10, F-MNIST, and SVHN to conduct model stealing attacks.
The adversary uses F-MNIST as the downstream task to evaluate the attack performance.
The x-axis represents different kinds of the target model.
The first line's y-axis represents the agreement of the model stealing attack. 
The second line's y-axis represents the accuracy of the model stealing attack.}
\label{fig:contrastive_imagenet_stl10}
\end{figure*}

\begin{figure*}[!t]
\centering
\begin{subfigure}{0.44\columnwidth}
\includegraphics[width=\columnwidth]{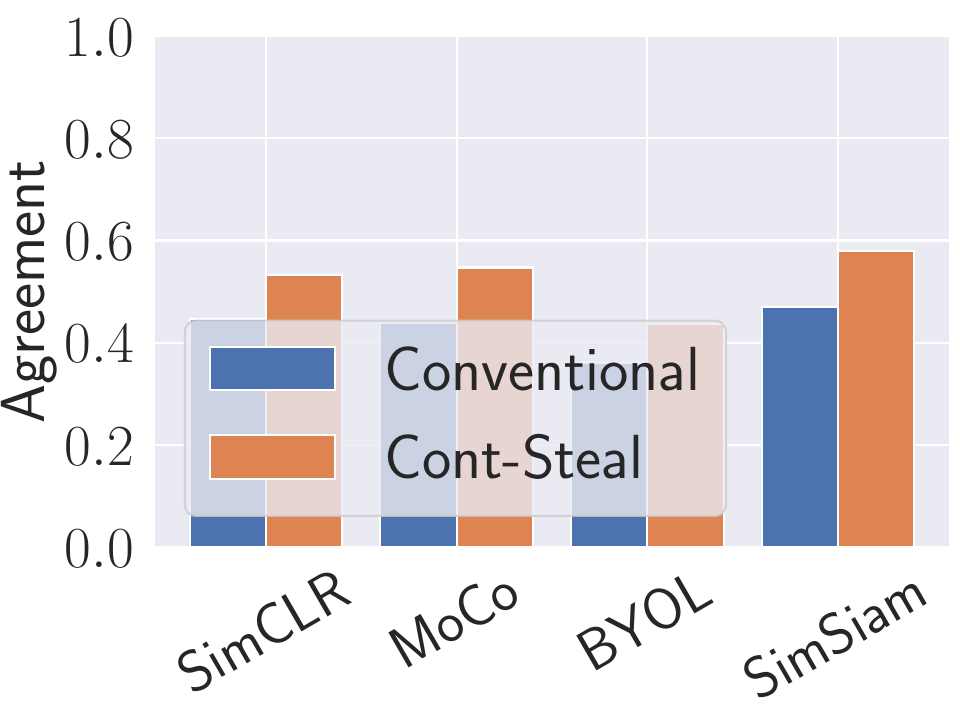}
\caption{CIFAR10}
\label{fig:contrastive_agreement_imagenet_svhn_cifar10}
\end{subfigure}
\begin{subfigure}{0.44\columnwidth}
\includegraphics[width=\columnwidth]{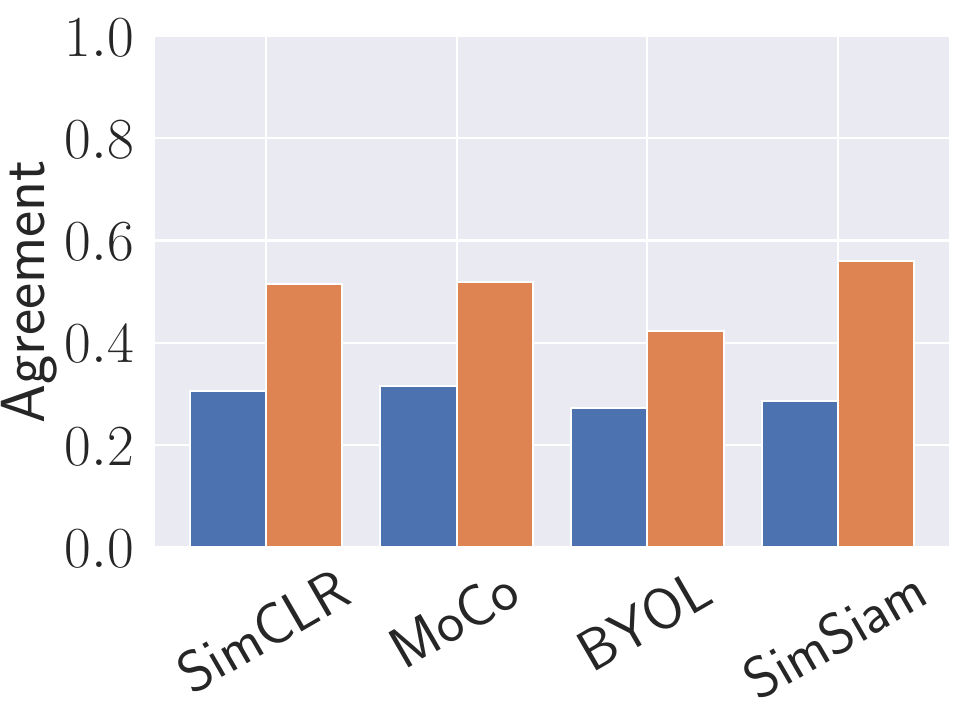}
\caption{STL10}
\label{fig:contrastive_agreement_imagenet_svhn_stl10}
\end{subfigure}
\begin{subfigure}{0.44\columnwidth}
\includegraphics[width=\columnwidth]{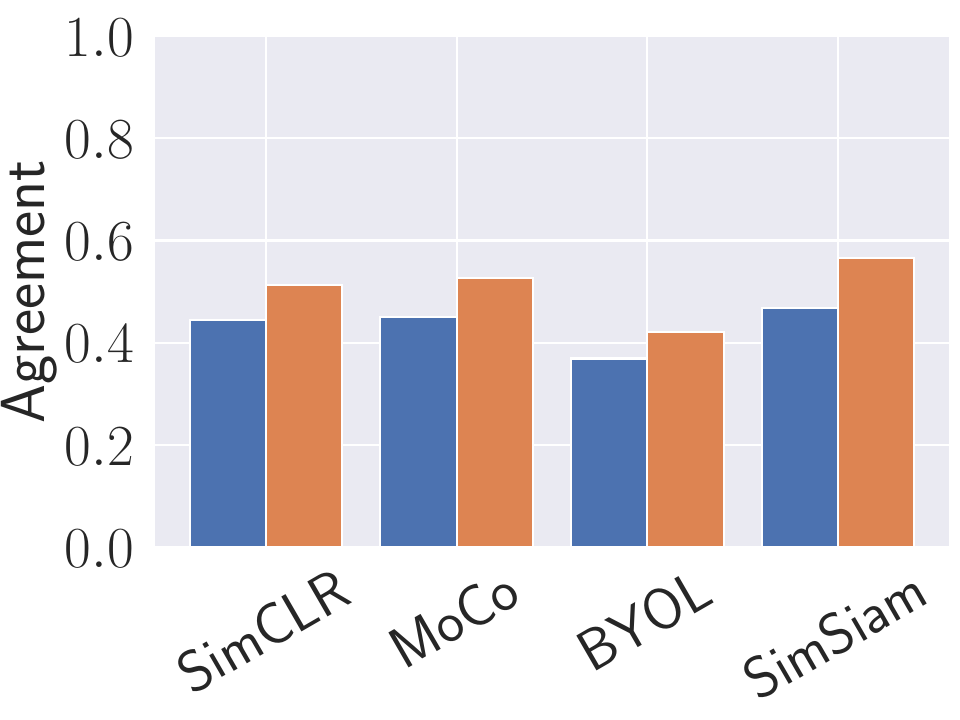}
\caption{F-MNIST}
\label{fig:contrastive_agreement_imagenet_svhn_mnist}
\end{subfigure}
\begin{subfigure}{0.44\columnwidth}
\includegraphics[width=\columnwidth]{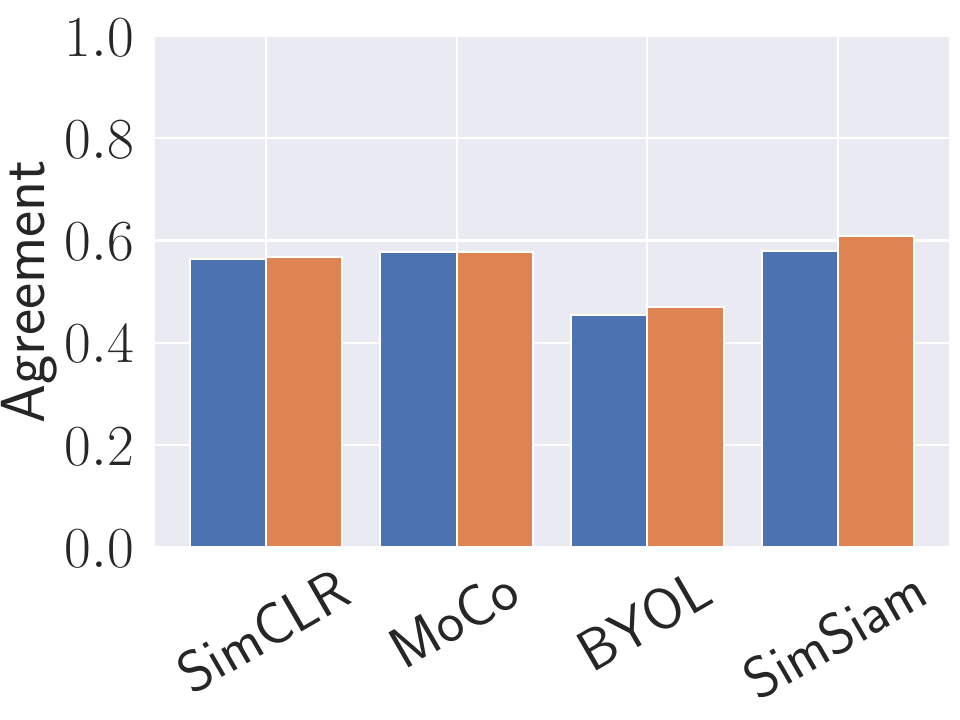}
\caption{SVHN}
\label{fig:contrastive_agreement_imagenet_svhn_svhn}
\end{subfigure}
\begin{subfigure}{0.44\columnwidth}
\includegraphics[width=\columnwidth]{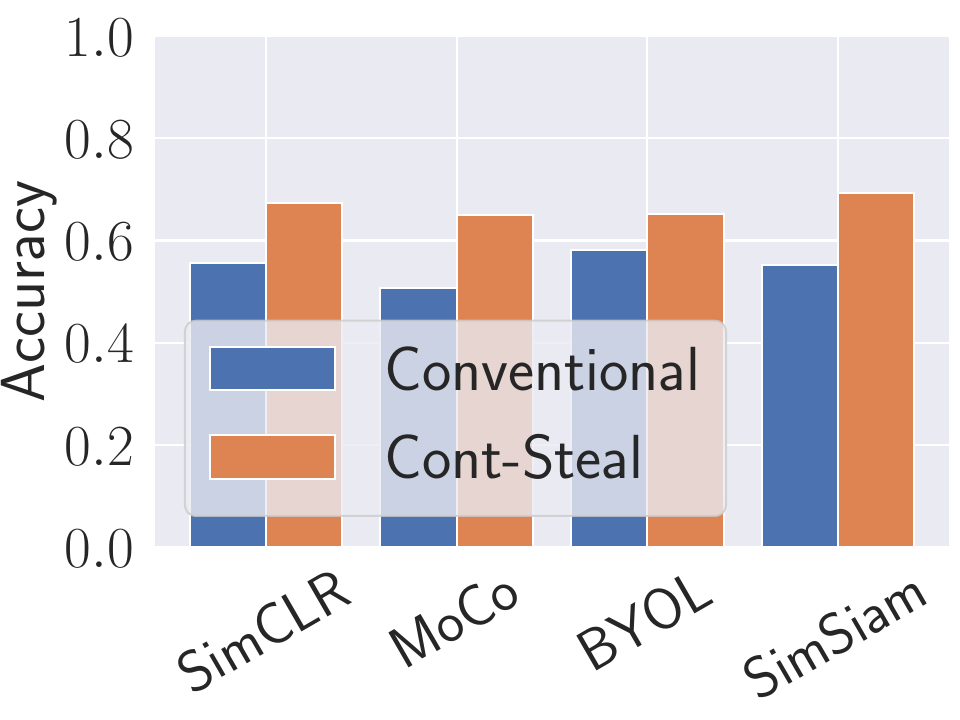}
\caption{CIFAR10}
\label{fig:contrastive_accuracy_imagenet_svhn_cifar10}
\end{subfigure}
\begin{subfigure}{0.44\columnwidth}
\includegraphics[width=\columnwidth]{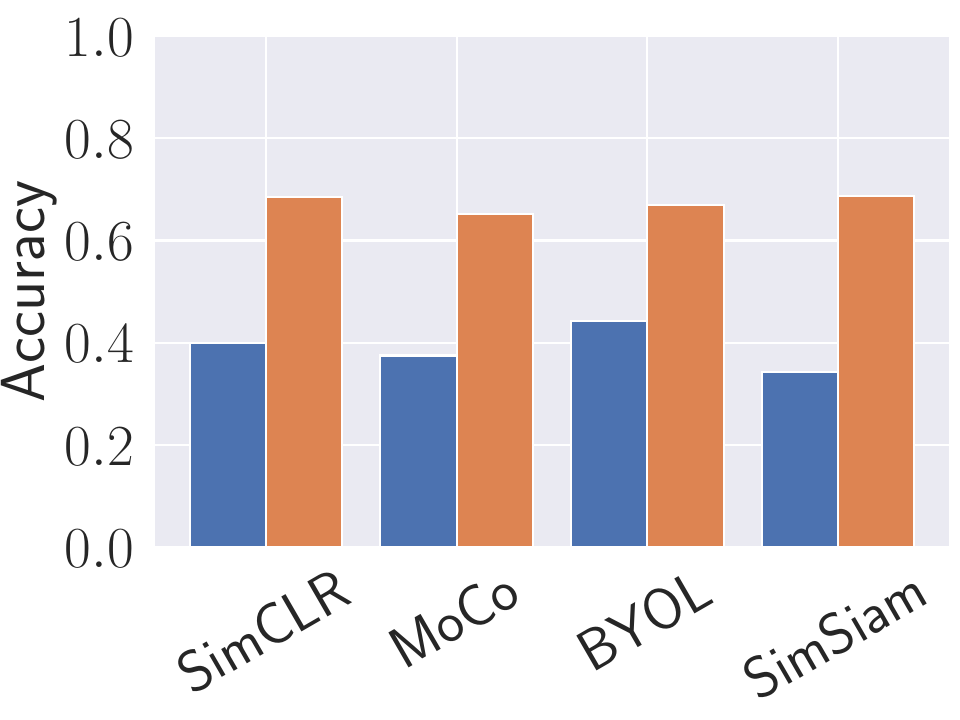}
\caption{STL10}
\label{fig:contrastive_imagenet_accuracy_svhn_stl10}
\end{subfigure}
\begin{subfigure}{0.44\columnwidth}
\includegraphics[width=\columnwidth]{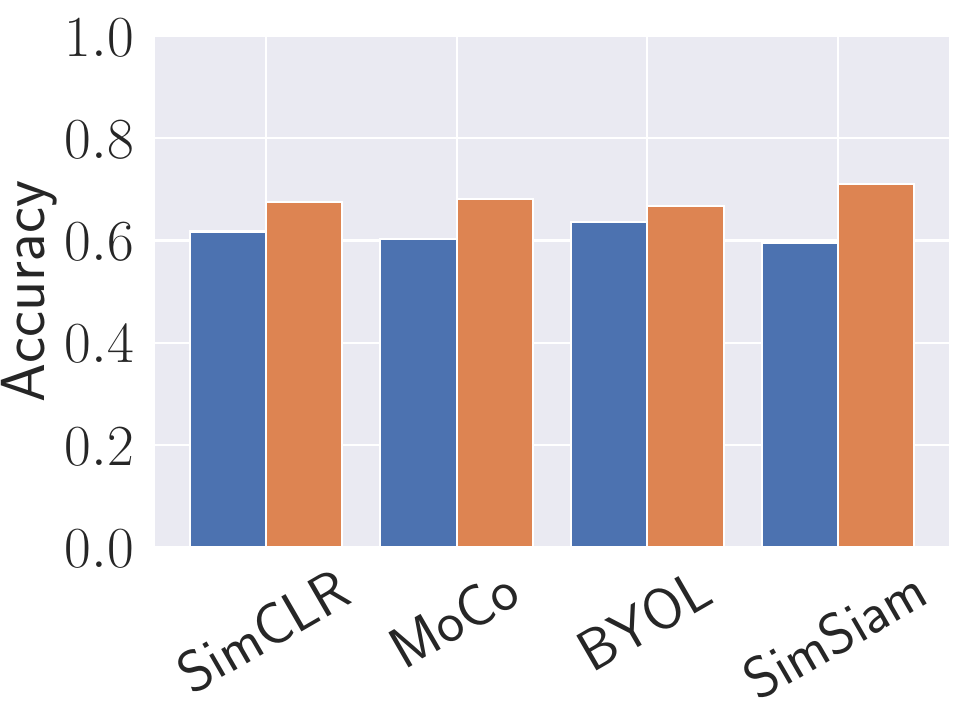}
\caption{F-MNIST}
\label{fig:contrastive_accuracy_imagenet_svhn_mnist}
\end{subfigure}
\begin{subfigure}{0.44\columnwidth}
\includegraphics[width=\columnwidth]{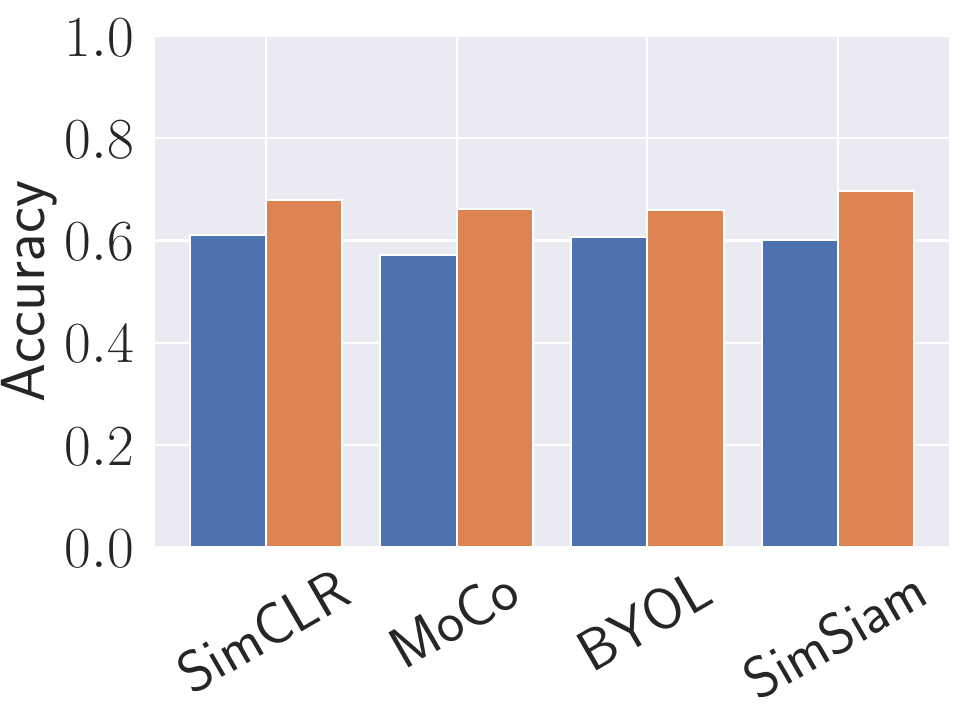}
\caption{SVHN}
\label{fig:contrastive_imagenet_accuracy_svhn_svhn}
\end{subfigure}
\caption{The performance of Cont-Steal and conventional attack against target encoders trained on ImagNet100.
The adversary uses CIFAR10, STL10, F-MNIST, and SVHN to conduct model stealing attacks.
The adversary uses SVHN as the downstream task to evaluate the attack performance.
The x-axis represents different kinds of the target model.
The first line's y-axis represents the agreement of the model stealing attack. 
The second line's y-axis represents the accuracy of the model stealing attack.}
\label{fig:contrastive_imagenet_svhn}
\end{figure*}

\autoref{fig:contrastive_cifar10_stl10}, \autoref{fig:contrastive_cifar10_mnist}, and \autoref{fig:contrastive_cifar10_svhn} show the results of the \ContSteal on target models whose encoders are pre-trained on CIFAR10 and downstream classifiers are trained on STL10, F-MNIST, and SVHN, respectively.
\autoref{fig:contrastive_imagenet_cifar10}, \autoref{fig:contrastive_imagenet_stl10}, \autoref{fig:contrastive_imagenet_svhn} show the results of the \ContSteal on target models whose encoders are pre-trained on ImageNet100 and downstream classifiers are trained on CIFAR10, STL10, and SVHN, respectively.

\begin{table}[b]
\centering
\caption{
The performance of \ContSteal and conventional attacks against state-of-the-art models.
Note that all of our target encoders are pre-trained encoders available online and downstream classifiers are trained on CIFAR10.
}
\label{table:steal-big-model}
\scalebox{0.7}
{
\begin{tabular}{c  c  c  c  c  c }
\toprule
\textbf{Surrogate Dataset} & \textbf{Metric} & \textbf{Attacks} & \textbf{ViT} & \textbf{MAE} & \textbf{CLIP} \\
\midrule
Original performance & Accuracy & NaN & 0.896 & 0.900 & 0.903  \\
\midrule
\multirow{2}*{CIFAR10} & Agreement & Conventional & 0.745 & 0.555 & 0.815  \\
~ & Agreement & \ContSteal & \textbf{0.967} & \textbf{0.712} & \textbf{0.889} \\
\midrule
\multirow{2}*{STL10} & Agreement & Conventional & 0.553 & 0.451 & 0.550  \\
~ & Agreement & \ContSteal & \textbf{0.942} & \textbf{0.624} & \textbf{0.905} \\
\midrule
\multirow{2}*{SVHN} & Agreement & Conventional & 0.587 & 0.419 & 0.578 \\
~ & Agreement & \ContSteal & \textbf{0.944} & \textbf{0.548} & \textbf{0.893} \\
\midrule
\multirow{2}*{F-MNIST} & Agreement & Conventional & 0.602 & 0.395 & 0.465  \\
~ & Agreement & \ContSteal & \textbf{0.696} & \textbf{0.501} & \textbf{0.598} \\
\bottomrule
\end{tabular}
}
\end{table}

\subsection{Attacks Performance on Other Visual Models}
\label{appendix:other-model-performance}

Apart from four contrastive models we tried in the paper, we also conduct our \ContSteal on other large, state-of-the-art models such as ViT and CLIP.
We show that \ContSteal can perform very well on ViT, MAE, and the image encoder of CLIP in~\autoref{table:steal-big-model}.
The results demonstrate the scalability of \ContSteal.

\subsection{Compare With Other Existing Works}
\label{subsection:compare}

Note that we are the first work to systematically propose model stealing attacks against image encoders.
There are also some parallel and follow-up works on this domain proposed after our work.
Here, we compare our works with other existing methods.
The main difference between our work and recent works is our designed contrastive steal loss and the usage of data augmentation.
Compared to StolenEncoder~\cite{LJLG22}, our loss focuses on the comparison of positive and negative samples, while StolenEncoder focuses on the combination of augmentation and non-augmentation loss.
The main difference between our work and the methods listed in~\cite{DDKGP22} is that 1) we leverage data augmentation as part of the methods, and 2) we design the loss function ourselves to consider more negative examples compared to the INFONCE loss.
We show in \autoref{table:compare-recent-work} that our method works better.
Note that KL divergence is also a loss function used by knowledge distillation.
As knowledge distillation is a similar task to model stealing, we also report the results of KL divergence.

\begin{table}[!t]
\centering
\caption{
Dataset inference performance on \ContSteal.
}
\label{table:defense}
\scalebox{0.7}
{
\begin{tabular}{ c  c  c  c  }
\toprule
\textbf{Model} & \textbf{Dataset} & \textbf{$S(\cdot, E_T)$} & \textbf{$C(\cdot, E_T)$} \\
\midrule
Target Encoder & CIFAR10 & 1.000 & 1.000 \\
\midrule
Surrogate Encoder & SVHN & 0.412 & 0.393 \\
\midrule
Surrogate Encoder (fine-tuning) & SVHN & 0.17 & 0.00 \\
\midrule
Independent Encoder & SVHN & 0.11 & 0.00 \\
\bottomrule
\end{tabular}
}
\end{table}

\subsection{More Defenses.}
\label{subsection:more-defense}

We implement the dataset inference defense in~\cite{DDKDGCBP22} (see \autoref{table:defense}).
$S(\cdot, E_T)$/$C(\cdot, E_T)$ represents the mutual information/cosine similarity between the given model and the target model (the higher, the more similar).
Note that the surrogate encoder will be fine-tuned for downstream tasks.
We find the fine-tuning process~\cite{YSBZ23,GSKGRF19} will disable the defense.
Normally, the open-source encoders are trained on very large public datasets instead of limited private datasets, which makes the defense less practical.

\begin{table}[!t]
\centering
\caption{
The comparison of \ContSteal and other existing works.
Both the target encoder and downstream classifier are trained on CIFAR10.
Note that our results are different from the original paper of~\cite{DDKGP22} because we test the surrogate encoder on the original task.
}
\label{table:compare-recent-work}
\scalebox{0.7}
{
\begin{tabular}{ c  c  c  c  c }
\toprule
~ & \multicolumn{2}{c}{\textbf{CIFAR10}}  & \multicolumn{2}{c}{\textbf{STL10}} \\
 \midrule
 ~ & Agreement & Accuracy & Agreement & Accuracy \\
 \midrule
 Baseline & 0.785 & 0.790 & 0.499 & 0.500 \\
\midrule
StolenEncoder & 0.811 & 0.808 & 0.766 & 0.767 \\
\midrule
KL Divergence  & 0.213 & 0.203 & 0.178 & 0.162 \\
\midrule
INFONCE  & 0.826 & 0.828 & 0.806 & 0.797\\
\midrule
\ContSteal  & \textbf{0.845} & \textbf{0.854} & \textbf{0.829} & \textbf{0.828} \\
\bottomrule
\end{tabular}
}
\end{table}


\end{document}